\documentclass[twocolumn]{aastex701}
\usepackage{hyperref}
\usepackage{extdash}

\shorttitle{FEEDBACK: M16 H~II Region}
\shortauthors{Karim et al.}

\def\hii{\hbox{{\rm H {\scriptsize II}}}}

\newcommand{\hone}{{\rm H}}
\newcommand{\htwo}{\ifmmode{{\rm H_2}} \else{H$_2$\/}\fi}
\newcommand{\nh}{\ifmmode{N(\hone)} \else{$N(\hone)$\/}\fi}
\newcommand{\nht}{\ifmmode{N(\htwo)} \else{$N(\htwo)$\/}\fi}
\newcommand{\ntot}{\ifmmode{N_\hone} \else{$N_\hone$\/}\fi}

\newcommand{\cii}{\ifmmode{{\rm [C\ II]}} \else{[C\ II]\/}\fi}
\newcommand{\oi}{\ifmmode{{\rm [O\ I]}} \else{[O\ I]\/}\fi}
\newcommand{\twcii}{\ifmmode{{\rm [^{12}C\ II]}} \else{[$^{12}$C\ II]\/}\fi}
\newcommand{\thcii}{\ifmmode{{\rm [^{13}C\ II]}} \else{[$^{13}$C\ II]\/}\fi}
\newcommand{\cp}{\ifmmode{{\rm C^{+}}} \else{C$^{+}$\/}\fi}
\newcommand{\twcp}{\ifmmode{{\rm ^{12}C^{+}}} \else{$^{12}$C$^{+}$\/}\fi}
\newcommand{\thcp}{\ifmmode{{\rm ^{13}C^{+}}} \else{$^{13}$C$^{+}$\/}\fi}

\newcommand{\twcoA}{\ifmmode{^{12}{\rm CO}} \else{$^{12}$CO\/}\fi}
\newcommand{\thcoA}{\ifmmode{^{13}{\rm CO}} \else{$^{13}$CO\/}\fi}
\newcommand{\ceighteenoA}{\ifmmode{{\rm C^{18}O}} \else{C$^{18}$O\/}\fi}

\newcommand{\jmton}[2]{\ifmmode{{\rm J=#1--#2}} \else{J=#1--#2\/}\fi}

\newcommand{\twco}{\ifmmode{^{12}{\rm CO(\jmton{1}{0})}} \else{$^{12}$CO(\jmton{1}{0})\/}\fi}
\newcommand{\thco}{\ifmmode{^{13}{\rm CO(\jmton{1}{0})}} \else{$^{13}$CO(\jmton{1}{0})\/}\fi}
\newcommand{\ceighteeno}{\ifmmode{{\rm C^{18}O(\jmton{1}{0})}} \else{C$^{18}$O(\jmton{1}{0})\/}\fi}
\newcommand{\twcott}{\ifmmode{^{12}{\rm CO(\jmton{3}{2})}} \else{$^{12}$CO(\jmton{3}{2})\/}\fi}
\newcommand{\thcott}{\ifmmode{^{13}{\rm CO(\jmton{3}{2})}} \else{$^{13}$CO(\jmton{3}{2})\/}\fi}
\newcommand{\vlsr}{\ifmmode{\rm V_{LSR}} \else{V$_{\rm LSR}$\/}\fi}
\newcommand{\nexpo}[2]{\ifmmode{#1 \times 10^{#2}}\else{$#1 \times 10^{#2}$}\fi}
\newcommand{\expo}[1]{\ifmmode{10^{#1}}\else{$10^{#1}$}\fi}

\newcommand{\Tex}{\ifmmode{{T_{\rm ex}}} \else{$T_{\rm ex}$\/}\fi}
\newcommand{\losD}{\ifmmode{1740~{\rm pc}} \else{1740~pc}\fi}
\newcommand\kms{\ifmmode{\,{\rm km~s^{-1}}}\else{{${\rm km~s^{-1}}$}}\fi}
\newcommand\solMass{\ifmmode{\,{\rm M_{\odot}}}\else{{${\rm M_{\odot}}$}}\fi}
\newcommand{\wrt}{with respect to}

\newcommand{\cm}{\ifmmode{\text{cm}} \else{cm}\fi}
\newcommand{\cmsq}{\ifmmode{\text{cm}^{-2}} \else{cm$^{-2}$\/}\fi}
\newcommand{\cc}{\ifmmode{\text{cm}^{-3}} \else{cm$^{-3}$\/}\fi}
\newcommand{\Kcc}{\ifmmode{\text{K~cm}~^{-3}} \else{K~\cc\/}\fi}
\newcommand{\pressure}[1]{\ifmmode{P_{\rm #1} / k_{\rm B}} \else{$\pressure{#1}$\/}\fi}

\graphicspath{{./}{figures/}}

\begin{document}

\title{SOFIA FEEDBACK Survey: The Eagle Nebula in \cii\ and Molecular Lines}

\author[0000-0001-8844-5618,gname='Ramsey',sname='Karim']{Ramsey L. Karim}
\affiliation{University of Maryland, Department of Astronomy, Room 1113 PSC Bldg. 415, College Park, MD 20742-2421, USA}
\email[show]{rlkarim@terpmail.umd.edu}

\author[0000-0002-7269-342X,gname='Marc',sname='Pound']{Marc W. Pound}
\affiliation{University of Maryland, Department of Astronomy, Room 1113 PSC Bldg. 415, College Park, MD 20742-2421, USA}
\email{mpound@umd.edu}

\author[0000-0003-0306-0028,gname='Alexander',sname='Tielens']{Alexander G.G.M. Tielens}
\affiliation{University of Maryland, Department of Astronomy, Room 1113 PSC Bldg. 415, College Park, MD 20742-2421, USA}
\email{tielens@strw.leidenuniv.nl}

\author[0000-0001-5540-2822,gname='Jelle',sname='Kaastra']{Jelle S. Kaastra}
\affiliation{SRON Netherlands Institute for Space Research, Niels Bohrweg 4, 2333 CA Leiden, The Netherlands}
\affiliation{Leiden Observatory, Leiden University, PO Box 9513, 2300 RA Leiden, The Netherlands}
\email{J.S.Kaastra@sron.nl}

\author[0000-0001-8081-9152,gname='Leisa',sname='Townsley']{Leisa K. Townsley}
\affiliation{Department of Astronomy \& Astrophysics, 525 Davey Laboratory, Pennsylvania State University, University Park, PA 16802, USA}
\email{lkt4@psu.edu}

\author[0000-0002-7872-2025,gname='Patrick',sname='Broos']{Patrick S. Broos}
\affiliation{Department of Astronomy \& Astrophysics, 525 Davey Laboratory, Pennsylvania State University, University Park, PA 16802, USA}
\email{patrick.broos@icloud.com}

\author[0000-0003-4260-2950,gname='Maitraiyee',sname='Tiwari']{Maitraiyee Tiwari}
\affiliation{University of Maryland, Department of Astronomy, Room 1113 PSC Bldg. 415, College Park, MD 20742-2421, USA}
\affiliation{Max-Planck Institute for Radioastronomy, Auf dem Hügel, D-53121 Bonn, Germany}
\email{maitraiyee.tiwari@gmail.com}

\author[0000-0002-0915-4853,gname='Lars',sname='Bonne']{Lars Bonne}
\affiliation{SOFIA Science Center, USRA, NASA Ames Research Center, M.S. N232-12, Moffett Field, CA 94035, USA}
\email{lmlbastro@gmail.com}

\author[0000-0002-7640-4998,gname='Umit',sname='Kavak']{Ümit Kavak}
\affiliation{Leiden Observatory, Leiden University, PO Box 9513, 2300 RA Leiden, The Netherlands}
\email{umitkavak34@gmail.com}

\author[0000-0003-0030-9510,gname='Mark',sname='Wolfire']{Mark G. Wolfire}
\affiliation{University of Maryland, Department of Astronomy, Room 1113 PSC Bldg. 415, College Park, MD 20742-2421, USA}
\email{mwolfire@umd.edu}

\author[0000-0003-3485-6678,gname='Nicola',sname='Schneider']{Nicola Schneider}
\affiliation{I. Physikalisches Institut, Universität zu Köln, Zülpicher Str. 77, D-50937 Köln, Germany}
\email{nschneid@ph1.uni-koeln.de}

\author[0000-0003-2555-4408,gname='Robert',sname='Simon']{Robert Simon}
\affiliation{I. Physikalisches Institut, Universität zu Köln, Zülpicher Str. 77, D-50937 Köln, Germany}
\email{simonr@ph1.uni-koeln.de}

\author[0000-0002-1708-9289,gname='Rolf',sname='Gusten']{Rolf Güsten}
\affiliation{Max-Planck Institute for Radioastronomy, Auf dem Hügel, D-53121 Bonn, Germany}
\email{rguesten@mpifr-bonn.mpg.de}

\author[0000-0001-7658-4397,gname='Jurgen',sname='Stutzki']{Jürgen Stutzki}
\affiliation{I. Physikalisches Institut, Universität zu Köln, Zülpicher Str. 77, D-50937 Köln, Germany}
\email{stutzki@ph1.uni-koeln.de}

\author[0000-0002-4903-9542,gname='Marc',sname='Mertens']{Marc Mertens}
\affiliation{Max-Planck Institute for Radioastronomy, Auf dem Hügel, D-53121 Bonn, Germany}
\email{mmertens@mpifr-bonn.mpg.de}

\author[0000-0002-2155-3259,gname='Oliver',sname='Ricken']{Oliver Ricken}
\affiliation{Max-Planck Institute for Radioastronomy, Auf dem Hügel, D-53121 Bonn, Germany}
\email{oricken@mpifr-bonn.mpg.de}

\author[0000-0003-4516-3981,gname='Friedrich',sname='Wyrowski']{Friedrich Wyrowski}
\affiliation{Max-Planck Institute for Radioastronomy, Auf dem Hügel, D-53121 Bonn, Germany}
\email{wyrowski@mpifr-bonn.mpg.de}

\author[0000-0002-8876-0690,gname='Lee',sname='Mundy']{Lee G. Mundy}
\affiliation{University of Maryland, Department of Astronomy, Room 1113 PSC Bldg. 415, College Park, MD 20742-2421, USA}
\email{lgm@umd.edu}

\begin{abstract}
    We characterize the physical conditions and energy budget of the M16 \hii\ region using SOFIA FEEDBACK observations of the \cii\ 158~$\mu$m line.
    The O stars in the $\sim \expo{4}~\solMass$ NGC~6611 cluster powering this \hii\ region have blown at least 2 cavities into the giant molecular cloud: the large M16 cavity and the small N19 bubble.
    We detect the spectroscopic signature of an expanding photodissociation region shell towards N19, and traces of a thin, fragmented expanding shell towards M16.
    Our \cii\ observations are resolved to 0.5~\kms\ and 15.5\arcsec\ and analyzed alongside similarly resolved CO \jmton{3}{2} observations as well as archival data ranging from the radio to X-ray tracing a variety of gas phases spanning dense $\sim$10~K molecular gas, \expo{4}~K photoionized gas, and million-K collisionally ionized plasma.
    With this dataset, we evaluate the coupling of energetic feedback from NGC~6611 and the O9 V star within N19 to the surrounding gas.
    Winds from NGC~6611 have blown a 20~pc radius cavity constrained in size along the major axis of the natal giant molecular filament, and much of the mechanical wind energy ($>$90\%) has escaped through breaches in the $\lesssim$\expo{4}~\solMass\ shell.
    Reservoirs of dense gas remain within a few parsecs of the cluster.
    N19, younger than M16 by $\gtrsim 10^6$~yr, is driven by a combination of mechanical wind energy and thermal pressure from photoionized gas and has swept up $\sim$\expo{3}~\solMass\ into neutral atomic and molecular shells.
\end{abstract}

\section{Introduction} \label{sec:introduction}
Massive stars form in massive complexes of dense gas \citep{Motte2018ARA&A..56...41M_review}, and upon illumination they inject vast amounts of radiative and mechanical energy back into those cloud complexes.
Interest in pre-supernova feedback has increased as it has become evident that a few million years of ionizing radiation can dissipate a cloud \citep{He2019MNRAS.489.1880H, Bonne2023A&A...679L...5B} or that a cluster can eject more than a supernova's worth of energy ($\sim$10$^{51}$~erg) in winds over its lifetime \citep{Tiwari2021ApJ...914..117T}.
Protostar, main-sequence, and evolved stellar feedback all set the stage for the effect the first supernova has on the surrounding cloud and intercloud environment, and understanding their effects is key to a complete understanding of the energetic life cycle of the interstellar medium.

The massive members of a cluster quickly reach their main sequence luminosities \citep{Zinnecker2007ARA&A..45..481Z} and ionize the surrounding gas which causes rapid pressure-driven Spitzer expansion \citep{Spitzer1978ppim.book.....S}.
These stars also blow supersonic winds which collide with the photoionized \hii\, injecting momentum directly through collision \citep{Geen2022MNRAS.509.4498G}, and adding further pressure as the shocked winds form a million-Kelvin plasma and fill an interior cavity inside the \hii\ region \citep{Weaver1977ApJ...218..377W}.
These three phenomena work together to inflate a multi-phase cavity around the cluster, leading to much debate and study as to whether winds or photoionization dominate this action \citep{Lopez2014ApJ...795..121L, Ngoumou2015ApJ...798...32N, He2019MNRAS.489.1880H, Pabst2019Natur.565..618P, Pabst2021A&A...651A.111P, Tiwari2021ApJ...914..117T, Bonne2022ApJ...935..171B_RCW36}.

Just beyond the edge of the \hii\ region lies a layer which receives no H-ionizing extreme ultraviolet (EUV; $h\nu > 13.6$~eV) radiation but still receives abundant far ultraviolet (FUV; $6 < h\nu < 13.6$~eV) radiation which can dissociate molecules like H$_2$ and CO and ionize atomic carbon.
The influence of FUV radiation upon these photodissociation regions (PDRs) sets them apart chemically from the further-away and poorly-illuminated neutral and molecular phases \citep{Hollenbach1997ARA&A..35..179H, Wolfire2022ARA&A..60..247W}.
PDR gas is heated photoelectrically as FUV radiation knocks electrons off large, complex organic molecules called polycyclic aromatic hydrocarbons (PAHs; \citealt{Tielens2008ARA&A..46..289T}).
Collisional excitation and radiative de-excitation of neutral oxygen (\oi\ 63 and 146~\micron) and ionized carbon (\cii\ 158~\micron) fine-structure lines, the dominant cooling lines in atomic gas around star forming regions, cool the gas \citep{Wolfire2022ARA&A..60..247W}.
With these far-infrared (FIR) transitions, the PDR is traced and its gas dynamics are resolved.
Due to FIR opacity from water in Earth's atmosphere, these PDR lines can only be detected high above the ground, either from space observatories like Herschel or from high altitude using balloons or airborne observatories such as the Stratospheric Observatory for Infrared Astronomy (SOFIA).

PDR tracers such as \cii\ are used to determine the conditions in the PDR, the FUV-illuminated interface between the photoionized \hii\ region and the largely unilluminated molecular gas behind it, and can shed light on whether feedback energy from within the cavity is coupling with the neutral and molecular gas outside it.
This leads to the greater question of how efficiently and in what particular ways stellar feedback is injected back into its environment.

This work focuses on M16, the Eagle Nebula, an \hii\ region driven by the $\sim$10$^4~M_{\odot}$ \citep{Pfalzner2009A&A...498L..37P} cluster NGC~6611 with a most massive member of type O3.5 V((f)) \citep{Stoop2023AA...670A.108S}.
The region, at a heliocentric distance of 1740~pc \citep{2019ApJ...870...32K}, is surrounded by several giant molecular clouds and giant molecular filaments (GMCs, GMFs; \citealt{Xu2019A&A...627A..27X, Zhan2016RAA....16...56Z}), at least one of which seems to have birthed the cluster \citep{Hill2012A&A...542A.114H, Nishimura2021PASJ...73S.285N}.
M16 is well studied across the electromagnetic spectrum, in part because it harbors the iconic Pillars of Creation, whose Hubble Space Telescope (HST) and JWST images are widely recognized \citep{Hester1996AJ....111.2349H}.
The bright PDRs associated with the Pillars were studied in \cii, \oi, and molecular lines by \citet{Karim2023AJ....166..240K}, and here we extend the same general methods to the wider M16 region.

We present a multiwavelength analysis using velocity-resolved and continuum observations of the M16 massive star-forming region, tracing multiple phases of gas from the $10^6$~K shocked wind plasma to the FUV-irradiated PDRs to cold, dense molecular gas.
Our velocity-resolved analysis centers on the brightest $\sim$30\arcmin-wide (15~pc) region of M16, and we contextualize these using archival degree-scale continuum images which reveal the faint outer reaches of NGC~6611's influence.
We introduce new and archival observations in Section~\ref{sec:observations} and discuss the structure of M16 using the archival data in Section~\ref{sec:m16-struct}.
In Section~\ref{sec:results}, we present our new \cii\ and CO line spectra and we discuss their implications for the 3-dimensional geometry of the region in Section~\ref{sec:geom}.
In Sections~\ref{sec:coldens} and \ref{sec:stars}, we derive column densities from the \cii\ and CO lines and then numerically estimate the feedback capacity of NGC~6611 based on observed catalogs of its members.
We compare stellar feedback capacity to the energies and pressures of the various phases of gas in Section~\ref{sec:energy} to evaluate how well stellar feedback has coupled to the gas.
We conclude the paper with a discussion in Section~\ref{sec:discussion} of whether stellar wind or photoionization appears to drive bubble expansion, and how M16 compares to other expanding \hii\ region bubbles in the Galaxy.

\section{Observations} \label{sec:observations}
\subsection{SOFIA} \label{sec:obs-sofia}
The \cii\ line was observed towards M16 between 2019 and 2022 on flights from Palmdale, California and Tahiti.
The 158~\micron\ $^{2}$P$_{3/2} \rightarrow$ $^{2}$P$_{1/2}$ transition was mapped on-the-fly using upGREAT\footnote{upGREAT and GREAT were developed by the MPI für Radioastronomie and the KOSMA/Universität zu K\"oln, in cooperation with the MPI für Sonnensystemforschung and the DLR Institut f\"ur Planetenforschung.} \citep{Risacher2018JAI.....740014R}, a 7-pixel heterodyne receiver with a fast Fourier transform spectrometer (FFTS) backend with 4~GHz instantaneous bandwidth and 0.244~MHz frequency resolution \citep{Klein2012A&A...542L...3K}.
Atmospheric calibration was done with the GREAT pipeline \citep{Guan2012A&A...542L...4G}

The nominal angular and spectral resolutions of the \cii\ data are 14.1\arcsec\ and 0.04~\kms, and here we use a \cii\ data cube with a spatial resolution of 15.5\arcsec, a grid of 5\arcsec, and a spectral binning of 0.5~\kms\ for increased signal-to-noise ratio.
The noise RMS in one channel is typically 1.0~K.
All spectra are presented on a main beam brightness temperature scale $T_{\rm MB}$ using an average main beam efficiency of $\eta_{\rm MB} = 0.65$.
The forward efficiency is $\eta_{\rm f}$ = 0.97.
\citet{Schneider2020PASP..132j4301S} present further observational details.

The map used in this study contains 6 more tiles (squares with 7.26\arcmin\ sides into which observations were divided) than the map presented by \citet{Karim2023AJ....166..240K}.
The observations for one of these tiles, to the west of the Pillars of Creation, were not fully completed due to SOFIA's decommissioning in late 2022.
The affected tile has a higher RMS noise, around 2.5~K, and contains stripe artifacts along the scanning axis.
This tile is coincident with the region within M16 where we search for an expanding shell signature.
The \cii\ signal from the shell is detected at $\sim$2--3~$\sigma$ due to the increased noise.

\subsection{APEX}
M16 was mapped in the \jmton{3}{2} transition of \twcoA\ and \thcoA\ using the LAsMA spectrometer on the APEX\footnote{APEX, the Atacama Pathfinder Experiment is a collaboration between the Max-Planck-Institut für Radioastronomie, Onsala Space Observatory (OSO), and the European Southern Observatory (ESO).} telescope \citep{Gusten2006A&A...454L..13G}.
We use data cubes with a 9.1\arcsec\ pixel and an 18.2\arcsec\ beam after gridding and 0.1~\kms\ spectral bins.
Spectra are calibrated in T$_{\rm MB}$ with a main-beam efficiency $\eta_{\rm MB} = 0.68$ at 345.8~GHz.
Further observational details are given by \citet{Karim2023AJ....166..240K}.

The OFF position is slightly contaminated near $\vlsr = 21.1~\kms$.
This makes a small divot in the spectra but does not affect our analysis and conclusions.

\subsection{CO (1--0) Line Observations}
We use the publicly available \twco, \thco, and \ceighteeno\ line observations made by \cite{Xu2019A&A...627A..27X} using the 13.7~m radio telescope at the Purple Mountain Observatory (PMO) in Delingha (data obtained via VizieR; \citealt{pmo_vizier:J/A+A/627/A27}).
The observations have a 53\arcsec\ beam, about 3$\times$ the size of the SOFIA \cii\ and APEX CO (\jmton{3}{2}) beams.
The 0.3~\kms velocity resolution, 0.2~K sensitivity, and half square degree field of view make it a valuable compliment to our observations.

We use the publicly available\footnote{\url{http://jvo.nao.ac.jp/portal/}} \twco, \thco, and \ceighteeno\ line observations from the FUGIN survey \citep{Umemoto2017PASJ...69...78U} made with the 45~m Nobeyama\footnote{The 45-m radio telescope is operated by the Nobeyama Radio Observatory, a branch of the National Astronomical Observatory of Japan.} radio telescope.
The observations have a 20\arcsec\ beam, 0.65~\kms\ velocity resolution, and $\sim$1~K sensitivity.
Scanning artifacts appear over the M16 region due to weather conditions during some of the observations.
The FUGIN survey covers $|b| < 1\arcdeg$, and M16 extends up to 1\fdg2.
However, the nearly 4 square degree field of view spanning all the way to the Galactic plane makes this a valuable asset alongside our wide-field continuum images.
The observations and their analysis are described in detail by \cite{Nishimura2021PASJ...73S.285N}.

We use both sets of CO (\jmton{1}{0}) observations because the PMO observations cover the central M16 region but do not cover the extended area including the rest of the GMF towards the Galactic plane, while the FUGIN observations cover a larger area including the GMF, but do not cover $b > 1\arcdeg$.

\subsection{Ancillary Data}
To analyze the shock-ionized plasma within wind-blown bubbles, we obtain the diffuse X-ray spectra and maps extracted by \citet{Townsley2014ApJS..213....1T} from the Chandra ACIS-I mosaics by \citet{Linsky2007ApJ...654..347L} and \citet{Guarcello2010A&A...521A..61G}.

We use a collection of publicly available continuum images in our analysis to trace a variety of other cold and warm phases of gas associated with \hii\ region bubbles.
We obtain 70--500~\micron\ Herschel PACS and SPIRE images from the Herschel Science Archive \citep{higal70_https://doi.org/10.26131/irsa27, higal160_https://doi.org/10.26131/irsa25, higal250_https://doi.org/10.26131/irsa24, higal350_https://doi.org/10.26131/irsa28, higal500_https://doi.org/10.26131/irsa26};
the GLIMPSE\footnote{\url{https://irsa.ipac.caltech.edu/data/SPITZER/GLIMPSE/}} 5.8 and 8~\micron\ Spitzer IRAC images and the MIPSGAL 24~\micron\ Spitzer MIPS image from the NASA/IPAC Infrared Science Archive (IRSA) \citep{glimpse_https://doi.org/10.26131/irsa210, mipsgal24_https://doi.org/10.26131/irsa259};
the WISE 3.4, 4.6, 12, and 22~\micron\ images from the IRSA;
90~cm VLA observations from the MAGPIS survey website\footnote{\url{https://third.ucllnl.org/gps}};
870~\micron\ APEX observations from the ATLASGAL survey \citep{Schuller2009A&A...504..415S, atlasgal_https://doi.org/10.18727/archive/20} from the ESO Science Archive Facility;
and the DSS2 red optical image (Region ER662, Plate ID A0Q1; \citealt{DSS_https://doi.org/10.26131/irsa441}).

\section{Structure of M16} \label{sec:m16-struct}
\begin{figure*}
    \centering
    \includegraphics[width=0.75\textwidth]{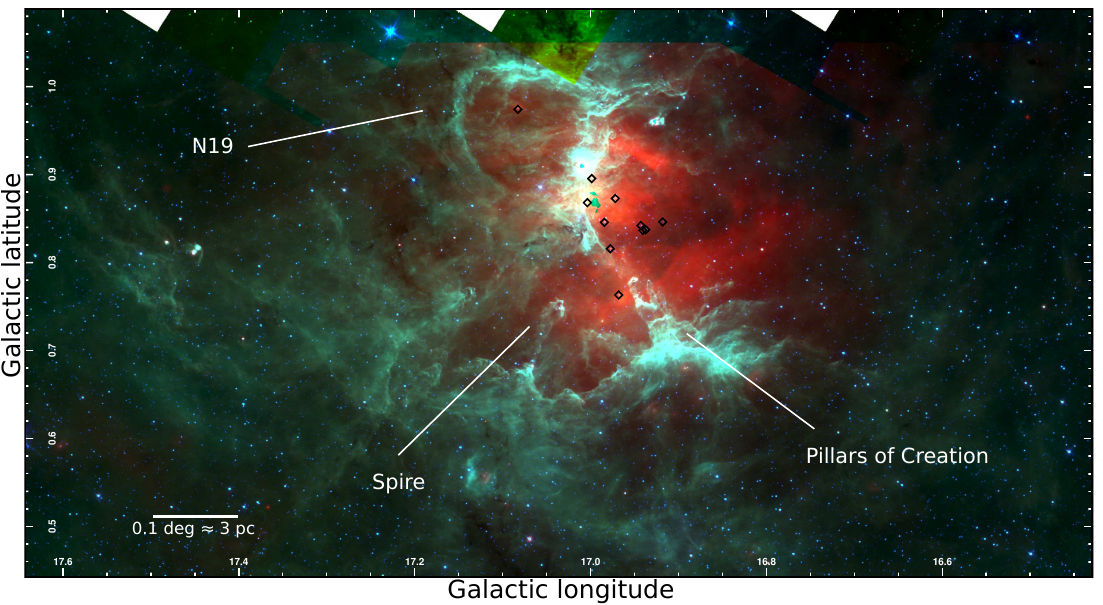}
    \caption{M16 in Spitzer IRAC 5.8 and 8.0~\micron\ (blue, green) and MIPS 24~\micron\ (red). The Pillars of Creation, Spire, and N19 ring are marked. The IRAC images are shown with a linear stretch and the MIPS band with a square root stretch. Black diamonds show early-type members of NGC~6611 (Section~\ref{sec:stars}).}
    \label{fig:m16_irac_rgb}
\end{figure*}

\begin{figure*}
    \gridline{\fig{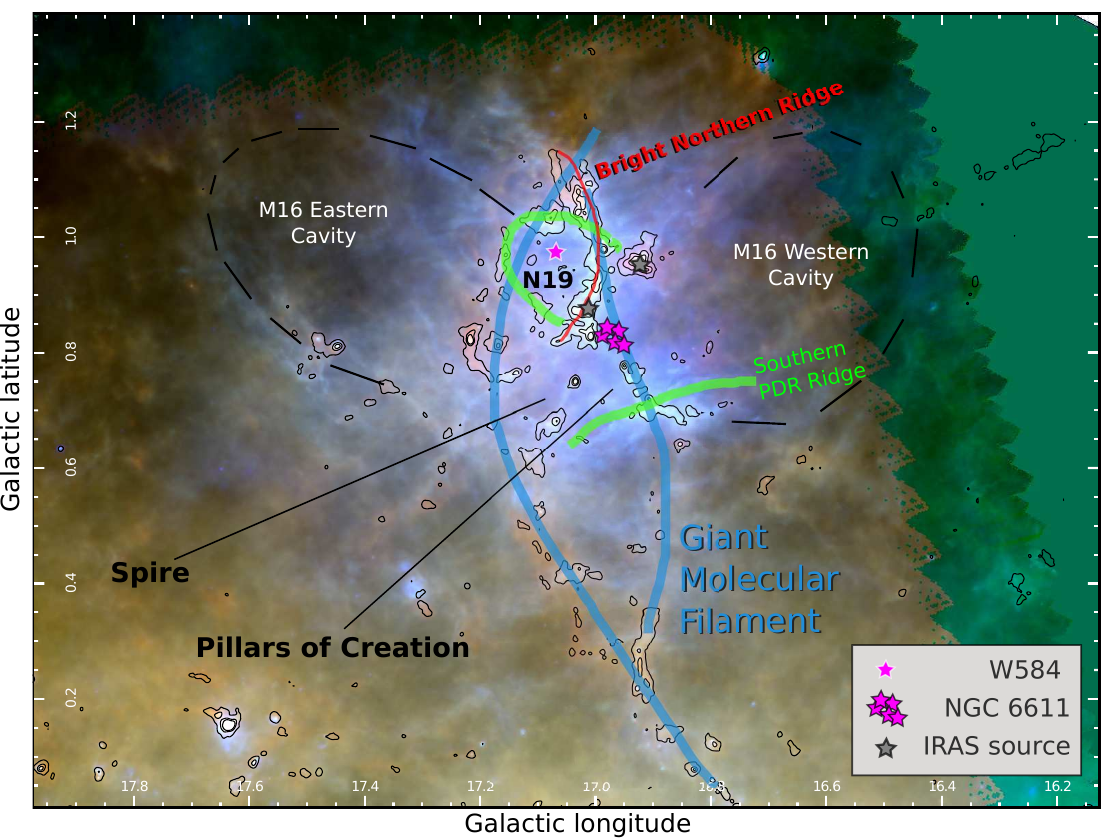}{0.95\textwidth}{(A)}}
    \gridline{\fig{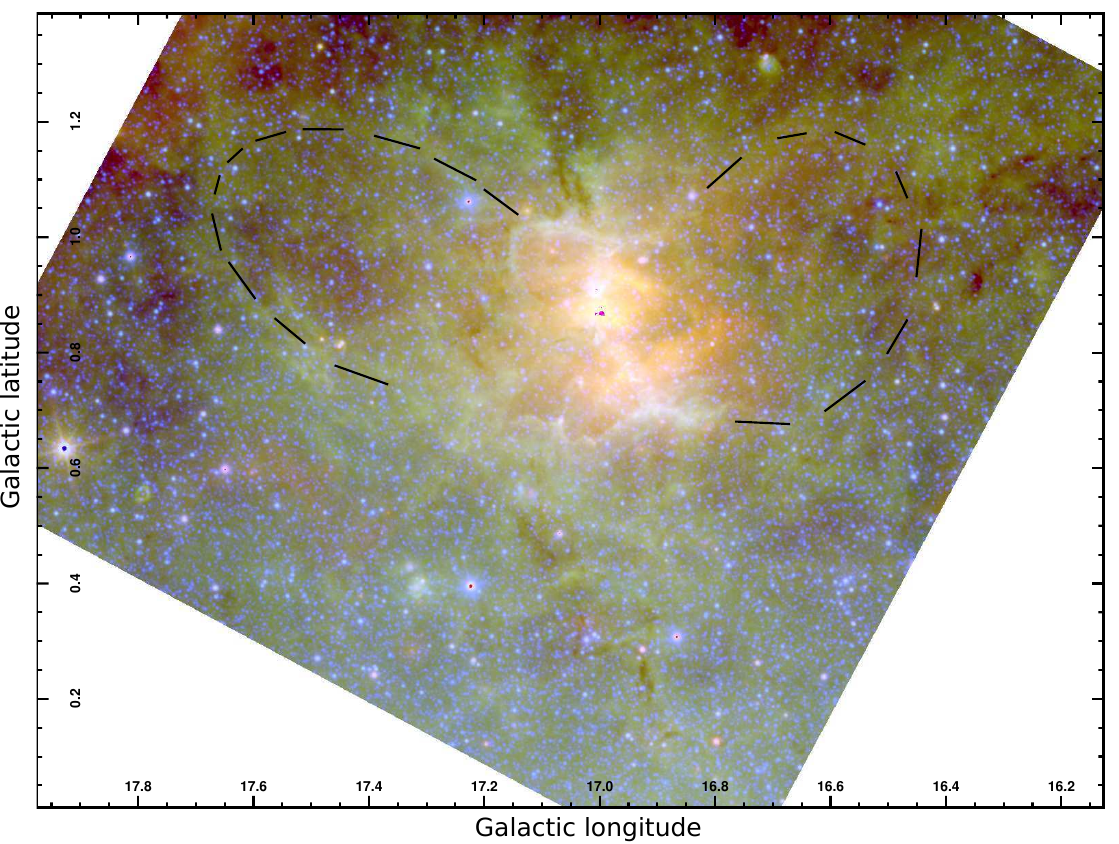}{0.48\textwidth}{(B)} \fig{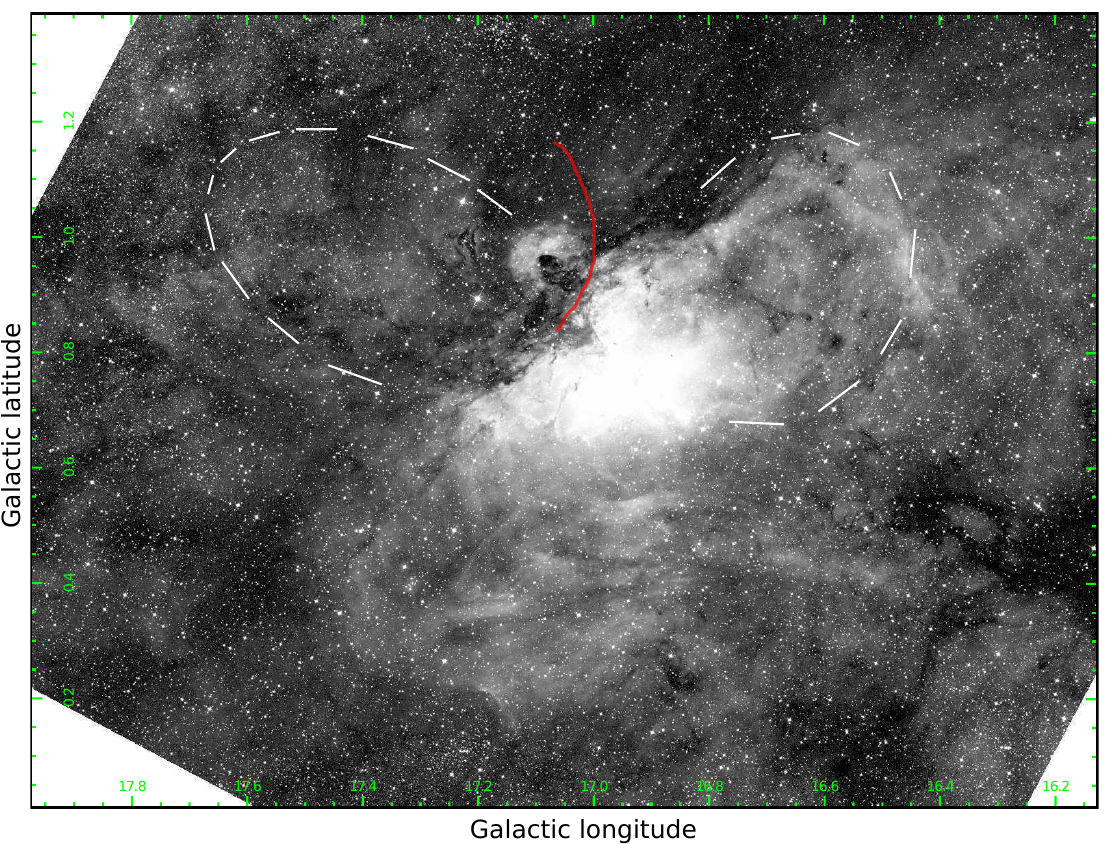}{0.48\textwidth}{(C)}}
    \caption{M16 in the far-IR, mid-IR, and optical. (A) Herschel 500, 160, 70~\micron\ in red, green, and blue, and 870~\micron\ in black contours. Features are marked with colored lines.
    The NGC~6611 cluster center is marked with a group of magenta stars, and W584 is marked separately.
    IRAS 18156--1343 (left) and IRAS 18152--1346 (right) are marked with grey stars.
    (B) WISE 22, 12, and 4.6~\micron\ in red, green, and blue, and (C) optical DSS2 R-band. Dashed black or white lines in all images outline the infrared lobes, and the Bright Northern Ridge is marked in red in the optical, where it is not visible.}
    \label{fig:m16_finder}
\end{figure*}

\begin{figure*}
    \centering
    \includegraphics[width=\textwidth]{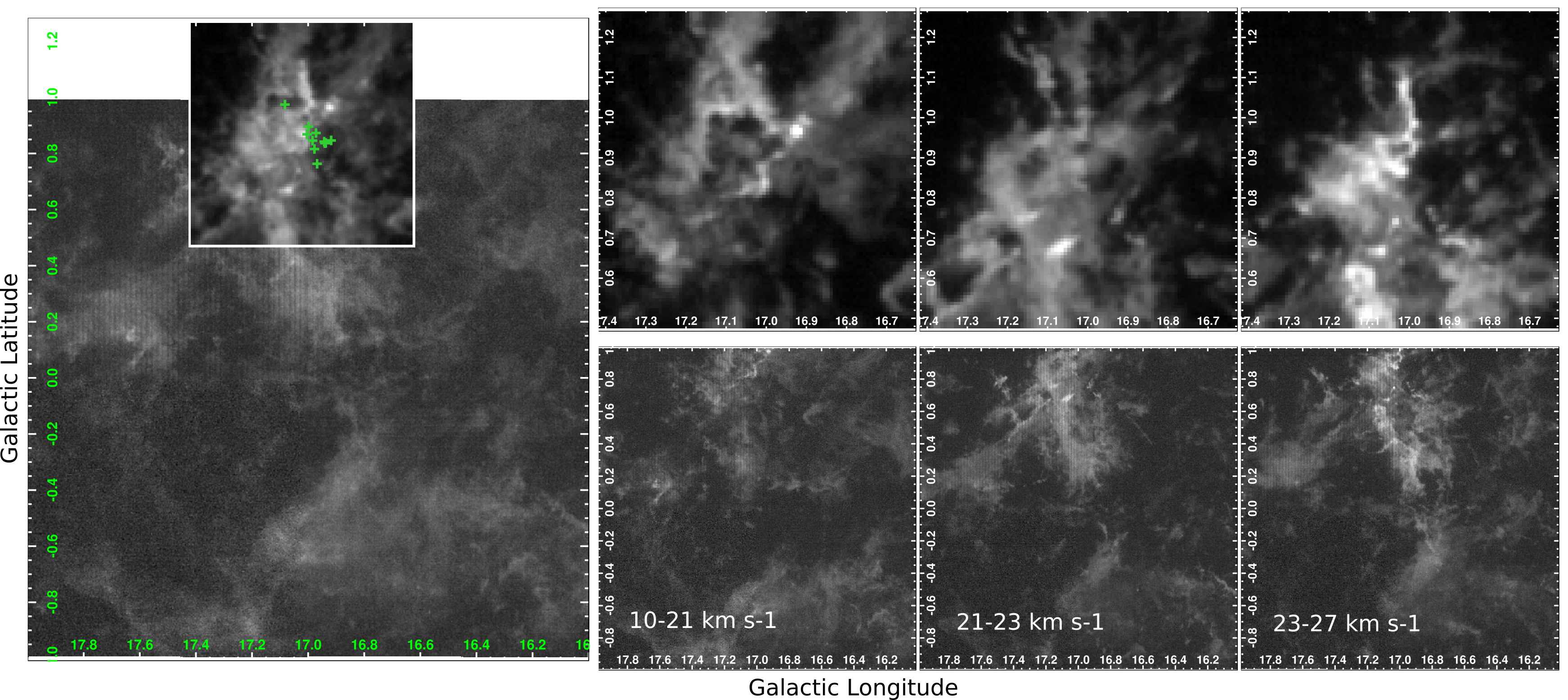}
    \caption{\twco\ integrated intensities in 4 different velocity intervals. The leftmost panel shows the PMO and FUGIN observations overlaid. The FUGIN observations cover $|b| < 1\arcdeg$, but M16 extends slightly above $b = 1\arcdeg$. The PMO observations cover a smaller area but reach up to $b \approx 1\fdg3$. Green crosses show early-type members of NGC~6611 (Section~\ref{sec:stars}). The two rows to the right show PMO observations on top and FUGIN observations on the bottom. The velocity intervals used for each column are marked on the bottom row. PMO observations are made with a 55\arcsec\ beam, and FUGIN survey observations using Nobeyama have a 20\arcsec\ beam.}
    \label{fig:PMO-NOB-3panels}
\end{figure*}

M16, shown in the mid-IR in Figure~\ref{fig:m16_irac_rgb}, is a well-known region featuring the Pillars of Creation and the Spire surrounded by a bright \hii\ region and PDRs.
In the following section, we draw from ancillary data to put this bright central region into context with the gas around it using wide-field images and spectra from archival sources.

\subsection{GMC W~37 and the GMF}
The 160--870~\micron\ images (Figure~\ref{fig:m16_finder}) and large-field FUGIN CO (\jmton{1}{0}) observations (Figure~\ref{fig:PMO-NOB-3panels}) trace a $\sim$40~pc long giant molecular filament (GMF) running perpendicular to the Galactic plane.
This GMF, hereafter ``the GMF,'' is marked with a blue curve in Figure~\ref{fig:m16_finder} following the red 500~\micron\ and contoured 870~\micron\ emission tracing dust within cold, dense gas.
The upper part $0\fdg5 \lesssim b \lesssim 1\arcdeg$ of the GMF widens out into a giant molecular cloud (GMC) W~37 \citep{Westerhout1958BAN....14..215W, Zhan2016RAA....16...56Z}.
The GMF intersects with M16 and is likely the birthplace of the cluster \citep{Hill2012A&A...542A.114H}.
CO line velocities place the GMF within $\vlsr = 21\Endash27~\kms$.
Parts of the GMF below the bright \hii\ region are seen in absorption in the mid-IR images in Figure~\ref{fig:m16_finder}.
These dark lanes are included in catalogs by \citet{Rygl2010A&A...515A..42R}, \citet{Peretto2016A&A...590A..72P}, and \citet{Eden2019MNRAS.485.2895E} and do not appear to be illuminated by the NGC~6611 cluster.

\subsection{Filaments within M16}
The GMF splits into two subfilaments where it intersects with M16 and the $\sim$$\nexpo{1.5}{5}~M_\odot$ GMC W~37 which dominates CO observations between $\vlsr = 10\Endash27~\kms$ \citep{Zhan2016RAA....16...56Z, Nishimura2021PASJ...73S.285N}.
The Spire, a pillar in M16, branches off the eastern filament, and the Pillars of Creation lie along the western filament where it protrudes into the \hii\ region in Figure~\ref{fig:m16_finder}.
\citet{Hill2012A&A...542A.114H} point out that the eastern filament is relatively unperturbed by stellar feedback, while the western filament was perhaps once connected and has been severed by stellar feedback as NGC~6611 formed within it.
We represent the western filament as continuous in Figure~\ref{fig:m16_finder}, though there is no dense gas bridging the Pillars of Creation and the Bright Northern Ridge across the NGC~6611 cluster center.

The Bright Northern Ridge north of NGC~6611 (Figure~\ref{fig:m16_finder}, marked in red) is bright across the IR and sub-mm and is kinematically associated with the higher velocity filamentary emission ($\vlsr = 23\Endash27~\kms$).
It seems to lie behind, and be obscured by, the lower velocity cloud we refer to below as the Northern Cloud, as the ridge does not appear in the optical despite its infrared brightness (Figure~\ref{fig:m16_finder}C).
Within a particularly bright piece of the ridge lies IRAS 18156--1343, associated with water maser emission \citep{Codella1994A&A...291..261C_theonewith13CII}.

We hereafter refer to the 23--27~\kms\ gas whose emission peaks around 25--26~\kms\ as the ``natal cloud'' of NGC~6611.
The natal cloud includes the filaments described by \citet{Hill2012A&A...542A.114H} as well as the Pillars, Spire, and Bright Northern Ridge and is kinematically associated with the GMF.

\subsection{Northern Cloud and N19}
\begin{figure}
    \centering
    \includegraphics[width=0.47\textwidth]{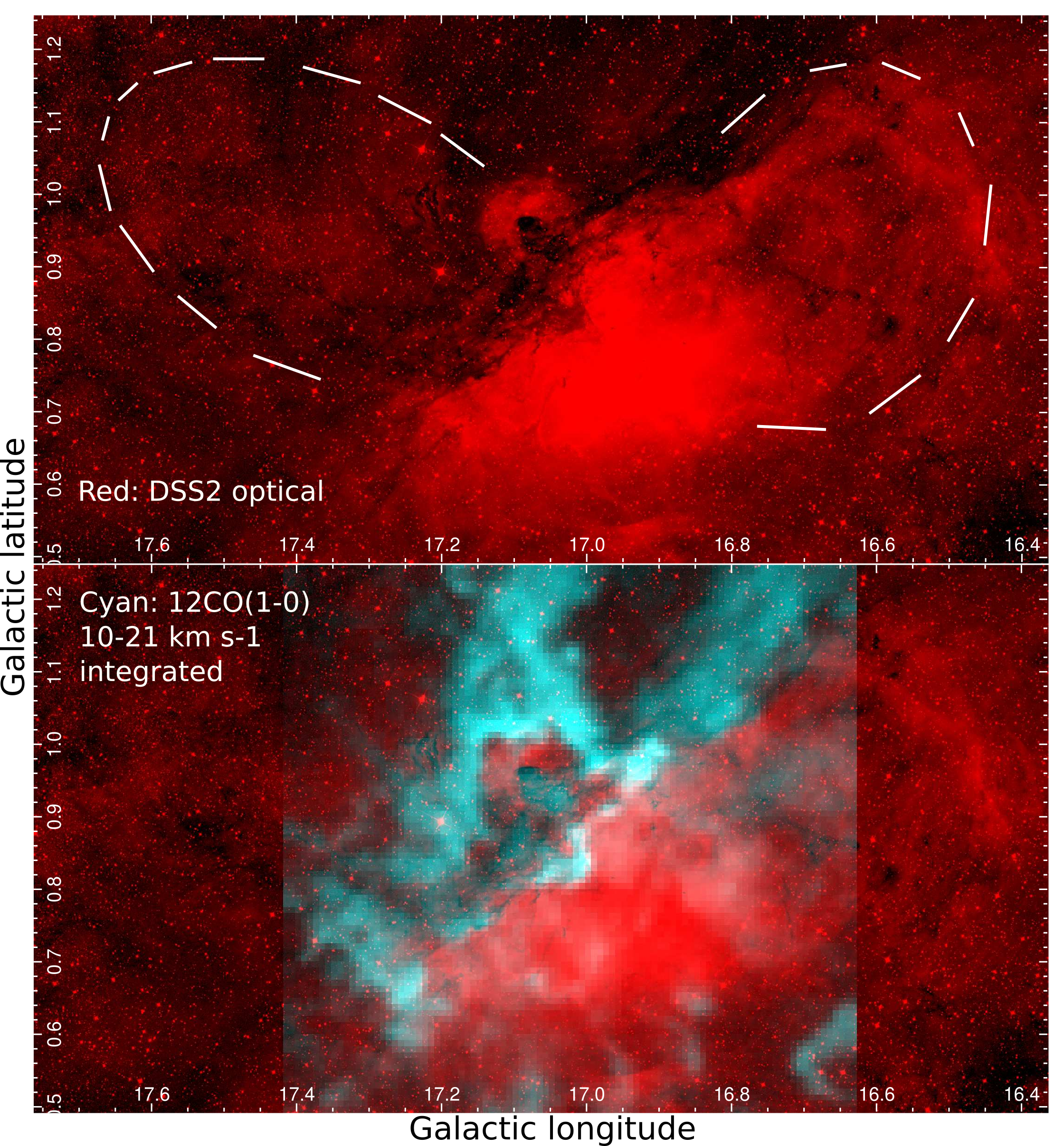}
    \caption{The Northern Cloud (\twcoA\ \jmton{1}{0}; cyan) obscures the eastern part of the M16 \hii\ region in the optical (DSS2 red optical filter; red). The PMO CO observations are integrated within $\vlsr = 10\Endash21~\kms$ to show the Northern Cloud. The dashed white lines on the top panel outline the eastern and western cavities (Section~\ref{sec:infraredlobes}). Note the anti-correlation between the Northern Cloud CO emission and the optical emission from the \hii\ region behind it.}
    \label{fig:northerncloud-optical-co}
\end{figure}

The northern, low velocity emission between 10--21~\kms\ (CO peak $\sim$19~\kms) seems to lie in front of the \hii\ region with respect to the observer.
Figure~\ref{fig:northerncloud-optical-co} shows low velocity CO maps overlaid on the optical.
This CO feature matches the optical absorption feature lying across the northern side of the \hii\ region.
We hereafter refer to the northern, low velocity cloud lying in front of the \hii\ region as the ``Northern Cloud.''
Together, the natal cloud and the Northern Cloud constitute the GMC W~37.

Within the Northern Cloud lies a bright infrared ring filled with optical emission.
The N19 ring \citep{Churchwell2006ApJ...649..759C} is bright between 8\Endash160~\micron\ but obscured by nearby bright emission at 24~\micron.
The N19 ring is likely the projected shell of a bubble driven by the O9 V star called W584 \citep{Hillenbrand1993AJ....106.1906H, Guarcello2010A&A...521A..61G} and is spatially and kinematically connected to the Northern Cloud in CO (see the 10\Endash21~\kms\ PMO map in Figure~\ref{fig:PMO-NOB-3panels}).

The source IRAS 18152--1346, identified by \citet{Indebetouw2007ApJ...666..321I} as an $\sim$8~$M_\odot$ massive young stellar object (MYSO), is a bright CO source \citep{Zhan2016RAA....16...56Z, Xu2019A&A...627A..27X} with a wide, asymmetric line profile and low velocity tail.
It is marked in Figure~\ref{fig:m16_finder}A by a grey star symbol and appears as a bright spot in the 10--21~\kms\ PMO map in Figure~\ref{fig:PMO-NOB-3panels} close to $(l,\, b) = (16.9\arcdeg,\, 1.0\arcdeg)$.
Its $\vlsr \sim 19~\kms$ peak associates it with the Northern Cloud.

\subsection{Bright PDR and \hii\ Interior of M16}
Emission at 24~\micron, 70~\micron, and 90~cm is particularly bright towards the Pillars, Spire, and Bright Northern Ridge, which are also bright PDR sources (8~\micron\ and \cii; Section~\ref{sec:results_cii_channel}).
The 70~\micron\ emission traces hot dust in the PDR, 90~cm traces ionized gas via free-free emission, and 24~\micron\ traces hot dust in both neutral and ionized gas \citep{Churchwell2009PASP..121..213C}.
The 24~\micron\ inner ring identified by \citet{Flagey2011A&A...531A..51F}, interior to the PDRs and perhaps created by the grinding of dust into very small grains by wind-driven shocks, can be seen at 22 and 24~\micron\ in red in Figures~\ref{fig:m16_finder}B and \ref{fig:m16_irac_rgb}, respectively.

\subsection{Extended Infrared Lobes} \label{sec:infraredlobes}
To the $\pm l$ sides of the bright central region lie $\sim$15~pc (approximately half-degree) diameter infrared lobes with faint edges and dark interiors in 8, 12, and 70--500~\micron\ images.
Their edges are most clearly traced by 8~\micron\ (green in Figure~\ref{fig:m16_irac_rgb}), 70~\micron\ (blue in the Figure~\ref{fig:m16_finder}A), and  12~\micron\ (green in Figure~\ref{fig:m16_finder}B) emission and marked with dashed lines in Figure~\ref{fig:m16_finder}.
They tilt slightly upwards towards higher Galactic latitude.

Two pillars with IR point sources within them sit along the clear bottom edge of the eastern ($+l$) lobe.
Their bright rims point back toward NGC~6611, indicating that the cluster's radiation influences dense gas along the edges of the lobes.
These pillars and the rest of the eastern lobe are completely obscured in the optical, likely by the Northern Cloud as demonstrated by the optical-CO overlay in Figure~\ref{fig:northerncloud-optical-co}.

\begin{figure*}
    \centering
    \includegraphics[width=\textwidth]{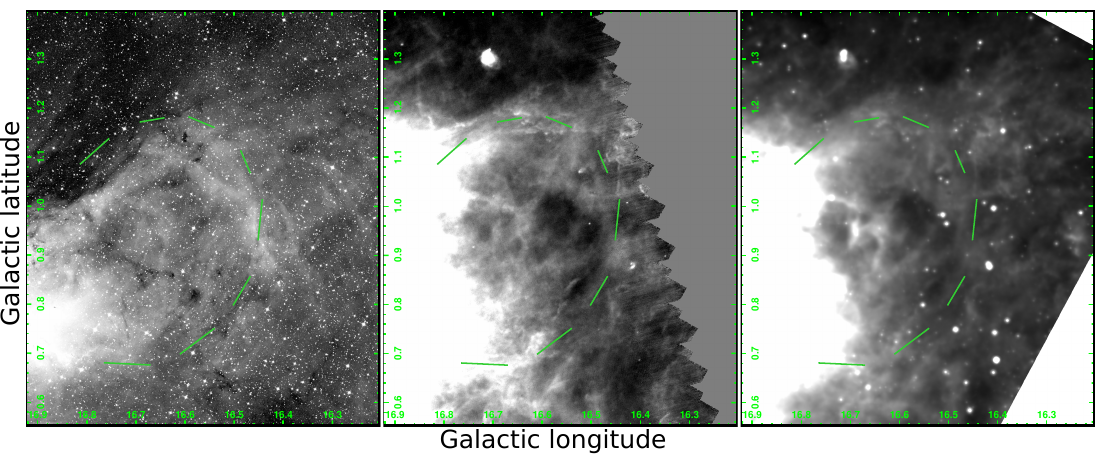}
    \caption{The edge of the western cavity described in Section~\ref{sec:infraredlobes}. From left to right: DSS2 red optical, 70~\micron, and 22~\micron. All three images show the same area of the sky. The optical map uses a square root color stretch; the others use a linear stretch. The edge of the western cavity, marked with green dashed lines, appears in emission at the center of each image.}
    \label{fig:westcavity-3panel}
\end{figure*}

The western lobe is filled with optical emission (Figures~\ref{fig:northerncloud-optical-co}).
The edge of the western lobe is traced by a bright optical feature which appears in the 24 and 70~\micron\ maps in Figure~\ref{fig:westcavity-3panel}.

\subsection{Summary of Structure}
The GMF is traced by CO at $\sim$25--26~\kms\ and 870~\micron\ continuum and extends all the way through W~37 and M16 to the Galactic plane.
NGC~6611 likely formed from this filament, as dense gas features hosting bright PDRs such as the Bright Northern Ridge, the Pillars, and the Spire share this velocity and are located along its subfilaments.
The cavity generated by feedback from NGC~6611 appears to have blown out of the sides of the filament in which the cluster formed.
The combined IR and sub-mm observations reveal a well-expanded \hii\ region consisting of two lobes bright in ionized medium tracers (e.g., 24~\micron\ dust emission) while lobe edges light up in PDR tracers like 70~\micron\ from warm dust and 8 and 12~\micron\ PAH emission.

The infrared lobes in Figures~\ref{fig:m16_irac_rgb} and \ref{fig:m16_finder} indicate limb brightening from a $\sim$40~pc diameter shell surrounding a cavity.
Radiation from NGC~6611 influences gas far from the bright $\sim$10~pc wide center of the \hii\ region, where gas and dust are hot and dense and shine brightly.
Within the central region, dense molecular gas is sculpted into parsec-scale pillars such as the Pillars of Creation and the Spire.
A bright PDR covers the surface of the Bright Northern Ridge, a dense molecular gas ridge which lies only a few parsecs in projection from the NGC~6611 cluster.
Diffuse X-ray-emitting plasma surrounds the cluster and also appears above and perhaps behind the surrounding molecular gas \citep{Townsley2014ApJS..213....1T}, indicating that shocked-wind plasma has escaped the central region.

The \hii\ region has expanded along $\pm l$, perpendicular to the GMF which extends along $\pm b$.
Simulations \citep{Fukuda2000ApJ...533..911F, Zamora-Aviles2019MNRAS.487.2200Z, Whitworth2021MNRAS.504.3156W} and at least one observational study \citep{Watkins2019A&A...628A..21W} indicate that \hii\ regions expand non-uniformly in the presence of dense gas, seeking out lower density paths.
The M16 \hii\ region's expansion is constrained along the filamentary axis and seems to have expanded out wherever the filament is not.

\section{Results from \cii\ and CO Observations} \label{sec:results}
\begin{figure*}
    \centering
    \includegraphics[width=0.8\textwidth]{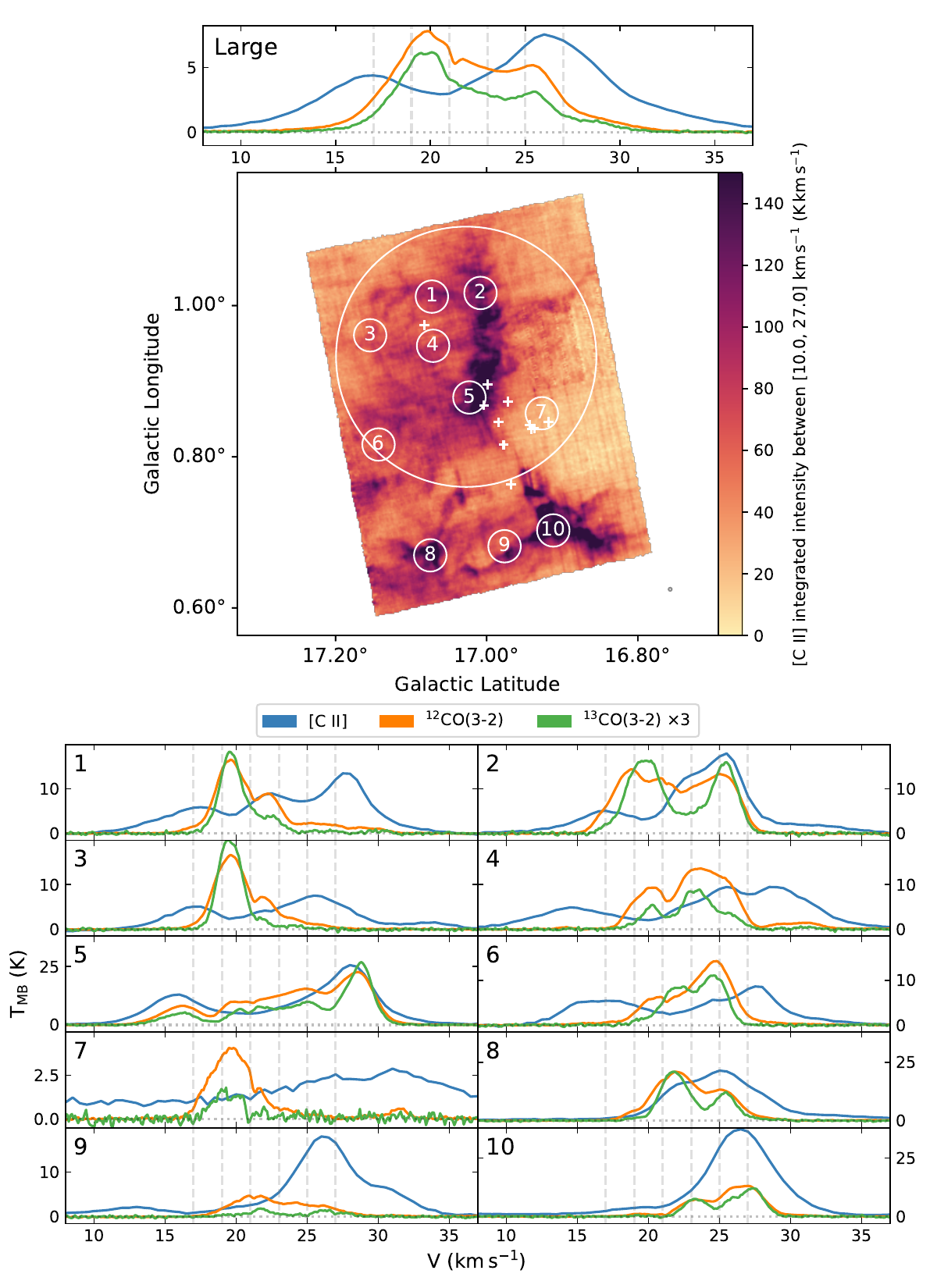}
    \caption{Averaged \cii, \twcott, and \thcott\ spectra from within regions outlined on the $\vlsr = 10\Endash27~\kms$ integrated \cii\ intensity image in the center. White crosses show early-type members of NGC~6611 and the 15.5\arcsec\ \cii\ beam is shown in the lower right corner. Spectra from the largest circle are shown above the image, and spectra from the numbered regions are shown in the panels below the image. \thcoA\ spectra are shown multiplied by 3 for better visibility. Vertical lines mark every 2~\kms\ between 17--27~\kms. This conveniently marks the Northern Cloud near 19~\kms, the diffuse CO component at 21--23~\kms, and the natal cloud emission near 25~\kms.}
    \label{fig:avg_spec}
\end{figure*}

Here we introduce our new \cii\ and CO (\jmton{3}{2}) observations of the bright central region of M16.
These velocity-resolved observations offer the capability to distinguish features kinematically as well as spatially, like we did with the CO (\jmton{1}{0}) spectral cubes in Section~\ref{sec:m16-struct}.
The CO (\jmton{3}{2}) transition lies at an equivalent temperature $E_{u}/k_{\rm B} = 33.2$~K above the ground state and has a critical density $n_{\rm cr} = \nexpo{2.1}{4}~\cc$, and is more sensitive than the (\jmton{1}{0}) to dense gas and higher temperatures and highlights gas structures which either retain high primordial densities or have been shock-compressed by feedback from the cluster.
The \cii\ observations unlock the capability to kinematically resolve PDRs.
The \cii\ transition ($E_u/k_{\rm B} = 91.25$, $n_{\rm cr} = \nexpo{3}{3}~\cc$) traces similar FUV-illuminated gas as FIR dust or mid-IR PAH tracers \citep{Pabst2017A&A...606A..29P, Pabst2021A&A...651A.111P}, but can be spectroscopically resolved.
The combination of \cii\ and CO observations is a powerful tool for determining the structure and kinematics of illuminated and un-illuminated gas in a star formation system and evaluating the effects of stellar feedback on the surrounding gas.

The typical \cii\ spectrum towards M16 is a combination of two velocity components: a bright component at the natal cloud velocity $\vlsr \approx 25\Endash26$~\kms, and a secondary peak at $\vlsr \approx 17\Endash18$~\kms\ close to the Northern Cloud velocity.
These components appear in the large-area average \cii\ spectrum at the top of Figure~\ref{fig:avg_spec} and lie between high and low velocity line wings.

Channel maps of \cii\ and \twcott\ in Figure~\ref{fig:channel_maps_cii_co} show that the 25--26~\kms\ emission traces the filamentary structure of the natal cloud and the $\sim$17--18~\kms\ emission traces N19 and the Northern Cloud.
The natal cloud \cii\ emission includes CO-emitting features like the Bright Northern Ridge, the Spire, the Pillars of Creation, and the southern PDR ridge at the base of the Pillars (near region 10 in Figure~\ref{fig:avg_spec}) as well as diffuse \cii\ emission with no CO counterpart that extends to $+l$ from the Bright Northern Ridge (behind N19, near region 4 in Figure~\ref{fig:avg_spec} for example).

We overlay \cii\ and \twcott\ intensity maps integrated within the same velocity intervals in the red-blue images in Figure~\ref{fig:cii_co_rb_overlay} and the velocity red-green-blue (RGB) images in Figure~\ref{fig:cii_co_rgbvelocity} to emphasize similarities and differences in what is traced by these transitions.

The footprints (i.e. projected areas covered by observations) of the \cii\ and CO (\jmton{3}{2}) observations are very similar, as seen in Figures~\ref{fig:channel_maps_cii_co}, \ref{fig:cii_co_rb_overlay}, and \ref{fig:cii_co_rgbvelocity}.
The footprint of the \cii\ observations is shown over the IRAC 8~\micron\ image in Figure~1 in the paper by \citet{Schneider2020PASP..132j4301S}.

All \cii\ emission is assumed to originate from the atomic, rather than the ionized, phase, an assumption that is reinforced by spatial correlation with other PDR tracers such as 8~\micron\ in M16 and other massive star-forming regions \citep{Pabst2021A&A...651A.111P} as well as \cii\ line profile similarity to other PDR tracers and dissimilarity to ionized gas tracers \citep{Pabst2024A&A...688A...7P}.
In the extragalactic context, \citet{Tarantino2021ApJ...915...92T} show that the ionized phase contributes only $\sim$10\% of \cii\ emission at large ($\sim$500~pc) scales.
This is in part because carbon tends to be doubly ionized by the hard UV field from such massive stars \citep{Kaufman2006ApJ...644..283K}.

\subsection{PDR Structure from \cii\ Channel Maps} \label{sec:results_cii_channel}

The lowest velocity \cii\ emission in the channel maps in Figure~\ref{fig:channel_maps_cii_co} appears around 7--11~\kms\ towards $-l$ and traces the gas along the interior of the right IR lobe as seen at 8--12 or 70--500~\micron.
From 13--21~\kms, \cii\ emission highlights the Northern Cloud and N19.
The Northern Cloud and N19 fade into confusion around 23~\kms\ with the line wings of the diffuse emission behind the Bright Northern Ridge.
Morphology clearly distinguishes the Northern Cloud and natal cloud components as separate features in \cii\ as we show in the RGB image in Figure~\ref{fig:cii_co_rgbvelocity}, and their peaks are well separated by $>5~\kms$ as seen in regions 1--4 in Figure~\ref{fig:avg_spec}, which is greater than the typical line width of 2--3~\kms\ in individual spectra.

The filamentary structure of the natal cloud---Pillars, Spire, and Bright Northern Ridge---appears in bright \cii\ emission between 23--27~\kms.
Beyond 27~\kms, the southern PDR ridge below the Pillars spreads out and highlights a different set of small ridges along the same lines of sight, indicating that the southern PDR ridge is a complex PDR surface viewed edge-on.
The highest velocity \cii\ emission $\vlsr \gtrsim 35~\kms$ appears toward the same region inside the right IR lobe as the lowest velocity emission, consistent with the signature of an expanding shell.

\subsection{Comparison of \cii\ to CO} \label{sec:results_both_channel}

The CO (\jmton{3}{2}) line generally traces the same components as CO (\jmton{1}{0}).
Regions that are bright in \cii\ are also bright in CO, but CO lines also trace diffuse W~37 emission between $\vlsr = 21\Endash23~\kms$.

\begin{figure*}
    \centering
    \includegraphics[width=\textwidth]{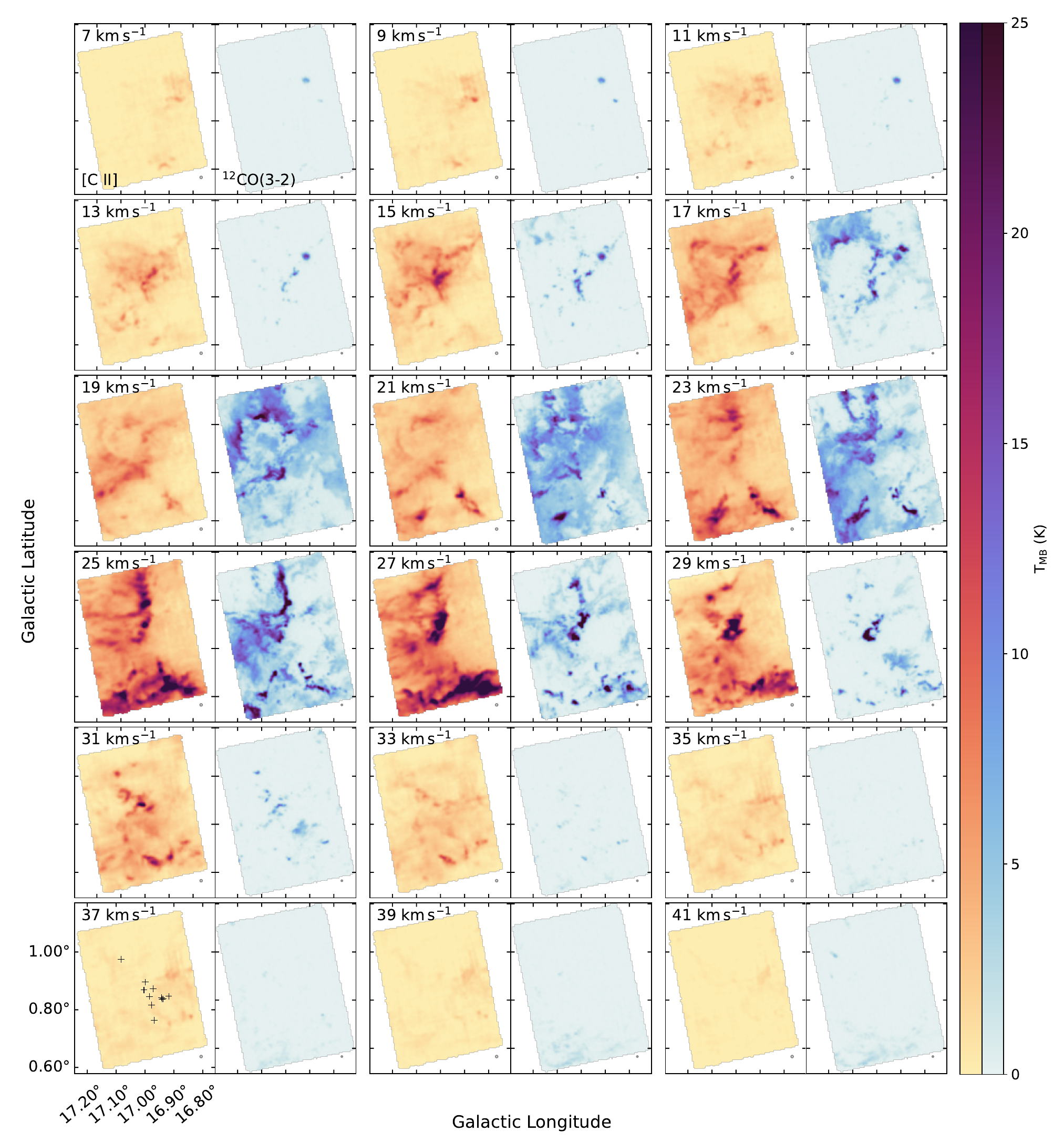}
    \caption{\cii\ and \twcott\ line channel maps. All maps are on the same linear scale shown on the colorbars to the right; two different color maps are used to distinguish \cii\ from CO. Black crosses in the bottom left panel show early-type members of NGC~6611 and the beam is shown in the lower right corner of each map. All data are binned to 2~\kms\ channels  to increase the signal-to-noise ratio. The \cii\ data are convolved here to a 30\arcsec\ beam. \twcott\ are presented at the original $\sim$20\arcsec\ resolution.}
    \label{fig:channel_maps_cii_co}
\end{figure*}

CO (\jmton{3}{2}) channel maps in Figure~\ref{fig:channel_maps_cii_co} show low velocity emission appearing along the flat bottom of the Northern Cloud between $\vlsr = 13\Endash17~\kms$.
One clump of CO emission at 9~\kms\ is coincident with \cii\ emission and traces a mid-IR and FIR clump, but besides that, we do not detect CO from the inside of the right IR lobe as we do with \cii.
Another clump of CO (\jmton{3}{2}) emission visible from 5--19~\kms\ is associated with the MYSO IRAS~18152--1346 and has a similar wide, asymmetric profile as in the CO (\jmton{1}{0}) observations.

N19 is well-outlined in CO 17--21~\kms\ and has a clear counterpart in \cii\ (Figure~\ref{fig:cii_co_rb_overlay}, top right), though the \cii\ ring is blueshifted by $\sim 1\Endash2~\kms$ with respect to the CO ring as illustrated by spectra towards the edge of N19 in Figure~\ref{fig:avg_spec}, regions 1--4.
CO (\jmton{3}{2}) and (\jmton{1}{0}) emission from the Northern Cloud extends further north/up than \cii; \cii\ emission here is faint, indicating that the far reaches of Northern Cloud are a poorly illuminated reservoir of dense gas.

The natal cloud component is prominent in the \cii, but not CO, spectra towards N19 around $\vlsr \sim 27~\kms$.
The difference in the spatial distribution of the line emission---the compact CO ridge versus the diffuse trail of \cii\ emission to the left of the ridge in Figure~\ref{fig:channel_maps_cii_co}---causes the $\sim27~\kms$ \cii\ component to appear without its CO counterpart.
CO emission tends to appear thinner and clumpier than \cii\ throughout the natal cloud interval $\vlsr = 23\Endash27~\kms$, particularly towards the Bright Northern Ridge and the southern PDR ridge below the Pillars of Creation.

The diffuse W~37 background component begins to appear in the CO channel maps below the Northern Cloud around 19~\kms\ and is strongest and most widespread between 21--23~\kms.
This component is largely undetected in \cii\ (see regions 5--7 in Figure~\ref{fig:avg_spec} and the bottom left panel in Figure~\ref{fig:cii_co_rb_overlay}), though we detect faint \cii\ emission in this interval towards some of the CO-bright locations.
The CO (\jmton{1}{0}) observations show that much of the surrounding, degree-scale W~37 CO emission lies at $\vlsr \sim 21\Endash23~\kms$.
The Galactic Sagittarius-Carina arm crosses around this velocity \citep{Kuhn2021A&A...651L..10K_galacticarms}, so some of this diffuse emission may also be unrelated foreground or background gas and therefore may not have a strong \cii\ counterpart.

The high velocity CO channel maps show a scattering of bright emission spots, many of which are spatially coincident with \cii\ emission in the same channels.
\cii\ and CO both trace a particularly bright clump along the Bright Northern Ridge close to the location of NGC~6611 around the 29~\kms\ channel (region 5 in Figure~\ref{fig:avg_spec}).
Just like in the low velocity channels, we do not detect high-velocity CO inside the right lobe like we do with \cii.

\subsection{Summary of Channel Maps} \label{sec:results-channel-summary}

\begin{figure*}
    \gridline{
        \fig{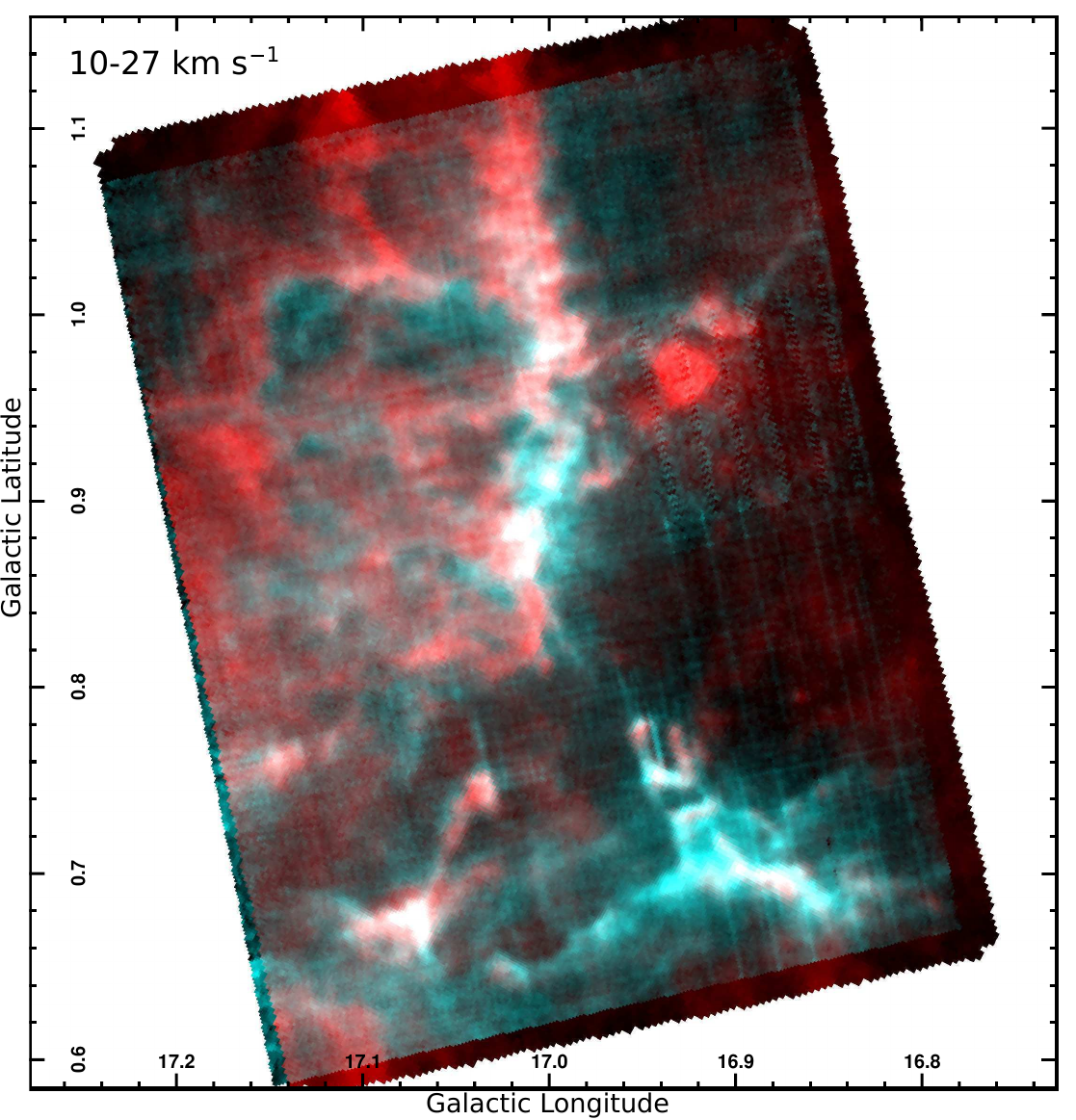}{0.5\textwidth}{}
        \fig{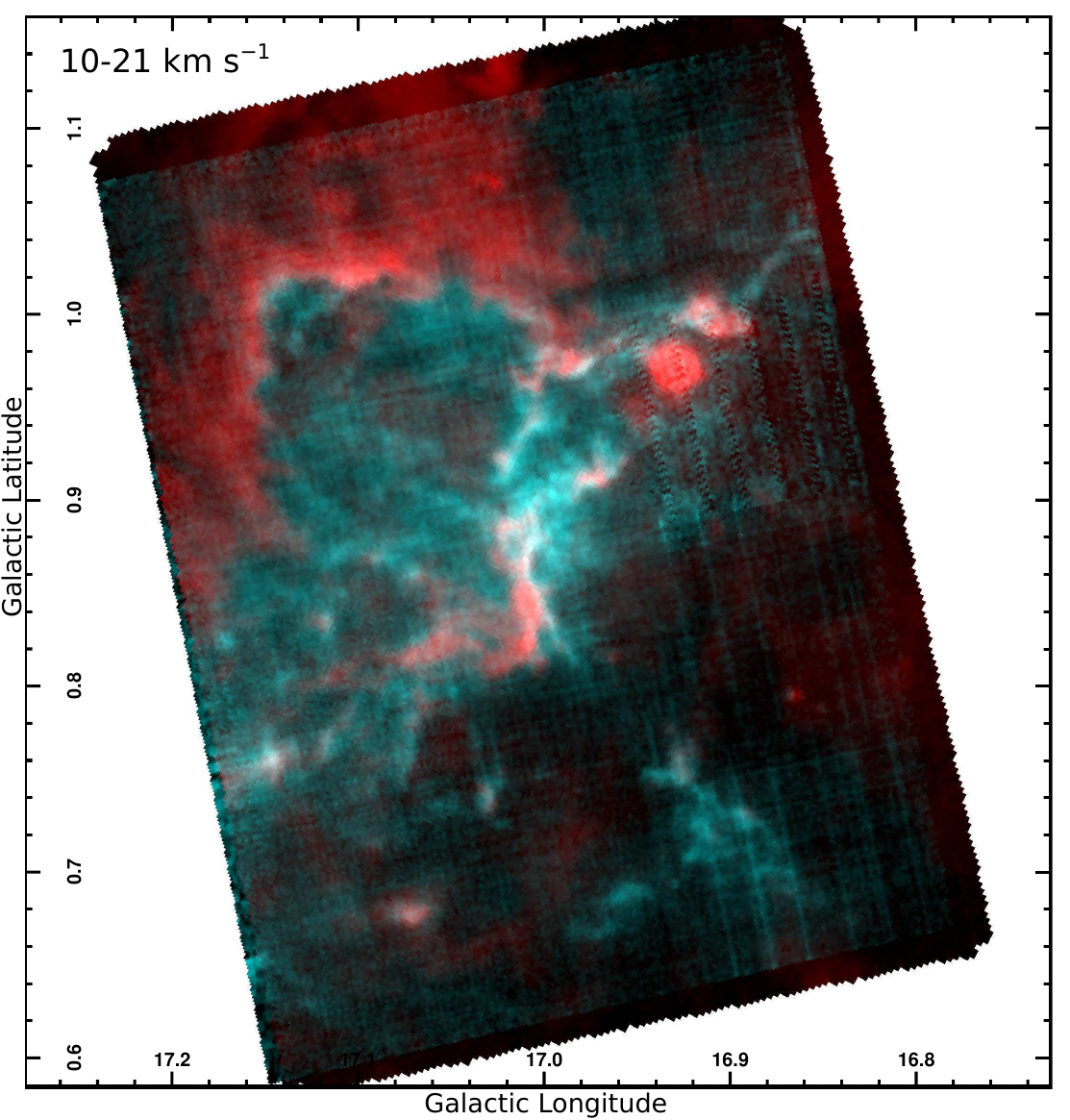}{0.5\textwidth}{}
    }
    \gridline{
        \fig{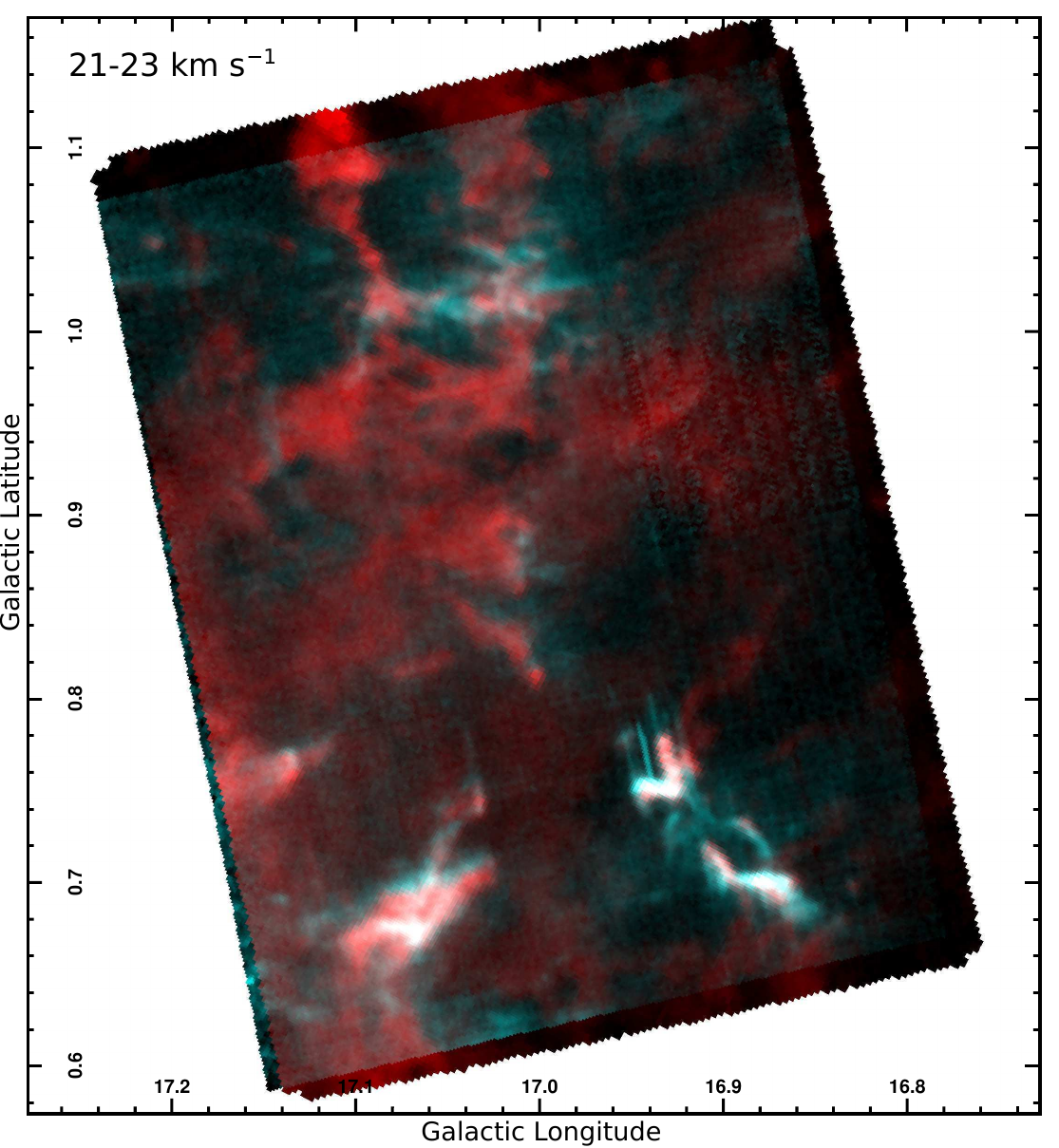}{0.5\textwidth}{}
        \fig{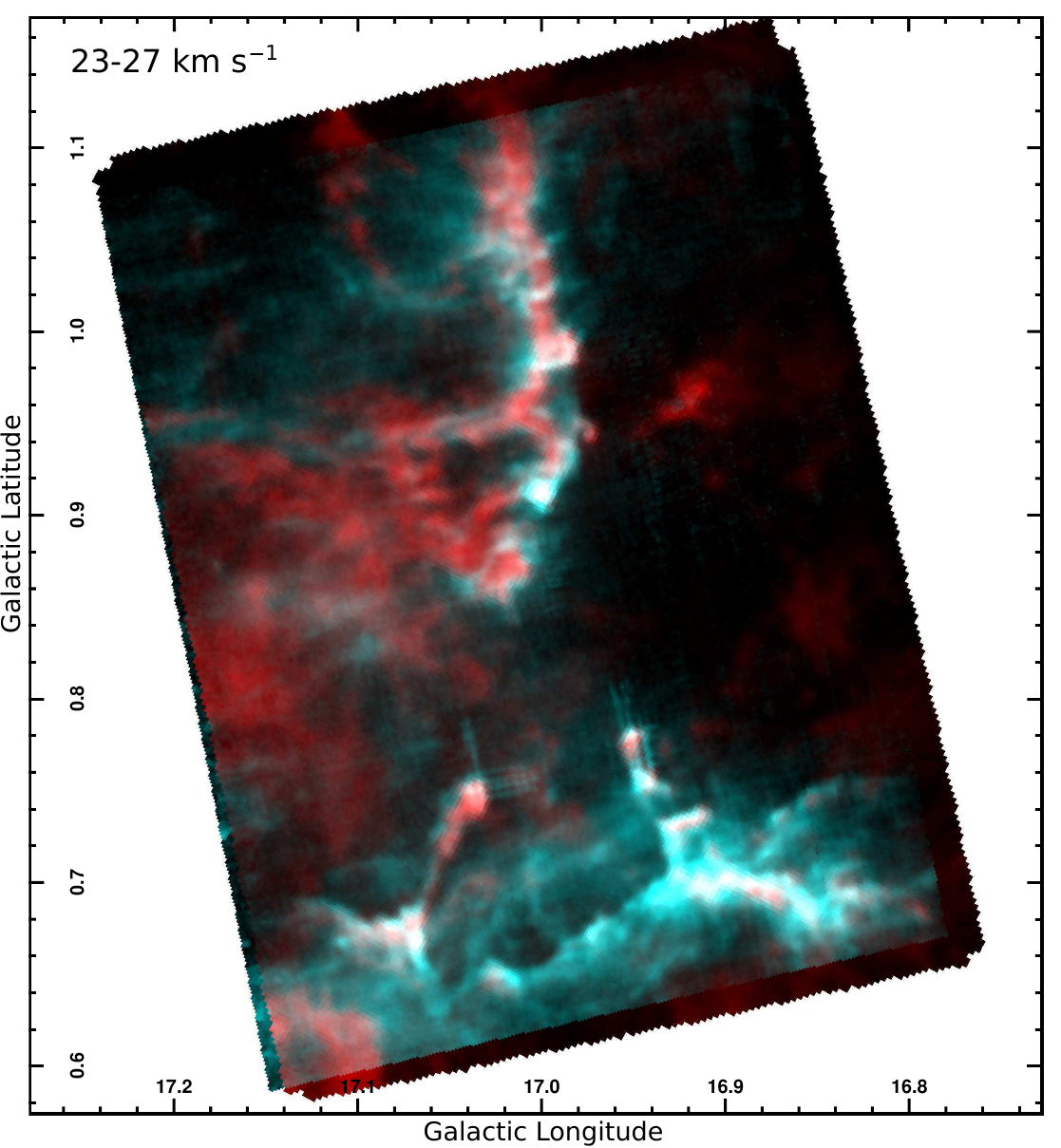}{0.5\textwidth}{}
    }
    \caption{Color composites showing \cii\ and \twcott\ integrated within four different velocity intervals. Velocity intervals are labeled in the top-left corners. CO is shown in red and \cii\ in cyan.}
    \label{fig:cii_co_rb_overlay}
\end{figure*}

\begin{figure*}
    \centering
    \includegraphics[width=\textwidth]{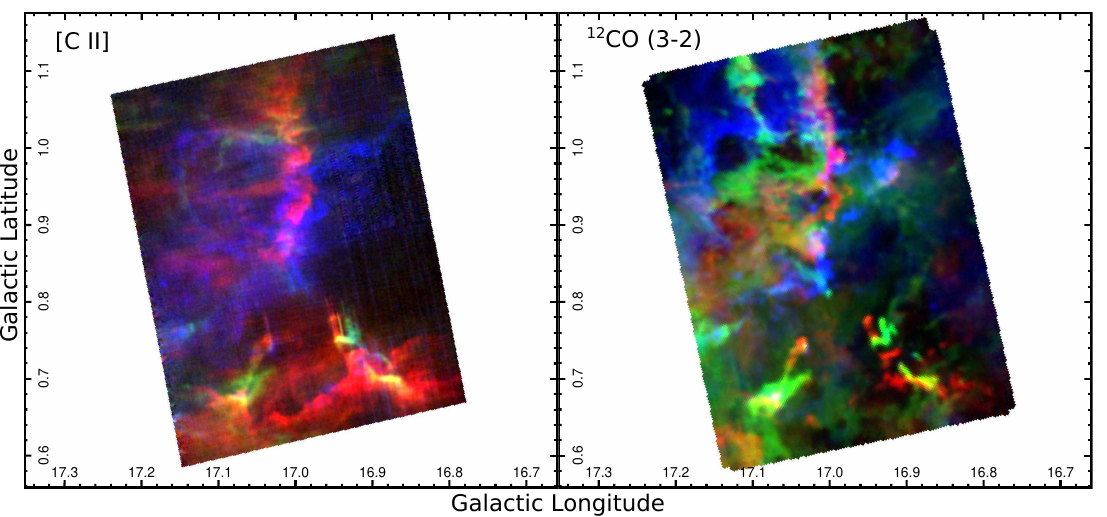}
    \caption{Velocity RGB composites made from \cii\ (left) and \twcott\ (right) observations. Blue, green, and red show $\vlsr = 10$--21, 21--23, and 23--27~\kms\ respectively.}
    \label{fig:cii_co_rgbvelocity}
\end{figure*}

In summary, \cii\ and CO trace the natal cloud and Northern Cloud, demonstrating that these two distinct molecular cloud substructures of W~37 are illuminated.
Some differences in relative brightness and traced morphology indicate that the PDR and molecular gas structures are similar but not one-to-one; the red-blue images in Figure~\ref{fig:cii_co_rb_overlay} reveal a complex interplay between the phases.
Gas behind the Bright Northern Ridge and along the southern PDR ridge is mostly atomic, and there are reservoirs of molecular gas attached to illuminated structures like the Bright Northern Ridge and Northern Cloud.
There is also a large reservoir of molecular gas between 21--23~\kms\ which lacks a bright PDR counterpart and may represent a portion of the W~37 gas reservoir which NGC~6611 has not influenced.

Optical absorption coincident with 17\Endash21~\kms\ CO and \cii\ emission, shown in Figure~\ref{fig:northerncloud-optical-co}, places the Northern Cloud in between the observer and NGC~6611/the natal cloud.
An arch-like segment along the southern edge of the Northern Cloud opens toward NGC~6611, near where the cloud extends into the bright \hii\ region in Figure~\ref{fig:northerncloud-optical-co}, indicating that the Northern Cloud is illuminated and influenced by the cluster.
N19 is a separate ring driven into the Northern Cloud by W584, the O9 star inside it \citep{Hillenbrand1993AJ....106.1906H, Guarcello2010A&A...521A..61G}.
The bottom edge of the Northern Cloud, which contains PDR and molecular gas, may be compressed between opposing feedback from W584 and NGC~6611.

\subsection{M16 Expanding Shell} \label{sec:m16-expanding-shell}
\begin{figure*}
    \centering
    \includegraphics[width=\textwidth]{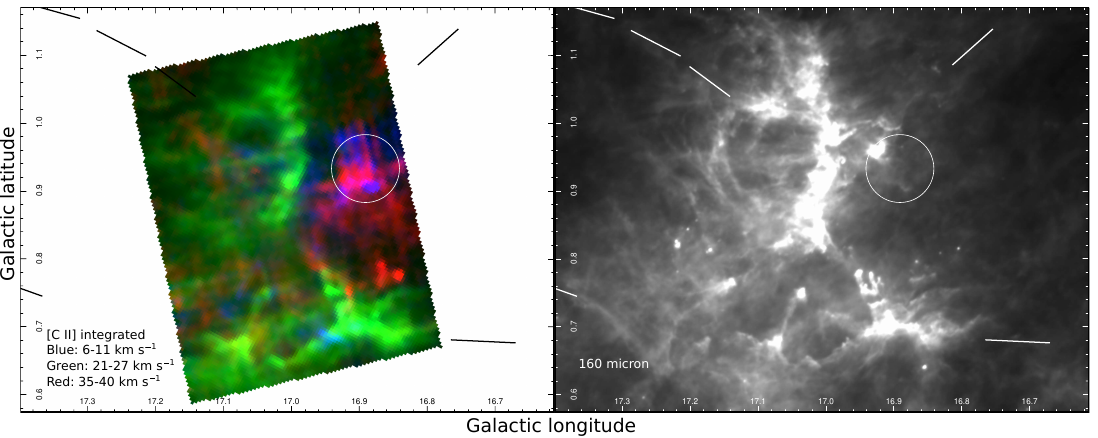}
    \caption{Left panel: Color composite showing \cii\ integrated intensities in three velocity intervals. Blue, green, and red show $\vlsr = 6$--11, 21--27, and 35-40~\kms, respectively. The blue and red velocity intervals highlight the detected foreground and background fragments of the expanding shell. Right panel: the same field at 160~\micron\ for reference. The lobes, which extend outside the field shown here, are marked in both panels with white or black dashed lines near the corners of the field. The white circle in both panels shows the area from which the spectrum in Figure~\ref{fig:diagram_expanding_shell} is extracted.}
    \label{fig:cii_m16shell_rgb}
\end{figure*}

\begin{figure*}
    \centering
    \includegraphics[width=\textwidth]{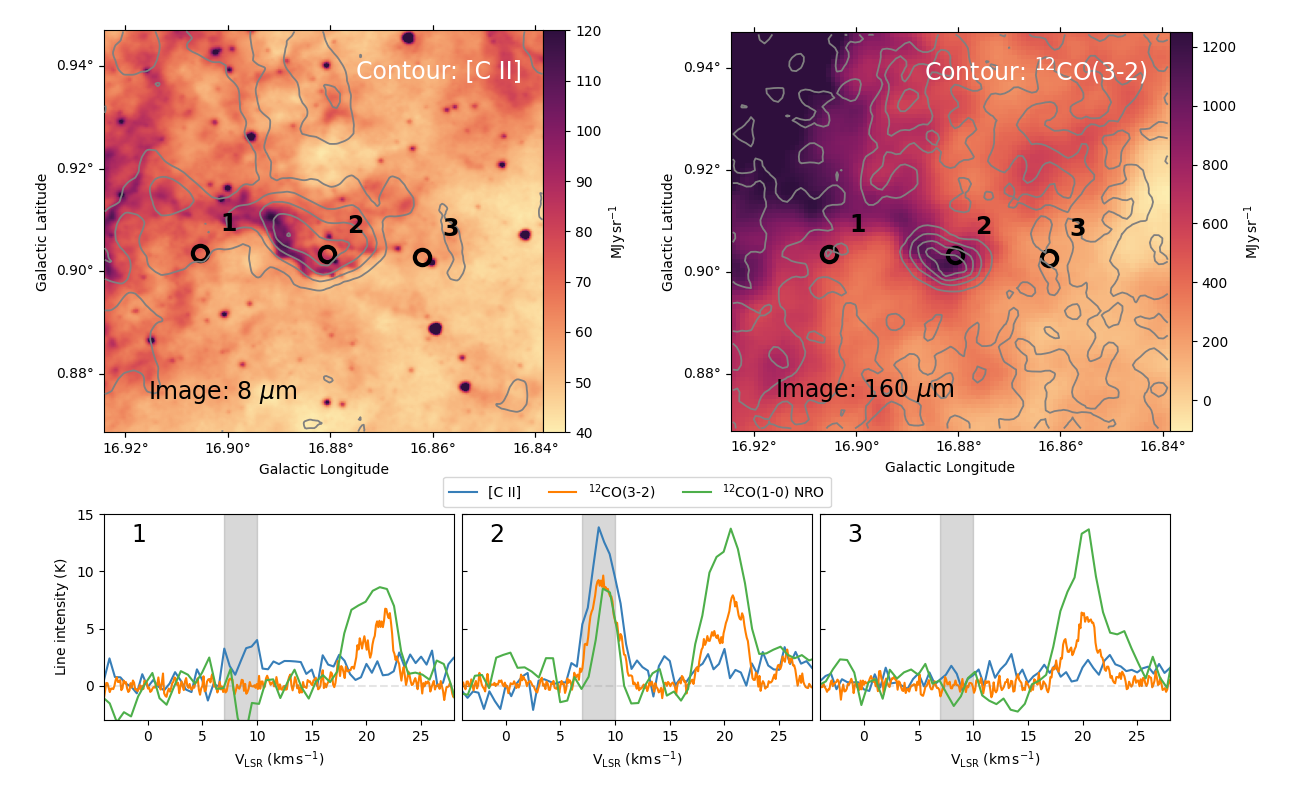}
    \caption{Infrared images and \cii\ and CO spectra highlighting the blueshifted clump dsecribed in Section~\ref{sec:m16-expanding-shell}, the only shell fragment detected in CO, on the edge of the opening of the western cavity.
    This clump links low-velocity emission from within the western cavity with a bright-rimmed cloud in high spatial resolution images, indicating that the clump is embedded within the foreground shell like those found in Orion by \citet{2020A&A...639A...1G_COglobOrion}.
    The two images on top show a zoom-in on the blueshifted clump at 8 and 160~\micron.
    The three circles mark three positions from which we extract spectra. One position lies towards the clump and two lie to each side for comparison.
    The contours are \cii\ (left) and \twcott\ (right) intensities integrated between $\vlsr = 7\Endash10~\kms$.
    The spectra on the bottom are extracted at the positions numbered 1--3.
    \cii\ spectra and \twcoA\ (\jmton{1}{0}) and (\jmton{3}{2}) spectra are shown.
    We show the Nobeyama CO (\jmton{1}{0}) observations since the 20\arcsec\ beam is similar to that of the APEX CO (\jmton{3}{2}) data.
    The \cii\ contours and spectra are both shown at the CO (\jmton{3}{2}) beam.
    A clear line associated with the blueshifted clump appears around 8--9~\kms, well separated from the diffuse 20--25~\kms\ emission.
    In Section~\ref{sec:mass} we measure the clump's atomic gas mass to be $\sim$20~$M_\odot$.}
    \label{fig:cii_blue_shell_clump}
\end{figure*}

High ($\vlsr \sim$35--40~\kms) and low ($\sim$10--15~\kms) velocity \cii\ emission relative to the bulk 19--26~\kms\ emission is detected towards the western opening of the M16 cavity, just west of the Pillars of Creation.
High velocity emission (red in Figure~\ref{fig:cii_m16shell_rgb}), is diffuse and fills part of the western opening while low velocity emission (blue in the same Figure) is concentrated in a few clumps.
The 9~\kms\ clump of \cii\ and CO emission mentioned in Section~\ref{sec:results_both_channel} is coincident with a bright-rimmed cloud (BRC) facing towards NGC~6611 in the IR, shown in Figure~\ref{fig:cii_blue_shell_clump}.
This is reminiscent of the CO globules observed by \citet{2020A&A...639A...1G_COglobOrion} to be embedded along the Orion Veil, which were also blueshifted by some $\sim$5\Endash10~\kms\ relative to the central emission, comparable to the 13~\kms\ expansion of the Veil bubble \citep{Pabst2019Natur.565..618P}.
The 8 and 160~\micron\ maps show that this blueshifted clump/BRC is nestled within a network of faint PDR structures near the northern end of the western cavity opening.
This PDR emission does not extend all the way across the opening, indicating a thin or broken foreground shell.

The western opening overlaps with the \cii\ map tile affected by the truncated observations mentioned in Section~\ref{sec:obs-sofia}, which cause the striped artifacts in Figure~\ref{fig:cii_m16shell_rgb} near the red and blue emission.
The noise RMS increases to $\sim$2.5~K in this tile so that the redshifted emission is typically detected at $\sim 2 \sigma$ and the blueshifted clump at $\sim 4 \sigma$.

We conclude that this faint PDR area on the northern edge of the western opening is a composition of limb brightened foreground and background shell expanding at no less than $\sim$10~\kms, the LOS-projected velocity separation from the $\vlsr \sim 25$~\kms\ natal cloud emission.
We do not detect significant high or low velocity emission in the \cii\ spectra towards the middle of the western opening, shown in orange in Figure~\ref{fig:cii_m16shell_rgb}, so the emission is faint if present at all.

\subsection{N19}
\begin{figure*}
    \centering
    \includegraphics[width=\textwidth]{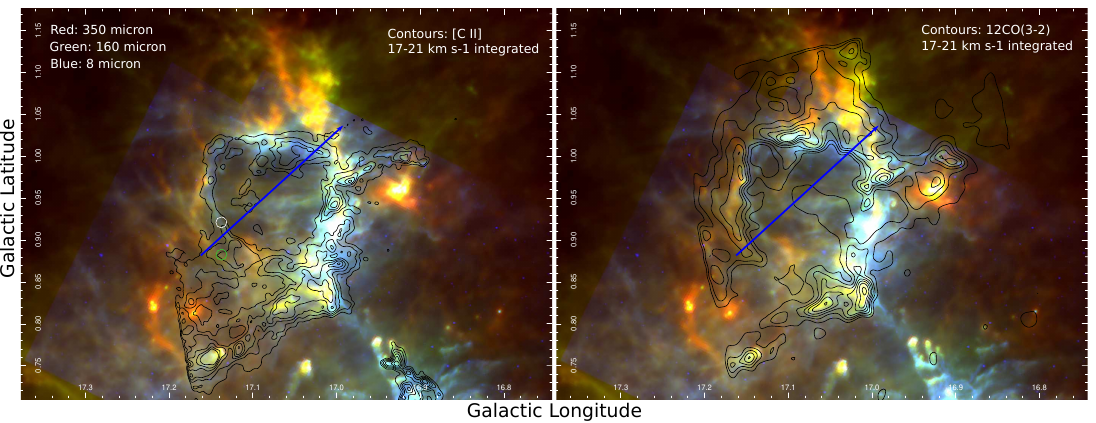}
    \caption{Infrared and millimeter view of the N19 ring. The color composite shown in both panels is 350~\micron\ in red, 160~\micron\ in green, and 8~\micron\ in blue. Contours show \cii\ (left) and \twcott\ (right) integrated between $\vlsr = 17\Endash21~\kms$. \cii\ contours mark 18, 24 30, 36, 42, and 48 K~\kms. CO contours mark 20, 30, 40, 50, 60, 70, and 80 K~\kms. The white and green circles in the left panel mark the areas from which the spectra in Figure~\ref{fig:n19-self-abs} are taken. The blue arrow in both frames shows the path for the PV diagram in Figure~\ref{fig:n19-pv}.}
    \label{fig:n19-fir}
\end{figure*}

% moved the first stuff to m16 intro section
The N19 ring appears as a thin $\sim$2~pc radius ring in Figure~\ref{fig:m16_irac_rgb} with a smooth interior in the mid-to-far IR, out to 160~\micron, and in \cii.
CO and 250--870~\micron\ trace a thicker ring with a jagged, clumpy interior (Figure~\ref{fig:northerncloud-optical-co}) with a radius $\sim$0.5 parsec larger than the mid-IR ring.
The wider molecular ring appears in absorption in the mid-IR (compare the 8~\micron\ emission to the FIR emission in Figure~\ref{fig:n19-fir}) and must be cold.

\begin{figure}
    \centering
    \includegraphics[width=0.45\textwidth]{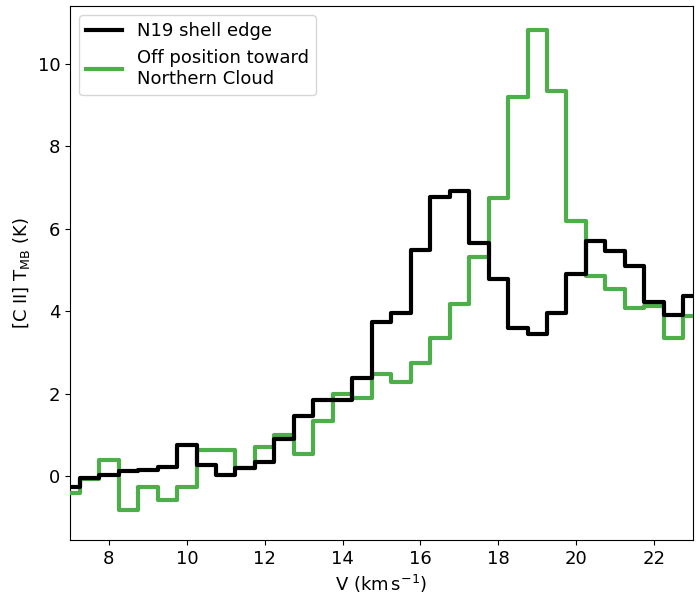}
    \caption{\cii\ spectra towards and off the edge of the N19 PDR shell showing possible signs of self-absorption. The black spectrum towards the shell edge dips near 19~\kms, and the nearby Northern Cloud spectrum in green peaks close to that velocity. The black and green spectra are taken from the white and green circles, respectively, marked in Figure~\ref{fig:n19-fir}.}
    \label{fig:n19-self-abs}
\end{figure}

The infrared ring is traced by the \cii\ line between $\vlsr = 17$--18.5~\kms\ and reappears at $\sim$21.5~\kms.
Spectra towards the northeast part of the ring show a self-absorption signature at 19~\kms\ as seen in Figure~\ref{fig:n19-self-abs}, while \cii\ spectra from nearby parts of the Northern Cloud peak at the same 19~\kms\ and the CO ring peaks around 19--20~\kms.
These observations can be explained by placing the observer behind the PDR, so that the colder, deeper layers absorb emission from the warmer surface layers along our line of sight.

\begin{figure*}
    \centering
    \includegraphics[width=0.75\textwidth]{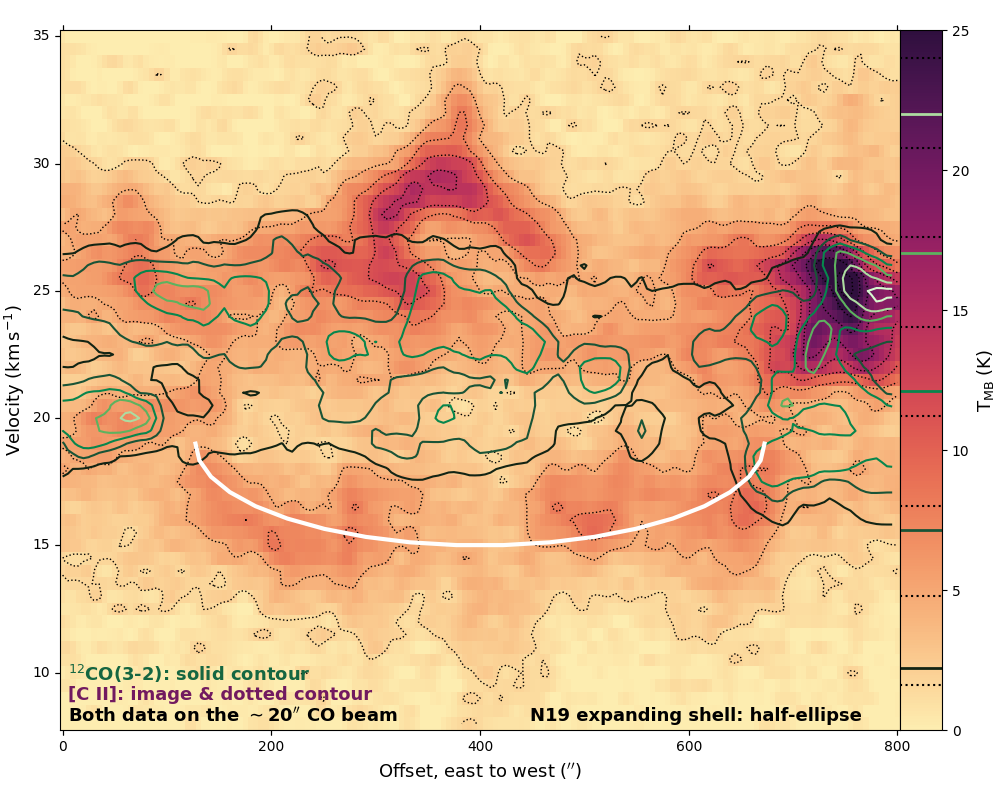}
    \caption{PV diagram showing \cii\ in color image and \thcott\ in contours. The N19 ring appears in \cii\ $\vlsr < 21~\kms$ and the expanding 4~\kms\ foreground shell signature is marked with a white half-ellipse. The path along which this PV cut is taken is shown as a black arrow in Figure~\ref{fig:n19-fir}. Emission $\vlsr > 23~\kms$ is mostly associated with the Bright Northern Ridge, and the $\vlsr > 30~\kms$ emission is unrelated to N19, as discussed in the text.}
    \label{fig:n19-pv}
\end{figure*}

The signature of half an expanding shell moving towards the observer appears inside the ring in \cii\ at $\vlsr \lesssim 17~\kms$ and reaches a velocity of $\sim$13~\kms\ at the center of the ring.
It appears in the position-velocity (PV) diagram in Figure~\ref{fig:n19-pv} as an arc deflecting downward, marked with a line.
A redshifted counterpart is not confirmed in either \cii\ or CO, but due to the confusion with the Bright Northern Ridge and other natal cloud emission, we cannot rule out its existence.
The PV diagram shows a high velocity $\vlsr \gtrsim 28~\kms$ feature co-located with the foreground shell, but the channel maps in Figure~\ref{fig:channel_maps_cii_co} reveal the feature to be $\sim$0.5~pc across and appear morphologically unrelated to the shell.
\citet{Faerber2025ApJ...990...30F} find an over-representation of \hii\ regions with a blueshifted signature but no redshifted signature in their sample, so it is also possible that some observational artifact obscures N19's redshifted signature.

\section{Geometry} \label{sec:geom}
\begin{figure*}
    \gridline{\fig{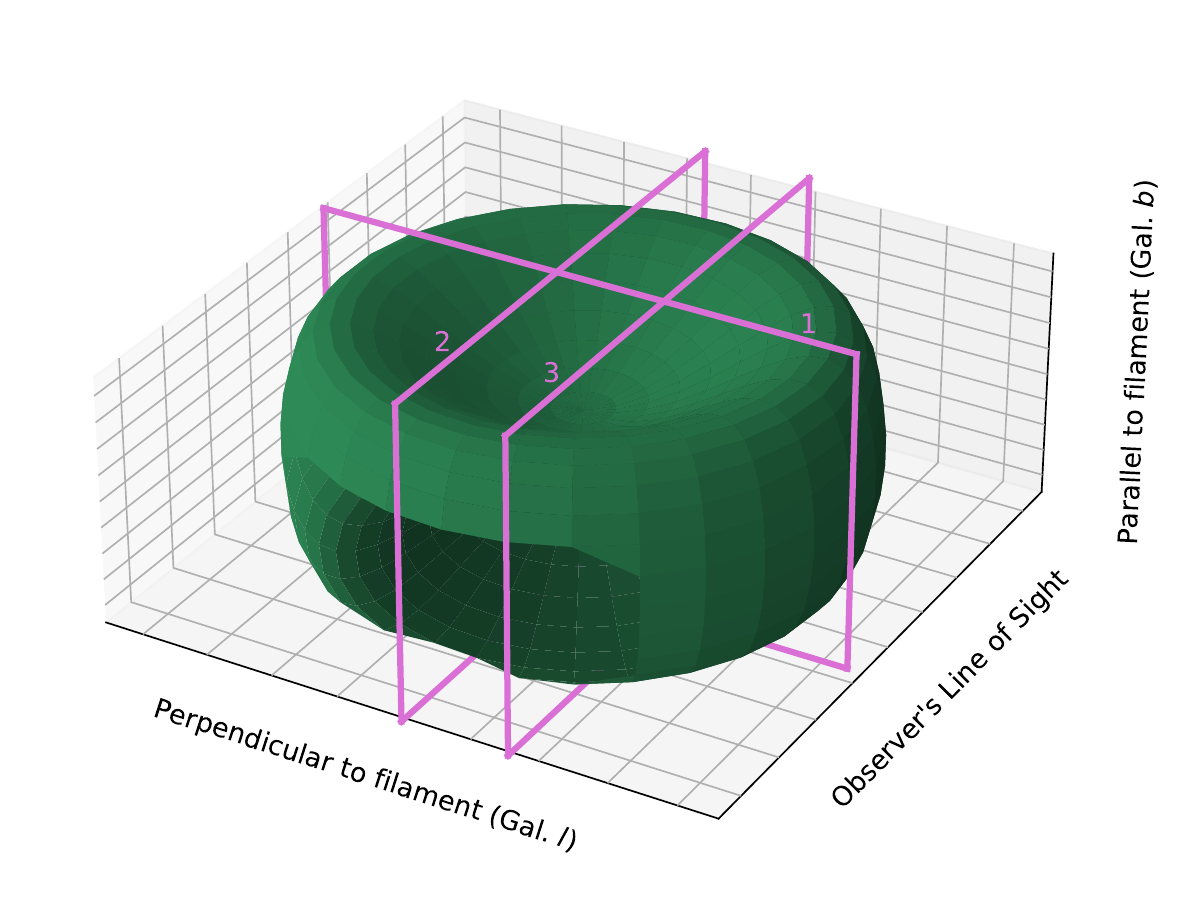}{0.5\textwidth}{(A)} \fig{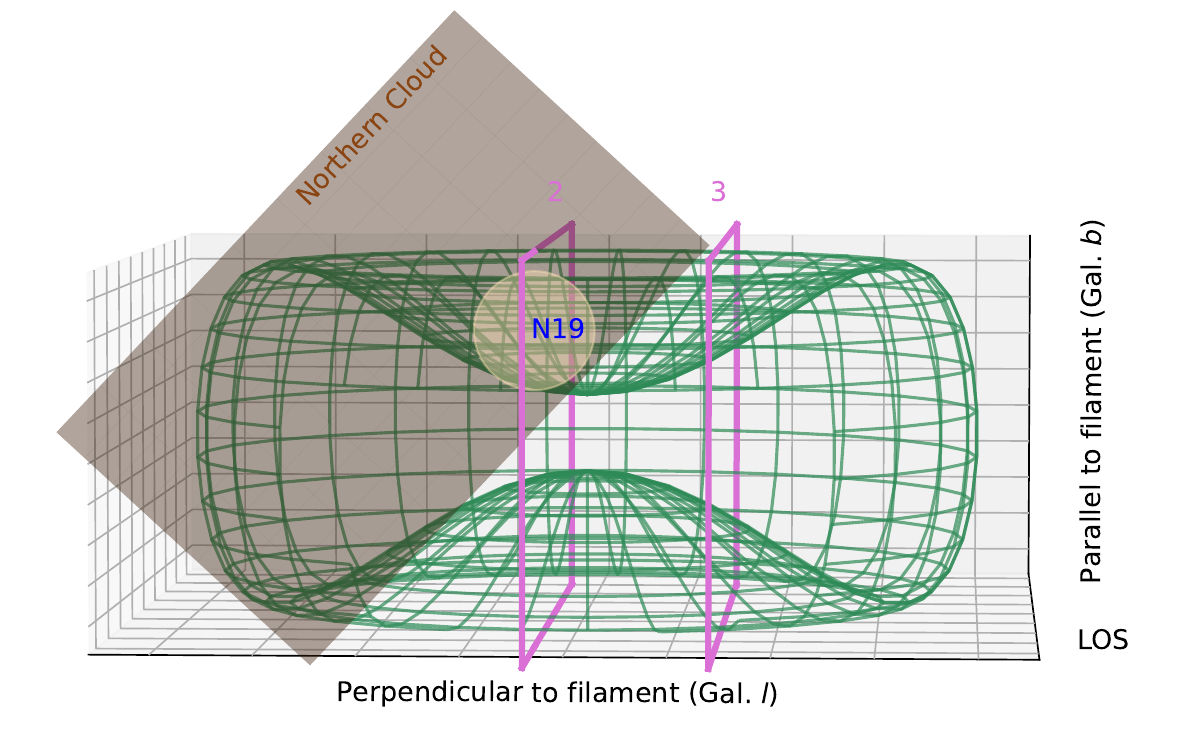}{0.5\textwidth}{(B)}}
    \gridline{\fig{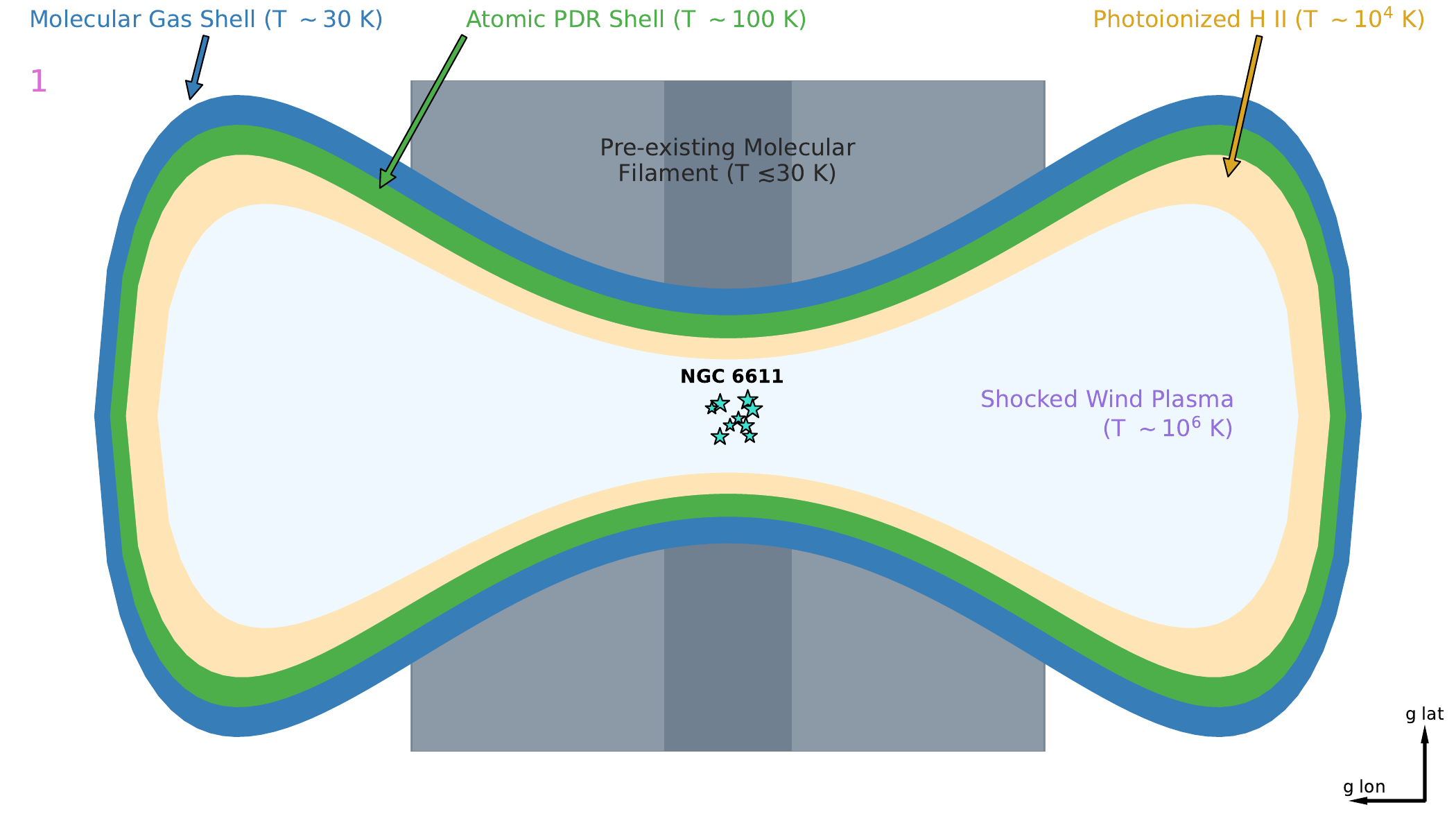}{0.5\textwidth}{(C)}
    \fig{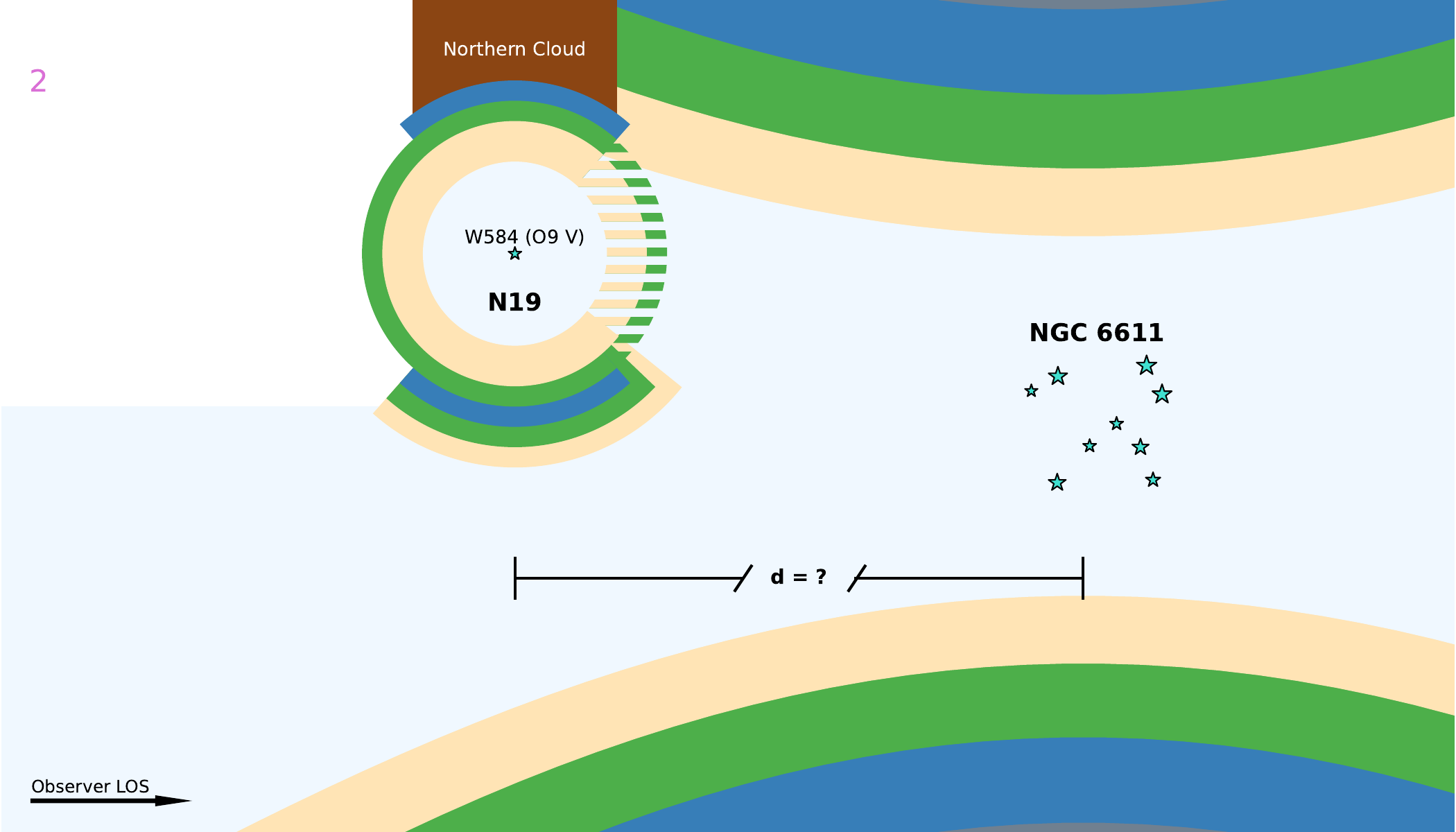}{0.5\textwidth}{(D)}}
    \caption{Four diagrams describing a simplified geometric picture of the shell. The top-left and top-right panels show a 3-dimensional schematic of the ``biconcave disc'' shape viewed from two different angles and with a solid and transparent wireframe surface respectively. The front side has been opened to show the inside of the shape as well as model how M16 may be broken open; observations suggest it is either broken or very thin towards the front and back. Pink frames numbered 1--3 outline the cross-cuts which are diagrammed in the lower-left (1) and lower-right (2) panels of this Figure and in the top panel of Figure~\ref{fig:diagram_expanding_shell} (3). The top-right panel shows the shell viewed approximately face-on from the observer's perspective and includes a semi-transparent rectangular overlay representing the Northern Cloud, which obstructs the left lobe of the cavity, and a circle representing N19, both of which appear in the \#2 cross-cut. The bottom-left panel shows cross-cut \#1 marked in the top-right panel, along the plane of the sky. Gas phases and the exciting cluster, NGC~6611, are labeled. The bottom-right panel shows cross-cut \#2, marked in the top two panels, which includes the Northern Cloud and N19. It is not determined whether N19 has an intact background shell, so that side is represented by a hashed surface. We do not detect a foreground or background molecular gas shell, so these are excluded from the diagram. The scale bar indicates that the physical separation between NGC~6611 and the Northern Cloud/N19 is unknown, though we estimate it in Section~\ref{sec:geom-nc-n19} to be $\lesssim \expo{2}$~pc.}
    \label{fig:diagram}
\end{figure*}

\begin{figure*}
    \gridline{\fig{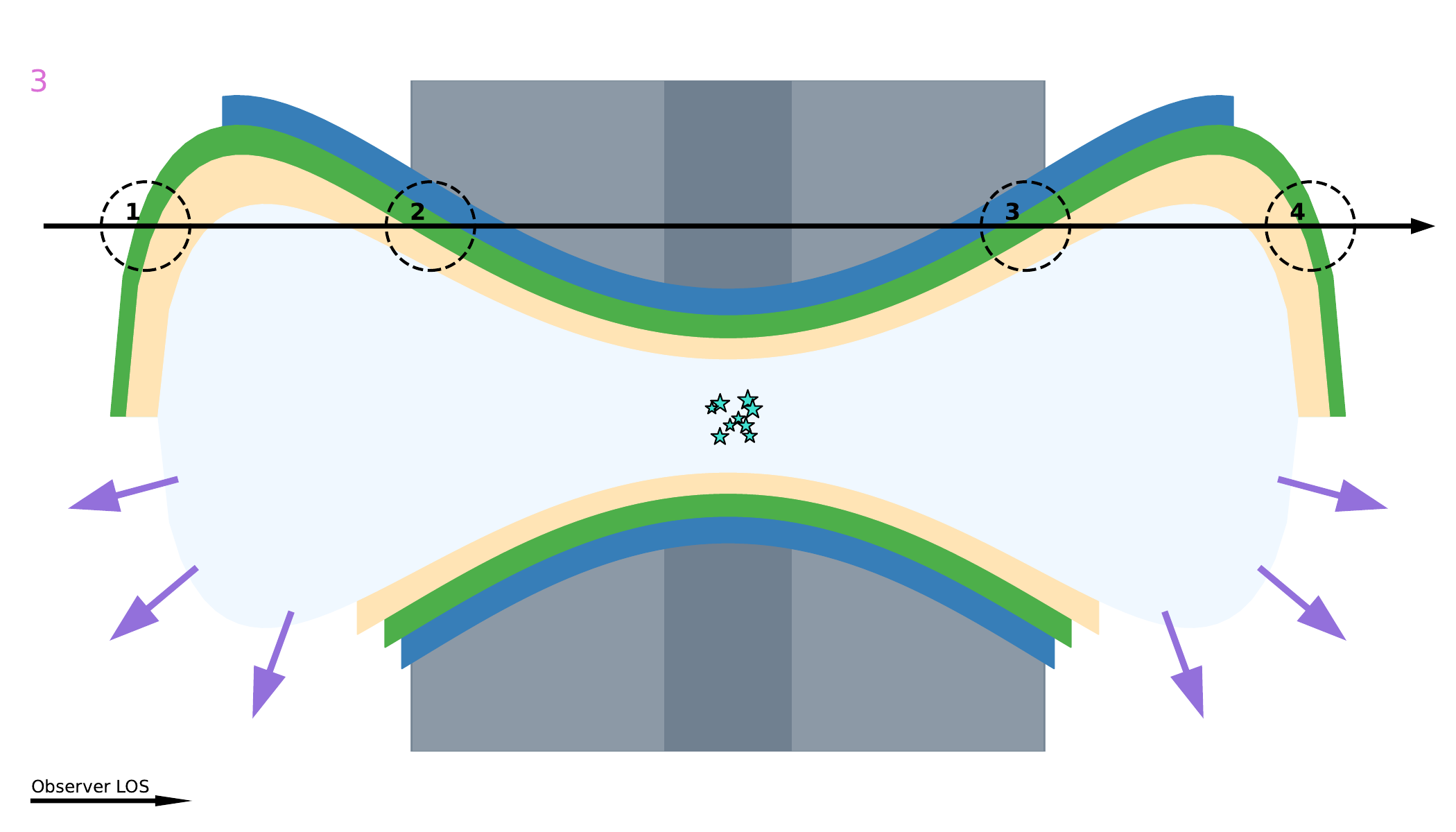}{0.7\textwidth}{(A)}}
    \gridline{\fig{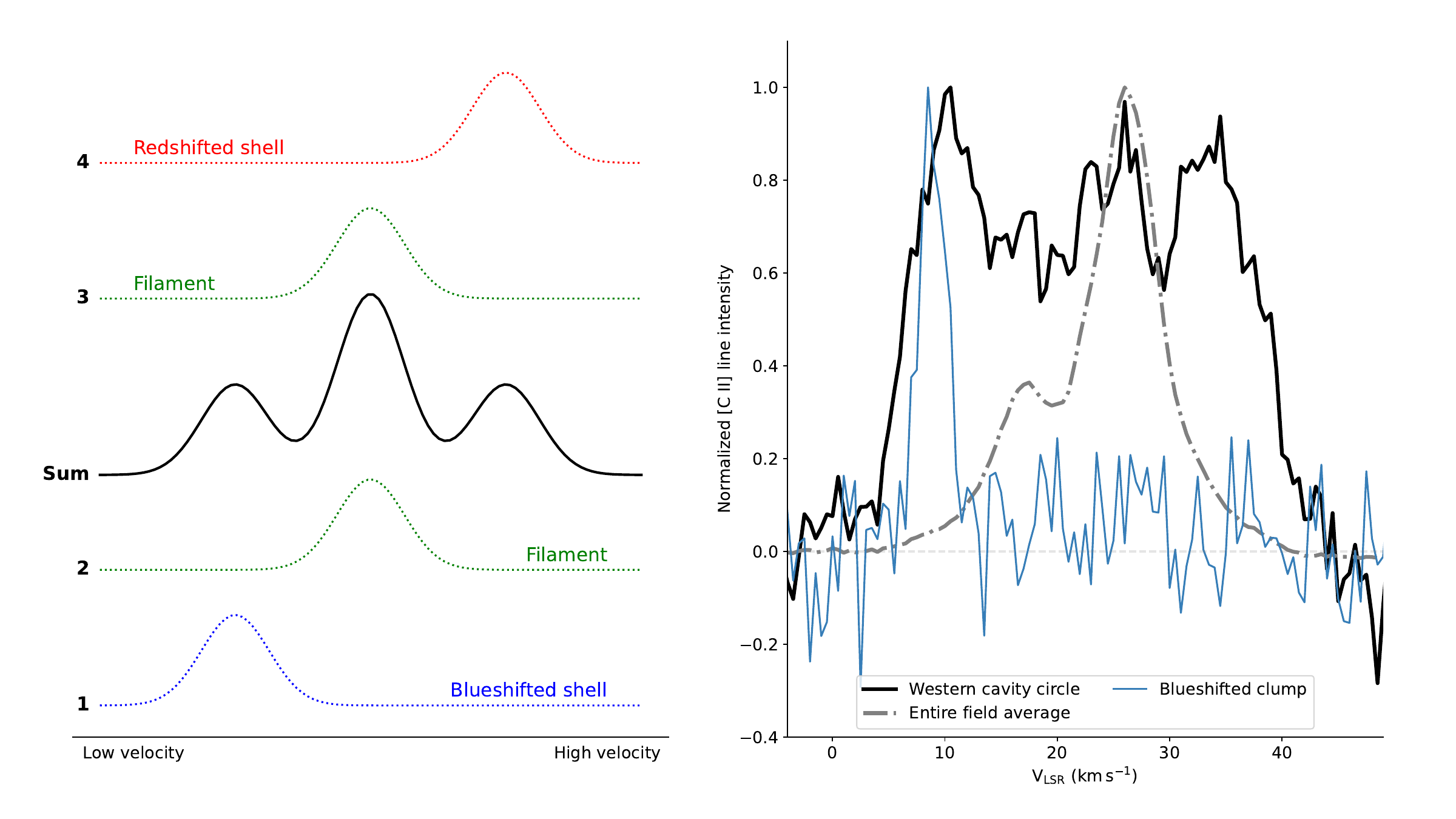}{0.7\textwidth}{(B)}}
    \caption{The top panel shows an observer's line of sight cross-cut through the M16 shell and cavity, marked as \#3 in the 3-dimensional diagrams in Figure~\ref{fig:diagram}.
    The front and back of the shell are depicted as broken open.
    The black horizontal arrow through the top of the shell shows an observer's line of sight, and numbered black dashed circles mark where the line of sight crosses neutral PDRs which would radiate \cii\ line emission.
    The bottom panels show a cartoon spectrum on the left and observed \cii\ spectra on the right.
    The line of sight PDR crossings circled in the top panel would contribute the colored dashed-line spectrum components, numbered accordingly, in the cartoon spectrum.
    The solid black cartoon spectrum is the sum of the four component spectra, representing what would be observed.
    This is described in more detail in Section~\ref{sec:geom-m16-filament}. The observed spectrum labeled ``Western cavity circle'' is from the circular region shown in Figure~\ref{fig:cii_m16shell_rgb} towards the shell fragments.
    We include a spectrum towards the blueshifted clump, a shell fragment with a clear IR counterpart (Section~\ref{sec:m16-expanding-shell}, Figure~\ref{fig:cii_blue_shell_clump}).
    The average \cii\ spectrum from the entire field is shown in dash-dotted gray to emphasize the natal cloud component near $\vlsr = 25\Endash26~\kms$.
    The observed spectra are consistent with our model.}
    \label{fig:diagram_expanding_shell}
\end{figure*}

\subsection{The Natal Cloud and M16 Cavity} \label{sec:geom-m16-filament}
In Section~\ref{sec:m16-struct} we discussed the infrared structure of M16 and its appearance as an expanded \hii\ region which is constrained along the axis of the GMF; this is summarized in the annotated chart in Figure~\ref{fig:m16_finder}.
This projected appearance may arise from either a 3-dimensional bipolar \hii\ region with two separate cavities (like RCW~36; \citealt{Bonne2022ApJ...935..171B_RCW36}), or a single large ellipsoidal cavity, and we argue here for the latter case.

We propose the following 3-dimensional model, shown in Figure~\ref{fig:diagram}, for the large cavity created by feedback from NGC~6611.
The GMF constrains cavity expansion towards and away from the Galactic plane, while relatively lower column density towards all other angles allows expansion.
The result is an ellipsoidal or toroidal cavity.
For the purpose of our diagram, we adopt a shape called the ``biconcave disc,'' used in cell biology to describe the shape of the typical red blood cell \citep{cells_biconcave} because it resembles our observations and has an analytical description that is easily plotted.
Figure~\ref{fig:diagram}(C) shows a plane-of-sky cross-cut through the biconcave disc so that the cross-cut resembles on-sky observations of the infrared lobes.
The full horizontal width of the cavity is $\sim$40~pc based on our observations, while the height of the cavity (at its tallest, off-center) is $\sim$12~pc; the diagrams are not to scale.

Numerical simulations reveal that density structures and inhomogeneities are key to the morphology created by stellar feedback in molecular clouds as they focus their expansion into regions of ``least resistance'' \citep{Zamora-Aviles2019MNRAS.487.2200Z}.
Simulations by \citet{Fukuda2000ApJ...533..911F} (see Figures~1 and 3 in their paper) produce a cavity compressed along one axis by a filament, while \citet{Whitworth2021MNRAS.504.3156W} suggest that recombinations along a dense surface can slow and perhaps even trap an encroaching ionization front.

Bipolar \hii\ regions have been observed and studied throughout the Galaxy and form within molecular gas sheets.
A separate cavity bursts out of each side of the sheet, while the \hii\ region expansion is constrained in the plane of the sheet (see the diagrams in Figures~1 and 2 in \citealt{Deharveng2015A&A...582A...1D}).
\citet{Deharveng2015A&A...582A...1D} describe one of the characteristic features of bipolar \hii\ regions as a ring of molecular gas in the plane of the sheet surrounding the cluster.
\citet{Bonne2022ApJ...935..171B_RCW36} observes such a ring in CO in the bipolar \hii\ region RCW~36.

Should we interpret the GMF as a sheet viewed nearly edge-on, we would expect a similar CO ring running vertically ($\pm b$) through the bright central region in front of and behind NGC~6611.
This ring would be viewed nearly edge on so that it appears as two spatially overlaid but kinematically distinct filaments in projection.
The foreground side of the ring would appear as an optical dark lane, but we detect no such ring towards M16.
Instead, the observer's line of sight is relatively unobstructed towards the exciting cluster.
While the 19~\kms\ Northern Cloud lies in front of the cluster, it is too spatially extended to be the foreground half of a ring created by NGC~6611 and must instead be a separate cloud of gas, as described in Section~\ref{sec:geom-nc-n19}.
The GMF is more likely a true filament, not a sheet, based on these observations.

Blueshifted \cii\ emission traces fragments rather than a complete shell (Section~\ref{sec:m16-expanding-shell}), so the foreground shell surface must be thin ($A_V \lesssim 0.5$; Section~\ref{sec:energ-m16}) or broken.
We illustrate the 3-dimensional shell model as broken open in the front in Figures~\ref{fig:diagram} and \ref{fig:diagram_expanding_shell} and we diagram an observer's line-of-sight cross-cut through the western cavity, where the \cii\ shell fragments are detected, in Figure~\ref{fig:diagram_expanding_shell}.
Purple arrows show escaping shocked-wind plasma in Figure~\ref{fig:diagram_expanding_shell}, whose large sound speed allows it to wrap around the region quickly compared to the dynamical timescale of the cavity.
The background side of the cavity is obscured in the spectroscopically unresolved images by the bright central emission, and it is faint in \cii\ like the foreground side.
For lack of more detailed information, we assume symmetry in our model.

A black arrow representing an observer's line of sight passes through the partial foreground/background of the shell in the top panel of Figure~\ref{fig:diagram_expanding_shell}.
This line of sight passes through the atomic PDR, colored green in the diagram, 4 times, marked with black dashed circles and numbered.
The \#1 and \#4 PDR crossings should be blueshifted and redshifted, respectively, since the cavity can easily expand in those directions.
The \#2 and \#3 PDRs are on the surface of the dense filamentary gas of the natal cloud and should not be so easily accelerated by stellar feedback due to the mass behind them, so we argue that both of these sections of PDR should share the velocity of the filament and therefore be kinematically indistinguishable from each other.
We diagram the resulting spectrum from these four PDR crossings in the bottom-left panel of Figure~\ref{fig:diagram_expanding_shell} and compare to the observed \cii\ spectrum through the western cavity in the bottom-right panel.
The observed spectrum has a middle velocity component at $\vlsr \sim 25~\kms$ which matches the natal cloud component which dominates the dash-dotted grey line in the same Figure.
We show integrated \cii\ intensity maps of the blue- and red-shifted shell fragments compared to the central component in the color composite in Figure~\ref{fig:cii_m16shell_rgb}, and we mark the lobes on the \cii\ composite and the 160~\micron\ image for reference.
The \cii\ shell fragments appear inside the cavity, but do not fill it, consistent with our conclusion that it has broken open.

The DSS2 red optical image in Figure~\ref{fig:m16_finder} shows optical emission running down from the center of the \hii\ region towards the Galactic plane approximately along the filament.
Should optical or UV emission be able to escape the cavity from the foreground break in the shell in Figure~\ref{fig:diagram_expanding_shell}, it could illuminate and/or reflect off filament-associated gas below the cavity.
The real filament would not terminate at a sharp edge as drawn in the diagram, but instead density would continue to drop off for some distance comparable to the extent of the cavity.

The Pillars of Creation, Spire, and a handful of other pillars are not depicted in our diagrams.
Their presence fits well into this model: gas near the central axis of the filament is dense and clumpy.
As dense-enough clumps within the filament are illuminated by the ionization front, they are sculpted into pillars whose tails face away from the illuminating stars.
Since these are remnants of the dense natal cloud, they are more likely to be found along the filamentary axis like the Pillars of Creation (which \citealt{Karim2023AJ....166..240K} suggest formed from an association of pre-existing density enhancements) and the Spire, both of which are connected to branches of the dense filament.
Close inspection of the 8~\micron\ map reveals a number of pillar-like structures which all point back towards the stars and are kinematically associated with larger associations of gas.

\subsection{Northern Cloud and N19} \label{sec:geom-nc-n19}
The Northern Cloud lies between the filament and the observer, obstructing optical emission from the \hii\ region as demonstrated in Figure~\ref{fig:northerncloud-optical-co}.
It spans some tens of parsecs, much of which is not illuminated by NGC~6611, as indicated by the PMO CO observations compared to the 8~\micron.
The CO (\jmton{1}{0}) observations, which cover a larger area than the \cii\ and CO (\jmton{3}{2}), show that it extends towards higher Galactic latitude.
The Northern Cloud, at $\sim$19~\kms, forms a curtain of molecular gas which blocks much of the \hii\ region from view in the optical but does not obstruct \cii, 8~\micron, 24~\micron\ and radio free-free emission, which trace PDRs and \hii\ gas behind the Northern Cloud.
The edge of the Northern Cloud facing the NGC~6611 cluster is itself a PDR, which indicates that the Northern Cloud is close to the cluster and influenced by its feedback as explained in Sections~\ref{sec:m16-struct} and \ref{sec:results-channel-summary}.

We summarize this geometry in panels B and D of Figure~\ref{fig:diagram}.
In panel B, we represent the Northern Cloud as a rectangle in front of the left lobe of M16.
We diagram a cross-cut through M16, the Northern Cloud, and N19 in panel D.
The Northern Cloud is simultaneously illuminated from within by N19's central star and from below and behind by NGC~6611; it is not a distant foreground feature, but rather a nearby cloud structure influenced by NGC~6611 feedback.

We determine that N19 is a bubble driven by feedback from W584 based on the expanding foreground shell signature in \cii\ and the projected ring structure in \cii, CO, and infrared maps.
Confusion with the Bright Northern Ridge and natal cloud emission prevents us from detecting or ruling out an expanding background shell or any PDR which could be the back face of the bubble.
We represent the back face of the bubble in Figure~\ref{fig:diagram}D with a hashed line.
The N19 ring is flat on the side facing NGC~6611, a sign that the cluster influences the shape of the bubble (Figure~\ref{fig:northerncloud-optical-co}).
The line-of-sight separation between NGC~6611 and the Northern Cloud/N19 is not constrained by our observations but they must be close enough that NGC~6611's influence is felt; we indicate this uncertainty using the scale bar in the left panel of Figure~\ref{fig:diagram}B.
The estimate of heliocentric distance to W584 by \citet{Stoop2023AA...670A.108S} is consistent (within $\sim \expo{2}$~pc) with heliocentric distance to the cluster, and we find the projected width of the M16 region to be $\sim$40~pc, so we bound N19's separation from the NGC~6611 cluster core to be $\lesssim \expo{2}$~pc.

\section{Column Densities and Masses} \label{sec:coldens}
\subsection{CO Column Densities} \label{sec:cocoldens}
We estimate \thco\ column densities using the observations of \cite{Xu2019A&A...627A..27X} and the local thermodynamic equilibrium (LTE) method used by \cite{Tiwari2021ApJ...914..117T} and \citet{Karim2023AJ....166..240K}, in which excitation temperature is obtained from the peak \twco\ brightness in the relevant velocity interval.
We adopt the isotopic ratio $\twcoA/\thcoA = 44.65$ derived for the galactocentric radius $D_{\rm GC} = 6.46$ of M16 by \citet{Karim2023AJ....166..240K}, use the abundance ratio $\twcoA/\htwo = \nexpo{8.5}{-5}$ \cite{tielens_2021} to convert CO column densities to \nht, and convert to mass using a mean molecular weight $\mu = 2.33$ \citep{2008A&A...487..993K}.

The N19 ring and the Bright Northern Ridge are particularly bright in CO(\jmton{3}{2}) and must be dense, so we use both CO transitions to derive column density and number density here.
As the CO \jmton{3}{2} and \jmton{1}{0} transitions have very different optical depths and critical densities, the estimate of column densities requires a coupled excitation and radiative transfer treatment.
We use the Radex radiative transfer software \citep{vanderTak2007A&A...468..627V} to model line emission from the \thcoA\ (\jmton{3}{2}) and (\jmton{1}{0}) and \ceighteeno\ transitions, freeing us from the assumption of LTE.
Excitation conditions are assumed to be constant throughout a given feature, so we must apply this method to coherent gas features.
Since we use \ceighteenoA, there is an additional constraint that this faint line must be detected.
The N19 CO ring and Bright Northern Ridge fulfill these criteria, so we derive column density and number density towards these two features using our Radex analysis.
We use the LTE \thco\ column density method everywhere else.

We use Radex to calculate emission for these transitions while varying the excitation conditions: kinetic temperature $T_{\rm K}$, total hydrogen column density \nht, and \htwo\ density $n$.
The transformations from \nht\ to \twcoA\ and \thcoA\ column density are given above.
We adopt the isotopic ratio \twcoA/\ceighteenoA\ $= 417$ \citep{Wilson1994ARA&A..32..191W}.
We compare an ensemble of measurements from the gas feature (either the N19 ring or the Bright Northern Ridge) at the appropriate velocity interval to modeled emission at a variety of excitation conditions.
A solution is determined for each set of measurements (the three CO transitions toward one line of sight, i.e. one pixel), and the median solution is taken from the ensemble of measurements (all pixels towards the gas feature).
The method is described in detail in Appendix~\ref{sec:appendix-radex}.

\begin{deluxetable*}{lccccc}
    \label{tab:cocoldens}
    \tablecaption{Molecular gas density and column density solutions.}
    \tablewidth{0pt}
    \tablehead{
        \colhead{Feature} & \colhead{$n$} & \colhead{\nht} & \colhead{\nht} & \colhead{\nht} & \colhead{\nht} \\
        \colhead{Name} & \colhead{Radex} & \colhead{Radex} & \colhead{\thco\ LTE} & \colhead{70, 160~\micron} & \colhead{160--500~\micron} \\
        \colhead{} & \colhead{(cm$^{-3}$)} & \colhead{(cm$^{-2}$)} & \colhead{(cm$^{-2}$)} & \colhead{(cm$^{-2}$)} & \colhead{(cm$^{-2}$)}
    }
    \decimalcolnumbers
    \startdata
        Bright Northern Ridge & $\nexpo{8.9_{-3.3}^{+3.7}}{3}$ & $\nexpo{7.1_{-2.1}^{+5.5}}{21}$ & $\nexpo{1.0\pm0.5}{22}$ & $\nexpo{3.1\pm2.2}{22}$ & $\nexpo{2.2\pm1.3}{22}$ \\
        N19 & $\nexpo{5.6_{-1.6}^{+2.3}}{3}$ & $\nexpo{1.1_{-0.4}^{+0.7}}{22}$ & $\nexpo{1.6\pm0.5}{22}$ & $\nexpo{2.2\pm1.5}{22}$ & $\nexpo{1.9\pm0.9}{22}$ \\
    \enddata
    \tablecomments{Density and column density for the Bright Northern Ridge and N19. Density and the leftmost column density are derived using Radex. Uncertainties from the Radex fit are asymmetric so the upper and lower error bars are given as superscripts and subscripts, respectively. The remaining columns list the mean $\pm$ standard deviation of the column densities derived toward each feature using the other methods discussed in the text. The mean and standard deviation are calculated from within the same mask as the Radex fits. All column densities are expressed as \nht. The Radex and CO LTE measurements are made at the PMO resolution, while the two FIR dust emission measurements are made at higher resolutions and so vary more within the mask.}
\end{deluxetable*}

The two \twcoA\ transitions are inconsistent with each other at any given kinetic temperature; $T_{\rm K}$ can be tuned to make either one agree with the three other measurements, but never both.
\twco\ measurements require $T_{\rm K}$ to be $\sim$5~K higher than \twcott, which in turn means that \twco\ leads to lower \nht\ and $n$ solutions than \twcott.
A combination of line-of-sight variability and differing optical depths cause the two optically thick transitions to be sensitive to different layers of gas with different excitation conditions.
Instead of choosing one of the optically thick \twcoA\ lines to set the temperature, we maintain $T_{\rm K} = 30$~K for all solutions since it is approximately consistent with both \twcoA\ measurements.
We tested the stability of the solutions under temperature variation between 20--40~K and find an inverse relationship between assumed $T_{\rm K}$ and the $n$ solution.
The column density solution is less sensitive.
Temperature-driven variation in both solution parameters is comparable to their estimated uncertainties when $T_{\rm K} = 30$~K is assumed.

Column densities derived from the \thco\ transition under LTE tend to agree with the column densities derived using the Radex grid method.
Both of these agree within a factor of $\sim$2 with column densities derived from FIR dust emission; dust-derived \nht\ are larger than CO-derived \nht.
Average column densities from each of these techniques are given for N19 and the Bright Northern Ridge in Table~\ref{tab:cocoldens}.

\subsection{\cp\ Column Densities}
We calculate the hydrogen column density \ntot\ associated with \cp\ using the \cii\ line using the method described by \citet{Karim2023AJ....166..240K}.

\begin{figure}
    \centering
    \includegraphics[width=0.5\textwidth]{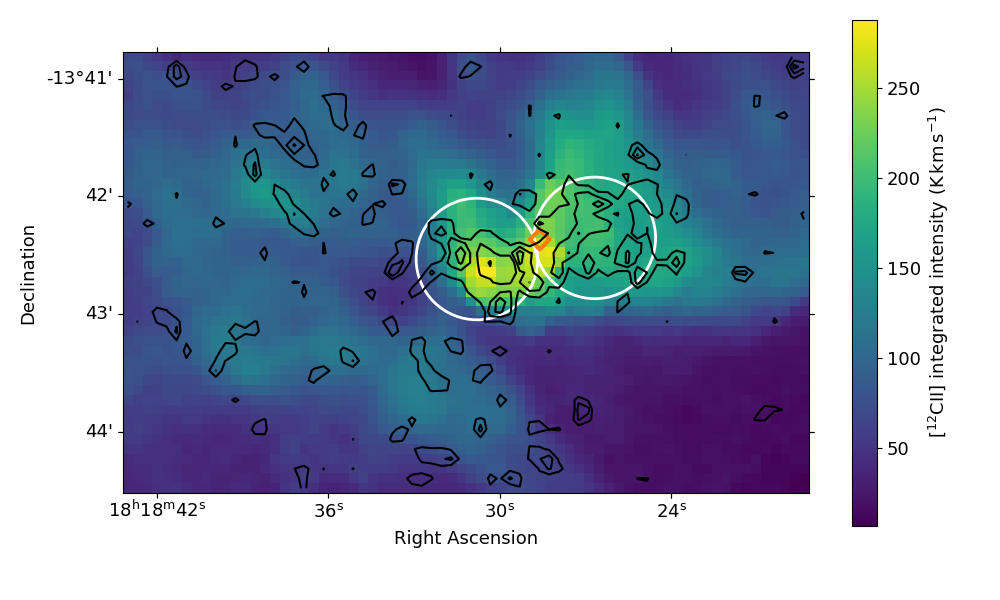}
    \caption{Integrated \cii\ (color image) and \thcii\ (contours) intensities between $\vlsr = 26\Endash31~\kms$ towards IRAS 18156--1343 (orange diamond) along the Bright Northern Ridge. Circles show the two regions in which spectra are averaged for Figure~\ref{fig:13cii-spec}. Each circle is 4 \cii\ beams across. The emission from within the left circle is $\sim$1~\kms\ higher velocity than that from the right circle. The brightest \cii\ emission locations are slightly offset from the \thcii\ emission locations.}
    \label{fig:13cii-contour}
\end{figure}

\begin{figure*}
    \centering
    \includegraphics[width=\textwidth]{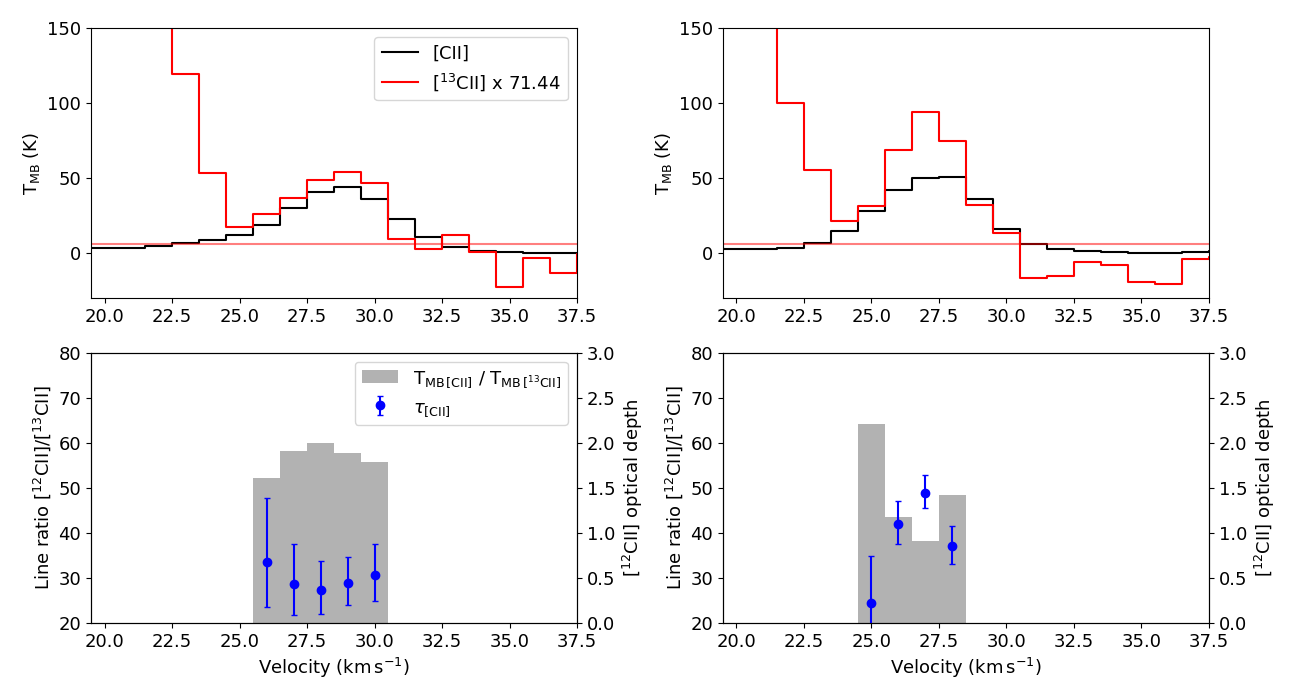}
    \caption{The top two panels show \cii\ (black) and \thcii\ (red) spectra averaged within the two circles in Figure~\ref{fig:13cii-contour}; the left circle's spectra are on the left here. The step plots show spectra binned to 1~\kms\ channels. The \thcii\ spectra are scaled up by the factor $\alpha / s_F$, described in the text, and shifted by $-11.2~\kms$ to account for the \thcii\ rest frequency. The horizontal line shows the 1~$\sigma$ level for \cii\ scaled up by $\alpha / s_F$ and adjusted for the number of averaged spectra. Note that the lower velocity side of the \thcii\ line overlaps with the relatively bright high velocity line wing of the \cii\ line. The bottom panels show the \cii / \thcii\ ratio (grey bars) and the \cii\ optical depth and associated uncertainty (blue points). We exclude channels where \thcii\ is below 1~$\sigma$ or where \cii\ $T_{\rm MB}$ exceeds the adjusted \thcii\ $T_{\rm MB}$. This figure is based on those by \citet{Guevara2020A&A...636A..16G}.}
    \label{fig:13cii-spec}
\end{figure*}

We detect the $F = 2-1$ \thcii\ line ($+$11.2~\kms\ relative to the \cii\ line; \citealt{Cooksy1986ApJ...305L..89C, Ossenkopf2013A&A...550A..57O, Guevara2020A&A...636A..16G}) only toward IRAS 18156--1343 within the Bright Northern Ridge.
The integrated \thcii\ intensities are shown in contour in Figure~\ref{fig:13cii-contour} and spectra are shown in red in Figure~\ref{fig:13cii-spec}.
The \thcii\ spectra are scaled up by the factor $\alpha / s_F$, where $\alpha$ is the assumed carbon isotopic ratio 44.65 \citep{Karim2023AJ....166..240K} and $s_F$ is the hyperfine transition strength coefficient 0.625 for the $F = 2-1$ line \citep[in their Table~1]{Guevara2020A&A...636A..16G}.
Estimated \Tex\ (described below) reaches nearly 150~K in a small area surrounding the source.

Since \thcii\ is only detected toward the IRAS source, that spectrum is likely not representative of the entire region.
The source is associated with higher optical depth \cii\ emission, up to $\tau = 2.2$ as shown in Figure~\ref{fig:13cii-spec}.
For the rest of the \cii\ emission, we apply the methodology developed by \citet{Okada2015A&A...580A..54O} and used by \citet{Karim2023AJ....166..240K}.
To calculate \Tex, we use the same $\tau < 1.3$ upper limit on optical depth as \citet{Karim2023AJ....166..240K}.
This is close to the typical upper limit calculated toward bright lines of sight south of the Pillars and along the Bright Northern Ridge.

We calculate \Tex\ for the Northern Cloud and natal cloud velocity intervals separately.
For the $\vlsr \sim 23\Endash27~\kms$ natal cloud emission, we assign $\Tex = 120$ based on the maximum \Tex\ calculated using $\tau = 1.3$ towards the southern PDR ridge as well as the Bright Northern Ridge.
We let $\Tex$ exceed 120~K in the small area surrounding the IRAS source, assigning to each pixel the $\Tex$ calculated from its peak line intensity assuming $\tau = 1.3$.
We assign $\Tex = 67$~K for gas between $\vlsr = 10\Endash21~\kms$ based on the maximum observed brightness towards the Northern Cloud and N19.

Column densities of \cp\ are converted to \ntot\ using the C/H abundance ratio $\nexpo{1.6}{-4}$ \citep{Sofia2004ApJ...605..272S}.
We find typical column densities $\ntot \sim 3\Endash6 \times 10^{21}$~cm$^{-3}$ towards most places and $\ntot \sim 1\Endash2 \times 10^{22}$~cm$^{-3}$ towards sites of bright \cii\ emission, including around IRAS 18156--1343.
We convert to mass using mean molecular weight $\mu = 1.33$ for atomic gas.

\subsection{FIR Column Densities}
Column densities from 70 and 160~\micron\ thermal dust emission are calculated directly following the method of \citet{Tiwari2021ApJ...914..117T} and \citet{Karim2023AJ....166..240K}.
We apply no zero-point offset to the 160~\micron\ map, and define the 70~\micron\ offset to be the average 70~\micron\ surface brightness where 160~\micron\ is close to 0~MJy/sr, such that both maps are close to zero at the same locations.
The 70~\micron\ zero-point offset, 300~MJy/sr, is $\lesssim$30\% of the typical $\sim$1000\Endash3000~MJy/sr surface brightness towards N19 and $\lesssim$10\% of the typical $\sim$3000-5000~MJy/sr towards the Bright Northern Ridge.

\citet{Hill2012A&A...542A.114H} made column density maps using SED fits to the 160, 250, 350, and 500~\micron\ dust emission maps.
We sample from these maps for additional comparison with the shorter wavelength FIR and the \cii\ and CO.
The shorter wavelength FIR is more sensitive to warmer dust compared to the longer wavelength FIR, which is more sensitive to high column density, cold dust.

Both sets of FIR column densities are presented in Table~\ref{tab:cocoldens}.

\subsection{Masses} \label{sec:mass}
We list the derived masses in Table~\ref{tab:mass}.
Atomic gas column density is derived using \cii\ and molecular gas is derived from \thco\ using the LTE method.
The Northern Cloud is defined between $\vlsr = 10\Endash21~\kms$.
The natal cloud is defined between $21\Endash27~\kms$ for the atomic gas and between $23\Endash27~\kms$ for the molecular gas.
Intermediate gas is defined between $21\Endash23~\kms$ for molecular gas only, as \cii\ emission in that interval is from line wings of Northern Cloud or natal cloud emission.
The N19 and Bright Northern Ridge masses are integrated from within appropriate masked regions, while natal cloud, intermediate gas, and Northern Cloud masses are all integrated over the entire observed \cii\ field.
The CO (\jmton{1}{0}) observations cover a wider field than the \cii, so we give molecular masses integrated over the \cii\ field only (third column) and over the entire CO observed field (fourth column).

We measure the atomic gas mass of the blueshifted clump in Figure~\ref{fig:cii_blue_shell_clump} to be $\sim$20~$M_\odot$ between 6\Endash11~\kms.

\begin{deluxetable*}{lccc}
    \label{tab:mass}
    \tablecaption{Atomic and Molecular Gas Mass.}
    \tablewidth{0pt}
    \tablehead{
        \colhead{Feature Name} & \colhead{Atomic} & \colhead{Molecular} & \colhead{Molecular} \\
        \colhead{} & \colhead{} & \colhead{} & \colhead{Entire field} \\
        \colhead{} & \colhead{($M_\odot$)} & \colhead{($M_\odot$)} & \colhead{($M_\odot$)}
    }
    \decimalcolnumbers
    \startdata
        Natal cloud & $\nexpo{4.1}{3}$ & $\nexpo{1.1}{4}$ & $\nexpo{2.1}{4}$ \\
        Bright Northern Ridge & $\nexpo{6.2}{2}$ & $\nexpo{1.2}{3}$ & ... \\
        Intermediate molecular gas & ... & $\nexpo{1.3}{4}$ & $\nexpo{2.4}{4}$ \\
        Northern Cloud & $\nexpo{1.9}{3}$ & $\nexpo{1.5}{4}$ & $\nexpo{2.9}{4}$ \\
        N19 & $\nexpo{6.5}{2}$ & $\nexpo{3.7}{3}$ & ... \\
    \enddata
    \tablecomments{Mass estimates towards features in M16 within their relevant velocity intervals.}
\end{deluxetable*}

\section{NGC~6611 Feedback Capacity} \label{sec:stars}
The NGC~6611 cluster, with a handful of O-type stars at its core, powers the M16 \hii\ region.
In order to estimate the ionizing radiation and stellar wind output of the cluster, we apply the {\tt scoby}\footnote{The code is archived at \doi{10.13016/dspace/2tgb-rqhz}; {\tt scoby} is also developed on \href{https://github.com/ramseykarim/scoby}{GitHub}} code \citep{Karim2023AJ....166..240K} to observed catalogs of NGC~6611 which we describe in Appendix~\ref{sec:appendix-stars}.
These feedback capacity estimates can be compared to observed gas motions to gauge the feedback coupling efficiency.

Table~\ref{tab:cluster-feedback} gives the total mass loss rate, mechanical energy injection, momentum transfer rate, FUV luminosity, and ionizing photon flux for the selected stars from each catalog.
Feedback capacity estimates from the two NGC~6611 catalogs agree fairly well.
The catalog compiled by \cite{Stoop2023AA...670A.108S} produces slightly larger mechanical and radiative luminosities on account of the earlier spectral types.
Binarity has a lesser effect since the companions tend to be late-O or B types.
We adopt the spectral types from the \citet{Stoop2023AA...670A.108S} catalog and the resulting feedback capacity values for the remainder of this study.

\begin{deluxetable*}{lccccc}
    \label{tab:cluster-feedback}
    \tablecaption{Stellar feedback capacity.}
    \tablewidth{0pt}
    \tablehead{
        \colhead{Selected stars} & \colhead{$L_{\rm bol}$ ($L_\odot$)} & \colhead{$L_{\rm FUV}$ ($L_\odot$)} & \colhead{$L_{\rm mech}$ ($L_\odot$)} & \colhead{$Q_0$ (s$^{-1}$)} & \colhead{$M_{\rm total}$ ($M_\odot$)}
    }
    \decimalcolnumbers
    \startdata
        NGC~6611 S & $\nexpo{2.33}{6}$ & $\nexpo{1.04}{6}$ & $\nexpo{4.06}{3}$ & $\nexpo{8.43}{49}$ & 385 \\
        NGC~6611 H & $\nexpo{1.53}{6}$ & $\nexpo{9.06}{5}$ & $\nexpo{1.94}{3}$ & $\nexpo{6.25}{49}$ & 235 \\
        N19 O9 V & $\nexpo{5.25}{4}$ & $\nexpo{3.14}{4}$ & 1.84 & $\nexpo{7.43}{47}$ & 18 \\
    \enddata
    \tablecomments{The first two rows give the estimates of the NGC~6611 cluster feedback capacity using the \cite{Stoop2023AA...670A.108S} and \cite{Hillenbrand1993AJ....106.1906H} catalogs under the filter criteria described in the text. The ``NGC~6611~S'' and ``NGC~6611~H'' rows in this table uses stars marked ``S'' and ``H'' respectively in the last column of Table~\ref{tab:stars} in Appendix~\ref{sec:appendix-stars}.
    The third row gives the estimates for the O9 V star driving N19. Columns list bolometric, FUV (6--13.6~eV), and mechanical (wind kinetic) luminosities, H-ionizing photon emission rate $Q_0$, and total stellar mass.}
\end{deluxetable*}

The total NGC~6611 cluster mass is approximately $\sim \nexpo{2}{4}~M_{\odot}$ \citep{Pfalzner2009A&A...498L..37P}, about half as massive as the Westerlund~2 cluster which powers RCW~49 \citep{Tiwari2021ApJ...914..117T, Zeidler2021AJ....161..140Z}.
Westerlund~2, over its similar $\sim$2~Myr lifespan, has emitted $\sim\nexpo{6}{51}$ ergs in its winds (or as little as 1/10$^{\rm th}$ if clumpy; \citealt{Puls2008A&ARv..16..209P}) and has a total FUV brightness of $\sim\nexpo{4}{6}~L_{\odot}$.

% We estimate extinction using the $E(B-V)$ values listed for some of the stars in the catalog by \cite{Stoop2023A&A...670A.108S}.
% Assuming an $R_V = 3.1$ reddening law (fix this, 3.57, and change Av numbers), the members with listed $E(B-V)$ measurements which are within 5\arcsec\ of the cluster center have mean and standard deviation $A_V = 3.34 \pm 0.92$ and median $A_V = 3.12$ (I have the new Av in my notes).

N19 is driven by a single O9 V star named W584 \citep{Guarcello2010A&A...521A..61G} whose type is given in the catalogs by \citet{Hillenbrand1993AJ....106.1906H} and \citet{Evans2005A&A...437..467E} (via \citealt{Stoop2023AA...670A.108S}).
This star is outside the 2.5~pc search radius from NGC~6611 and so is not double-counted in our feedback estimates.
We list its position and catalog identification in Appendix~\ref{sec:appendix-stars} and its feedback capacity estimates in Table~\ref{tab:cluster-feedback}.

\section{Energetics} \label{sec:energy}
We present estimates of various pressures and energies for both M16 and N19, all compiled in Table~\ref{tab:energetics}.

\subsection{M16} \label{sec:energ-m16}

The western lobe of the M16 cavity, which opens up west of the Pillars of Creation, is associated with shell fragments traced by \cii\ and expanding at $\pm 10$~\kms\ \wrt\ the natal cloud $\vlsr \approx 25\Endash26$~\kms\ emission.
Integrated \cii\ emission maps in Figure~\ref{fig:cii_m16shell_rgb} show these fragments located inside the infrared lobes.
We assume that the infrared lobes and the $\pm$10~\kms\ \cii\ shell fragments all trace the same large shell based on our geometrical picture of the region discussed in Section~\ref{sec:geom-m16-filament}.
These fragments are detected in \cii\ but not in CO, with the exception of the blueshifted clump (Figure~\ref{fig:cii_blue_shell_clump}).

The line-of-sight expansion velocity estimate of $\sim$10~\kms\ suggests that, over the $\sim$2~Myr lifespan of the cluster, the cavity could have expanded $\sim$20~pc from the cluster core.
Since the inclination of these shell fragments is not known, their projected line-of-sight velocities are lower limits.
Nevertheless, this matches the $\sim$20~pc on-sky projected $\pm l$ widths of the infrared lobes.
The lobes are smaller in the $\pm b$ direction, $\sim$10~pc; this is also the direction along which the GMF extends.
We suggest that the cavity is expanding at $\sim$10~\kms\ in all directions perpendicular to the $\pm b$-extended filament and is restricted along the filament.

We place an upper limit on the mass in this large cavity shell using our non-detection of \cii\ and CO towards the western cavity and an assumption about its geometry.
\cii\ is only detected towards a few fragments in this area, so we use the 1~K sensitivity limit to estimate a column density detection limit.
This limit is $\ntot \sim \expo{21}~\cmsq$ ($A_V \sim 0.5$) assuming a typical line width of 3~\kms\ and $\Tex = 60$~K based on the shell fragment emission.
To estimate surface area, we use a 3-dimensional compressed ellipsoidal cavity rather than a biconcave disc (Section~\ref{sec:geom-m16-filament}) for simplicity; the difference between the surface areas of these shapes is not important for an order-of-magnitude estimate.
We use two equal semimajor axes of 20~pc and a semiminor axis of 6~pc, and we assume that \nht\ through any part of the shell surface is $< \expo{21}~\cmsq$.
The mass in the shell must be $\lesssim 10^4~M_\odot$.
This is similar to the mass of the cluster ($\sim \nexpo{2}{4}~M_\odot$, \citealt{Pfalzner2009A&A...498L..37P}) and large gas clouds in M16 such as the Northern Cloud or the natal cloud gas near M16 (Section~\ref{sec:coldens}), and is smaller than the $\sim 10^5~M_\odot$ GMC W~37 \citep{Zhan2016RAA....16...56Z}.

We estimate the shell thickness using the $\sim$0.5~pc on-sky width of the edges of the lobes in the 8 and 160~\micron\ images.
There are multiple overlaid edges, probably from a wavy shell surface viewed edge on.
Our 0.5~pc estimate is the width of the smallest resolved edges and is consistent at both wavelengths.
Using the $\expo{21}~\cmsq$ upper limit on the column density through the foreground shell, we place an approximate upper limit $\lesssim$700~\cc\ on the average shell density.

\subsubsection{M16 Pressure and Energetics}
We estimate the pressure and energy within the M16 PDR shell, photoionized \hii\ region, and collisionally-ionized plasma.
Using our 700~\cc\ upper limit on number density, the upper limit on thermal pressure for a typical 100~K PDR is $\pressure{therm} \lesssim \nexpo{6}{4}~\Kcc$.
To properly estimate turbulent pressure, we would need a \cii\ line width towards the shell.
While we detect \cii\ emission towards the few shell fragments shown in Figure~\ref{fig:cii_m16shell_rgb}, their properties may not be representative of the entire shell and we do not detect emission towards the center of the cavity as shown in that Figure.
We also do not have observations constraining the magnetic field in the PDRs in the greater M16 region, and magnetic fields tend to provide a significant amount of nonthermal support within PDRs \citep{Pellegrini2007ApJ...658.1119P, Pellegrini2009ApJ...693..285P, Nakamura2008ApJ...687..354N, Hennebelle2019FrASS...6....5H_BfieldH2cloudreview, Karim2023AJ....166..240K}.
Therefore we assume pressure equipartition so that $P_{\rm therm} = P_{\rm turb} = P_{\rm B}$.
The total pressure in the shell would be $\pressure{tot} = (P_{\rm therm} + P_{\rm turb} + P_{\rm B}) / k_B = \lesssim \nexpo{2}{5}~\Kcc$.
Magnetic support, typically approximated as a pressure, is $P_{\rm B} = B^2 / 8 \pi$.
Under our assumptions, $B \lesssim 15~\mu {\rm G}$ in the M16 PDR shell.

The upper limit on the kinetic energy associated with the $\lesssim 10^4~M_\odot$ shell expanding at 10~\kms\ is $\lesssim 10^{49}$~erg.
We also estimate an upper limit on the thermal energy inside the shell $E_{\rm therm} < \frac{3}{2}\,k_B T\ (M/\mu m_{\rm H}) \sim \expo{48}$~erg.
Our analysis of stellar feedback capacity of NGC~6611 places its mechanical wind luminosity over the last 2~Myr at $\sim 10^{51}$~erg (or $10^{50}$~erg if winds are clumpy; \citealt{Puls2008A&ARv..16..209P}).
The column densities of the detected shell fragments are low ($\ntot \lesssim 10^{21}~\cmsq$ and only a few tens of $M_\odot$, equivalent to $\lesssim 10^{47}$~erg at 10~\kms) and so their kinetic energy is insignificant compared to the feedback capacity of NGC~6611.

The ionized gas pressure analysis by \citet{Pattle2018ApJ...860L...6P} is appropriate for the ambient ionized gas within the M16 cavity.
\citet{Hester1996AJ....111.2349H} used photoevaporative flows from the Pillars of Creation to estimate the density in the ambient (not part of the flow) ionized gas $n_{\hone} \approx 29~\cc$ within a few parsecs from the Pillars, and we assume that $n = n_\hone + n_e = 2n_\hone = 58~\cc$.
We adopt the same $T_{\rm K} = 8000$~K as \citet{Hester1996AJ....111.2349H} and \citet{Pattle2018ApJ...860L...6P}, so $\pressure{therm} = \nexpo{4.6}{5}~\Kcc$.
\citet{Higgs1979AJ.....84...77H} derives the three-dimensional turbulent velocity $\sigma_{\rm v,\,3d} \approx 12~\kms$ from H recombination lines towards the bright center of the M16 region; there is some risk that these measurements are biased towards the dense gas, but \citet{Karim2023AJ....166..240K} found it unlikely that the high-density photoevaporative flow, whose bright emission can dominate the line of sight, is turbulent.
At least one of the individual positions observed by \cite{Higgs1979AJ.....84...77H} does not include emission from the Pillars or Bright Northern Ridge, and the linewidth does not vary significantly.
We estimate turbulent support in the ionized gas $\pressure{turb} = \nexpo{2.3}{5}~\Kcc$ using the one-dimensional turbulent velocity $\sigma_{\rm v,\,1d} = \sigma_{\rm v,\,3d} / \sqrt{3} \approx 7~\kms$.
In regions close to where the bubble has broken open, pressure will be smaller because the gas is flowing out.
We neglect magnetic support in the ionized gas.

We describe in Appendix~\ref{sec:appendix-xray} the extraction and analysis of Chandra ACIS spectra between 0.5--7~keV.
From our analysis, we determine that only $E_{\rm therm} = \nexpo{3}{48}$~erg of thermal energy is contained within the extraction region, and no more than $\nexpo{8}{49}$~erg is present if the entire ellipsoidal cavity were filled with such emission.
Both figures fall short of the $\expo{51}$~erg of mechanical energy injected into the cavity by winds from NGC~6611 over the 2~Myr cluster lifetime (Table~\ref{tab:energetics}).
The M16 bubble must have vented a significant fraction of its collisionally-ionized plasma when it burst, thereby reducing the amount of stellar wind energy available to be injected into the dense gas.

%%%%%%%%%%%%%%%%%%%%%%%%%%%%%%%%%%%%%%%%%%%%%%%%%%%%%%%%%%%%%%%%%%%%%%%%%%%%%%%%%%%%% N19

\subsection{N19} \label{sec:energ-n19-topsection}
The N19 cavity has an associated foreground expanding shell moving at $\sim$4~\kms\ towards the observer projected along the line of sight.
The on-sky projected radius of N19 is $\sim$2~pc \citep{Churchwell2006ApJ...649..759C, Jayasinghe2019MNRAS.488.1141J}.
The dynamic age of the cavity is $\sim$0.5~Myr, not accounting for acceleration (Section~\ref{sec:paper2-discussion-expansion}).
This is less than the estimated $\sim$2~Myr age of NGC~6611, so either the N19 bubble has accelerated so that its average expansion velocity over time has been 1--2~\kms, or the N19 cavity was formed after NGC~6611.

\subsubsection{N19 PDR Shell}
Since we detect only the foreground half of an expanding PDR shell associated with N19, we model the PDR shell as a hemispherical shell of inner radius 1.8~pc and outer radius 2.3~pc based on its on-sky projected size in the 8~\micron\ image.
The integrated \cii\ column densities between $\vlsr = 10\Endash21~\kms$ toward the N19 ring suggest 650~$M_\odot$ of PDR gas (Table~\ref{tab:mass}).

The column density of the limb brightened edge of the PDR shell is $\ntot \sim 10^{22}~\cmsq$ based on the \cii\ line measurements.
The limb brightened path tangent to the inner sphere through a simple concentric-sphere configuration is $l = 2(r_2^2 - r_1^2)^{1/2}$ for inner and outer radii $r_1,\ r_2$; for the hemispherical shell model, we remove the factor of 2 and estimate a 1.4~pc limb brightened path through the shell.
Using the limb brightened column density through the PDR shell, we estimate number density $n \sim 2300~\cc$.
We also estimate number density using the 650~$M_\odot$ PDR mass and the estimated shell volume $\sim$13~pc$^3$, which yields $n \sim 1500~\cc$.

The \cii\ line brightness towards the N19 cavity is just a few Kelvin.
Our estimated RMS noise of 1~K suggests a column density detection threshold $\sim\nexpo{1}{21}~\cc$ for a typical 3~\kms\ linewidth and $\Tex \sim 60~K$ (same assumptions as for the M16 shell).
If we take a column density upper limit of $\ntot < \nexpo{5}{21}~\cmsq$ through the assumed 0.5~pc thick foreground shell, density is implied to be $n \lesssim 3200~\cc$.
Should the shell density be the $n \sim 1500~\cc$ estimated above, then the column density through the foreground shell is $\ntot \sim \nexpo{2.3}{21}~\cmsq$ which is consistent with our observations.
This is equivalent to $A_V \sim 1\Endash2$ \citep{Bohlin1978ApJ...224..132B_NH_AV}.

A measured $E(B-V) = 1.55$ is listed for W584 within N19; this yields $A_V = 5.5$ for assumed reddening $R_V = 3.56$ \citep{Kumar2004MNRAS.353..991K, Stoop2023AA...670A.108S}, which is $\sim$1--2 standard deviations ($\sigma_{A_V} \approx 1.1$) greater than the mean $A_V \approx 3.8$ of the NGC~6611 core early-type members listed in Appendix~\ref{sec:appendix-stars} using $A_V$ values from \citet{Stoop2023AA...670A.108S}.
The extra $A_V \sim 1\Endash2$ towards the N19 star is consistent to order-of-magnitude with the $A_V \sim 1$--2 that we estimate through the N19 foreground shell; however, this assumes (1) the cluster-core members see no additional extinction and (2) the N19 foreground shell is uniform, and integrated \cii\ intensity maps suggest that it is not uniform.
We determine that all observations are consistent with a PDR shell density $n \sim 1500~\cc$ and mass 650~$M_\odot$.

\subsubsection{N19 Molecular Gas Shell}
We use the CO and 160--500~\micron\ images to estimate a molecular gas shell inner radius $\sim$2~pc and outer radius $\sim$3~pc; the overlap with the PDR shell outer radius is not important in this approximation.
Although we do not detect a foreground or background molecular gas shell in CO, we assert that the ring is associated with the N19 shell on account of the spatial and kinematic coincidence.

Column densities calculated from \thco\ observations and integrated around the N19 ring indicate the presence of $\sim$3700~$M_\odot$ of molecular gas.
Line diagnostics described in Section~\ref{sec:cocoldens} suggest column density $\nht \sim \nexpo{1.1}{22}~\cmsq$ and density $n \sim 5600~\cc$ towards the molecular shell.

The line of sight path $\nht / n \approx 0.6$~pc is smaller than the 4.5~pc limb brightened path through the edge of a spherical shell with our adopted dimensions.
We suggest that the filling factor varies along the line of sight through the limb-brightened shell.
This could be caused by a wavy surface, so that the line of sight repeatedly passes in and out of the shell, or density variation along the line of sight.
The projected shell we see has a wavy surface, as seen in the CO and the FIR in Figure~\ref{fig:n19-fir}, and the higher resolution FIR in that image shows plenty of clumpiness not resolved in the PMO CO observations.

\subsubsection{N19 Pressure and Energetics} \label{sec:energ-n19}
A typical PDR temperature $T \sim 100$~K and our estimated PDR density $n \sim 1500~\cc$ imply a pressure $\pressure{therm} \sim \nexpo{1.5}{5}~\Kcc$.
Using a typical \cii\ line width of 3--4~\kms, the turbulent pressure must be $\pressure{turb} \sim \nexpo{5}{5}~\Kcc$.
Assuming equal turbulent and magnetic pressures, the magnetic field strength may be $\sim$40~$\mu$G.
The sum of the thermal, turbulent, and magnetic pressures in the PDR is $\pressure{tot} \sim \expo{6}~\Kcc$.
For the $650~M_\odot$ PDR shell expanding at 4~\kms, we estimate the kinetic and thermal energies to be $\sim \nexpo{1}{47}$~erg and $\sim \nexpo{1.2}{46}$~erg, respectively.

For our CO line diagnostics, we assumed a kinetic temperature $T \sim 30$~K in the molecular gas, giving thermal pressure $\pressure{therm} \sim \nexpo{1.7}{5}~\Kcc$.
CO line velocity dispersions are $\sigma \approx 1~\kms$ towards N19, so turbulent pressure $\pressure{turb} \sim \nexpo{1.6}{6}~\Kcc$ and magnetic field strength may be 70~$\mu$G.
The sum of the thermal, turbulent, and magnetic pressures in the molecular gas is $\pressure{tot} \sim \nexpo{3}{6}~\Kcc$, which is in agreement with the total pressure in the PDR given the approximate nature of these estimates.
We do not detect the expansion of the molecular shell along the line of sight, so it may be moving in the plane of the sky anywhere $\lesssim 4$~\kms, since its mass may slow expansion in those directions, and may carry $\lesssim \nexpo{6}{47}$~erg of kinetic energy.

We do not have direct measurements which probe photoionized gas conditions toward N19.
The ionized gas is expected to be in pressure equilibrium with the PDR and the molecular cloud and the derived pressure $\sim$1--$\nexpo{4}{6}~\Kcc$ of these implies an electron density $\sim$200~\cc\ in the ionized gas and therefore $\sim \expo{47}$~erg in thermal energy, disregarding any bubble volume occupied by a shock-ionized plasma phase.
This thermal energy is comparable to the kinetic energy in the shell.

\begin{deluxetable}{lccc}
    \label{tab:energetics}
    \tablecaption{Summary of M16 and N19 shell energetics.}
    \tablewidth{0pt}
    \tablehead{
        \colhead{Property} & \colhead{M16} & \colhead{N19} & \colhead{Unit}
    }
    \decimalcolnumbers
    \startdata
        $Q_0$ & $\nexpo{8.43}{49}$ & $\nexpo{7.43}{47}$ & s$^{-1}$ \\
        $E_{\rm wind}$ & $\nexpo{9.82}{50}$ & $\nexpo{1.12}{47}$ & erg \\
        Age & 2 & $\sim$0.5 & Myr \\\hline \\
        Neutral shell \\ \hline
        Mass & $< 10^4$ & 650 & $M_\odot$ \\
        $n$ & $<700$ & 1500 & cm$^{-3}$ \\
        $T$ & 100* & 100* & K \\
        $P_{\rm th}$ & $< \nexpo{6}{4}$ & $\nexpo{1.5}{5}$ & K~cm$^{-3}$ \\
        $P_{\rm turb}$ & ... & $\nexpo{5}{5}$ & K~cm$^{-3}$ \\
        $P_{\rm tot}$ & $< \nexpo{2}{5}$ & $\nexpo{1}{6}$ & K~cm$^{-3}$ \\
        $E_{\rm th}$ & $< \expo{48}$ & $\nexpo{1.2}{46}$ & erg \\
        $E_{\rm kin}$ & $< 10^{49}$ & $\nexpo{1}{47}$ & erg \\\hline \\
        Molecular shell \\ \hline
        Mass & ... & 3700 & $M_\odot$ \\
        $n$ & ... & 5600 & cm$^{-3}$ \\
        $T$ & ... & 30 & K \\
        $P_{\rm th}$ & ... & $\nexpo{1.7}{5}$ & K~cm$^{-3}$ \\
        $P_{\rm turb}$ & ... & $\nexpo{1.6}{6}$ & K~cm$^{-3}$ \\
        $P_{\rm tot}$ & ... & $\nexpo{3}{6}$ & K~cm$^{-3}$ \\\hline \\
        Photoionized gas \\ \hline
        $n$ & 58 & $\sim$200 & cm$^{-3}$ \\
        $T$ & 8000 & ... & K \\
        $P_{\rm th}$ & $\nexpo{4.6}{5}$ & ... & K~cm$^{-3}$ \\
        $P_{\rm turb}$ & $\nexpo{2.3}{5}$ & ... & K~cm$^{-3}$ \\
        $P_{\rm tot}$ & $\nexpo{6.9}{5}$ & ... & K~cm$^{-3}$ \\\hline \\
        X-ray plasma \\ \hline
        $n$ &  1.1 & ... & cm$^{-3}$ \\
        $T$ &  $\nexpo{1.7}{6}$ & ... & K \\
        $P_{\rm th}$ & $\nexpo{1.9}{6}$ & ... & \Kcc \\
        $E_{\rm th}$ (observed) & $\nexpo{3}{48}$ & ... & erg \\
        $E_{\rm th}$ (est.) & $<\nexpo{8}{49}$ & ... & erg \\
    \enddata
    \tablecomments{Top rows give star/cluster H-ionizing photon emission rate, wind energy injection over the age of the system, and the age. Gas pressures are expressed as $P / k_B$, as indicated by their units. Due to lack of relevant measurements, no molecular gas estimates are given for the M16 shell and no ionized gas estimates are given for N19. The atomic gas temperatures marked with (*) are assumed based on typical atomic PDR temperatures. See Section~\ref{sec:energy} for further details.}
\end{deluxetable}

\section{Discussion} \label{sec:discussion}
\subsection{Wind-Driven or Thermally Expanding?} \label{sec:paper2-discussion-expansion}

\subsubsection{Thermal Expansion} \label{sec:thermalexp}
\begin{figure*}
    \centering
    \includegraphics[width=0.75\textwidth]{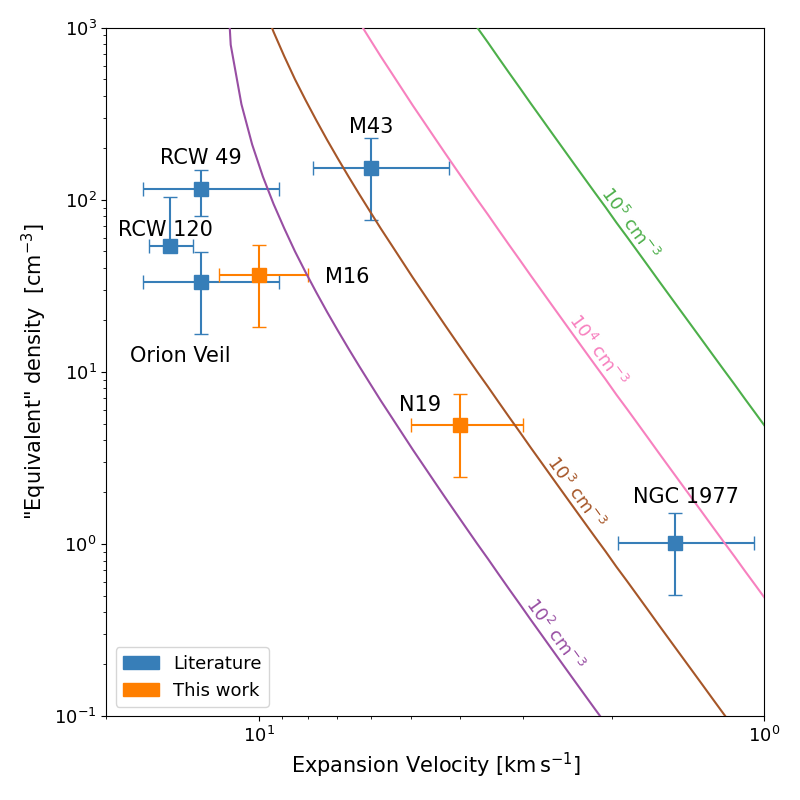}
    \caption{Equivalent shell density plotted against shell expansion velocity for a few \hii\ regions including M16 and N19 from this work. The mass of the shell is translated to an ``equivalent'' density $\frac{Q_0}{\beta_B} \frac{\mu m_{\rm H}}{M_{\rm shell}}$ where the recombination rate coefficient $\beta_B = \nexpo{2.3}{-13}~{\rm cm}^3~{\rm s}^{-1}$ \citep{Tielens2010pcim.book.....T}. We use PDR shell masses for N19 and M16, neglecting any molecular gas. The equivalent density and expansion velocity are related through the initial density of the medium. Contours show several initial, pre-bubble densities (cm$^{-3}$) \citep{Spitzer1978ppim.book.....S}. Regions which fall to the left of all contours have expanded too rapidly to be purely thermally driven and must have been wind-blown. Note that expansion velocity decreases to the right, so that over time a shell would advance to the right-hand side along a contour.}
    \label{fig:spitzer-exp}
\end{figure*}

The description of the expansion of an \hii\ region follows the discussion in Chapter 12.2 in \citet{Tielens2010pcim.book.....T}.
We consider a massive star with an ionizing photon luminosity $Q_0$ turning on in a homogeneous hydrogen cloud with density $n_0$.
Initially, the ionization front will race through the cloud ionizing a region with size $R_0$ given by the Str{\"o}mgren relation
\begin{equation} \label{eq:stromgren}
Q_0 = \frac{4\pi}{3} R_0^3 n_0^2 \beta_B
\end{equation}
with $\beta_B$ the recombination coefficient to all levels with principle quantum number $\geq 2$.
Ionization will raise the temperature to some $\expo{4}$~K and the overpressure will drive a shock into the surrounding cloud that sweeps the cloud up into a shell.
High shell density allows the shell to cool rapidly and remain thin.
As the \hii\ region expands, \hii\ gas density will drop and the ionization front will eat into the swept-up shell, increasing the mass of ionized gas.
At time $t$, the expansion velocity and radius of the \hii\ region are given by
\begin{equation} \label{eq:expansionvel}
v_s(t) = \frac{dR_s(t)}{dt} = c_{i} \Bigg(\frac{R_0}{R_s(t)}\Bigg)^{3/4}
\end{equation}
\begin{equation} \label{eq:R_t}
\frac{R_s(t)}{R_0} = \Bigg( 1 + \frac{7}{4}\frac{t}{t_0} \Bigg)^{4/7}
\end{equation}
where $t_0 = R_0 / c_{i}$ and $c_{i}$ is the sound speed in the ionized gas.
Assuming that the shell of swept up gas is very thin, the mass of the shell is given by
\begin{equation}
M_{\rm shell}(t) = \frac{4\pi}{3} R_s^3(t) \mu m_{\rm H} \big(n_0 - n(t)\big)
\end{equation}
where $n(t)$ is related to the radius of the \hii\ region through the Str{\"o}mgren relation.
Realizing that
\begin{equation}
\Bigg( \frac{R_s(t)}{R_0} \Bigg)^3 = \Bigg( \frac{n_0}{n(t)} \Bigg)^2
\end{equation}
and
\begin{equation}
\frac{v_s(t)}{c_i} = \Bigg( \frac{n(t)}{n_0} \Bigg)^{1/2}
\end{equation}
we can rewrite this as
\begin{equation}
M_{\rm shell}(t) = \frac{\mu m_{\rm H}}{\beta_B Q_0} \Bigg(\frac{c_i}{v_s(t)}\Bigg)^4 \Bigg( 1 - \Bigg(\frac{v_s(t)}{c_i}\Bigg)^2 \Bigg)
\end{equation}
We can then define an ``equivalent'' density as
\begin{equation} \label{eq:equivdens}
\mathcal{N}(t) = \frac{Q_0}{\beta_B} \frac{\mu m_{\rm H}}{M_{\rm shell}(t)}
\end{equation}
Using the ``equivalent'' density allows us to compare the results for \hii\ regions powered by stars with very different ionizing luminosities.
This results in
\begin{equation} \label{eq:equivdens_over_initdens}
\frac{\mathcal{N}(t)}{n_0} = \Bigg(\frac{v_s(t)}{c_i}\Bigg)^4  \Bigg( 1 - \Bigg(\frac{v_s(t)}{c_i}\Bigg)^2 \Bigg)^{-1}
\end{equation}
Equations~\ref{eq:equivdens} and \ref{eq:equivdens_over_initdens} relate the measured mass of the shell to the measured velocity of the expansion.
Equations~\ref{eq:expansionvel} and \ref{eq:R_t} relate the expansion velocity to the age.

As emphasized in \citet{Bisbas2015MNRAS.453.1324B} and \citet{Raga2012RMxAA..48..149R}, taking the inertia of the expanding shell into account changes the analysis slightly.
The shell velocity (Equation~\ref{eq:expansionvel}) is now given by Eq.~12 in the paper by \citep{Bisbas2015MNRAS.453.1324B}
\begin{equation} \label{eq:inertia_expansionvel}
v_s(t) = c_{i} \sqrt{\frac{4}{3} \Bigg(\frac{R_0}{R_s(t)}\Bigg)^{3/2} - \frac{\mu_i T_0}{\mu_0 T_i}}
\end{equation}
where $\mu$ and $T$ are the molecular weights and temperatures of the ionized and neutral phases.
The second term on the right-hand side is typically very small ($\nexpo{3}{-4}$ for molecular clouds).
The two terms become equal when $R_s(t)/R_0 \approx 270$ with $R_0 = 0.4\Endash1.2$~pc, for spectral types between O9 and O4 and densities of $\expo{3}~\cc$.
As an example, when $R_s(t)/R_0 = 100$, the shell velocity is $v_s(t) \approx 0.35~\kms$, close to the sound speed in molecular gas ($0.22~\kms$), and the shock driven by the expanding shell becomes very weak, the shock becomes thick and the approximations in our analysis break down.
However, this is not relevant for the \hii\ regions considered here and the second term can be ignored.
This boils down to assuming that the pressure of the surrounding gas is negligible; \hii\ region pressures are $>\nexpo{5}{6}~\Kcc$ compared to molecular cloud pressures of $\sim \nexpo{3}{4}~\Kcc$.
Hence, taking the shell inertia into account results in a change in the expansion velocity from the Spitzer solution by a factor $\sqrt{\frac{4}{3}} \approx 1.15$ (compare Equations~\ref{eq:expansionvel} and \ref{eq:inertia_expansionvel}), and the shell will expand slightly faster.
The radius of the shell is, accordingly, larger
\begin{equation}
R_s(t)/R_0 = \Bigg( 1 + \Bigg(\frac{7}{4}\Bigg)\Bigg(\frac{4}{3}\Bigg)\frac{t}{t_0} \Bigg)^{4/7} = \Bigg(1 + \frac{7}{3}\frac{t}{t_0}\Bigg)^{4/7}
\end{equation}
which differs from Equation~\ref{eq:R_t} by the factor $4/3$.
When $t \gg t_0$, the radius will be a factor $(4/3)^{4/7} \approx 1.17$ larger.
It should be noted that this does not affect the relation between equivalent density and expansion velocity (Equations~\ref{eq:equivdens} and \ref{eq:equivdens_over_initdens}).
The only effect is on the timescale of the expansion and the size of the \hii\ region.

Figure~\ref{fig:spitzer-exp} shows the relation between expansion velocity and the mass of the shell translated to this equivalent density, as defined above, for a few relevant initial densities of the medium.
The expansion velocity drops rapidly with time as the mass of the swept-up shell increases.
We note that \hii\ regions powered by earlier spectral types than approximately O9 have observed expansion speeds which exceed the sound speed in the ionized gas and these cannot be driven by thermal expansion of the ionized gas in a homogeneous medium.
The $10~\kms$ \cii\ expansion velocity observed toward the M16 shell fragments rules out thermal expansion for the M16 bubble.

We can estimate the kinematic age and expansion velocity of N19 under this model of thermal expansion.
Rearranging Equation~\ref{eq:R_t} for $t$ yields
\begin{equation}
t = \frac{4}{7}\frac{R_0}{c_i} \Bigg[\Bigg(\frac{R_s(t)}{R_0}\Bigg)^{7/4} - 1\Bigg]
\end{equation}
where the Str{\"o}mgren radius $R_0^3 = (3/4\pi)Q_0 / (n_0^2 \beta_B)$ from Equation~\ref{eq:stromgren} and the observed shell radius $R_s(t) = 2$~pc.
Recombination coefficient to the first excited electronic state $\beta_B = \nexpo{2.6}{-13}$~cm$^3$~s$^{-1}$ \citep{Tielens2010pcim.book.....T}, and we adopt $c_i = 10~\kms$.
For an O9~V star such as W584, ionizing photon emission rate $Q_0 = \nexpo{7.4}{47}$~s$^{-1}$ (Table~\ref{tab:energetics}).
The initial density may have been as high as $\expo{3}$~\cc\ in the past, and we estimate from pressure equilibrium in Section~\ref{sec:energ-n19} that the photoionized gas must be $\sim$200~\cc\ at present.
Using these numbers as limits, we estimate that the age of N19 is between 0.2--0.5~Myr assuming a purely thermally-driven expansion, with higher density yielding greater age.

Applying the same density limits to Equation~\ref{eq:expansionvel}, we find that initial densities $\lesssim$300~\cc\ produce velocities $>4~\kms$ and $n_0 = 1000~\cc$ yields $\sim$2.5~\kms\ expansion.
These are consistent with the observed $4~\kms$ \cii\ expansion signature.

\subsubsection{Wind-Driven Expansion} \label{sec:windexp}
\begin{figure*}
    \centering
    \includegraphics[width=0.75\textwidth]{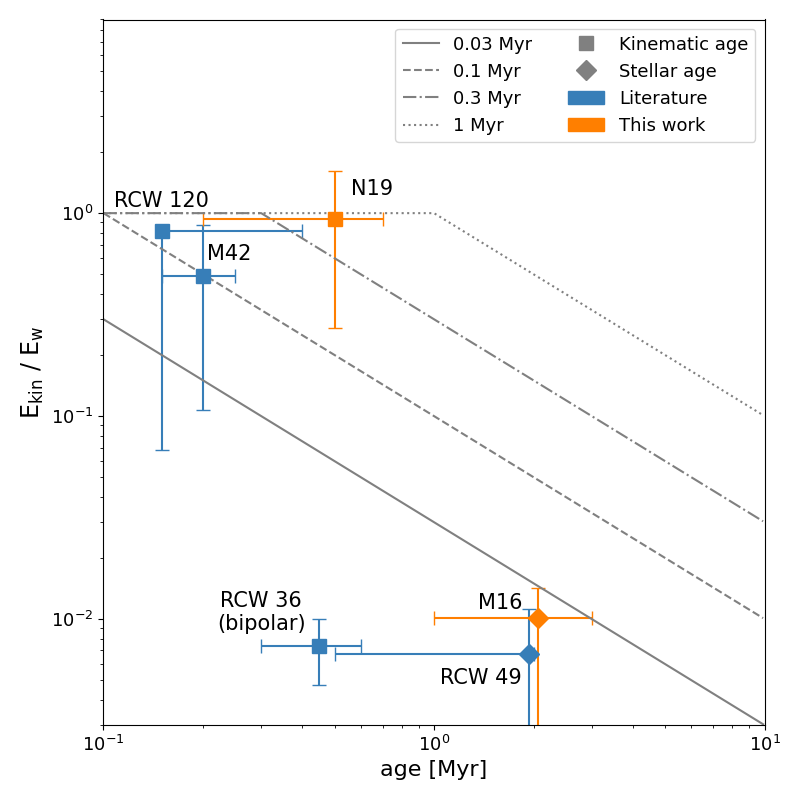}
    \caption{Ratio of the shell kinetic energy to the cluster's mechanical energy (mechanical luminosity $\times$ age) plotted against the age of the cavity for a few \hii\ regions including M16 and N19 from this work. Lines represent idealized evolutionary tracks following a burst, and subsequent decoupling of gas motion from mechanical input, at the age given in the legend. The diagram and evolution within it are discussed further in the text of Section~\ref{sec:paper2-discussion-expansion}.}
    \label{fig:energy-age}
\end{figure*}

We use the model of wind-driven expansion by \citet{Weaver1977ApJ...218..377W} to estimate kinematic age and expansion velocity of N19 and compare to the age and velocity obtained in Section~\ref{sec:thermalexp} using the thermal expansion model and the observed $4~\kms$ expansion velocity.
The expression for bubble age under wind-driven expansion given the same initial density $n_0$, mechanical luminosity from W584 $L_{\rm mech} = \nexpo{7}{33}~$erg~s$^{-1}$, and shell radius $R_s(t)$ can be rearranged from Eq.~51 by \citet{Weaver1977ApJ...218..377W}:
$$ t_6 = (27)^{-5/3} n_0^{1/3} L_{36}^{-1/3} R_s(t)^{5/3} $$
where age $t_6$ is expressed in Myr, $L_{36}$ in $\expo{36}$~erg~s$^{-1}$, $n_0$ in \cc, and $R_s(t)$ in parsecs.
Shell velocity is given by Eq.~52  by \citet{Weaver1977ApJ...218..377W}:
$$ v_s(t) = 16n_0^{-1/5} L_{36}^{1/5}t_6^{-2/5} $$
where $v_s(t)$ is in \kms.
Using the same 200--1000~\cc\ initial density range as in Section~\ref{sec:thermalexp}, we find wind-driven expansion ages of 0.4--0.7~Myr given the observed radius of N19, where higher density yields greater age.
Expansion velocities would be 2--3~\kms, with higher density yielding lower velocity.
The observed $4~\kms$ \cii\ expansion signature of N19 exceeds these wind-driven theoretical estimates, but not by much.
We do not apply this spherical model to the anisotropic expansion of M16.

The diagram in Figure~\ref{fig:energy-age} groups \hii\ regions together by the similarity between their time-integrated mechanical luminosities and the kinetic energy carried by their expanding shells.
A perfect coupling between gas motion and mechanical input from stellar wind would yield $E_{\rm kin}/E_{\rm W} = {\rm constant}$ for all time and make a horizontal track in this diagram.
\citet{Weaver1977ApJ...218..377W} finds that more than half the mechanical energy ends up as thermal energy in the collisionally-ionized plasma.
At later times, thermal conduction between the collisionally-ionized plasma and the photoionized \hii\ gas causes the mechanical-to-kinetic energy transfer to reduce over time, and \citet{Lancaster2021ApJ...914...89L, Lancaster2021ApJ...914...90L, Lancaster2024ApJ...970...18L} find that mechanical energy is further lost to radiative cooling facilitated by turbulence at the plasma-photoionized gas interface.
These inefficiencies would cause the evolutionary track to start lower ($E_{\rm kin}/E_{\rm W} < 1$) on the diagram.
Other expansion drivers not accounted for in this diagram, such as thermal pressure from the ionized gas or radiation pressure, would increase $E_{\rm kin}$.

Should the bubble burst and release pressurized gas into the environment, effectively decoupling the mechanical input from the gas motion so that $E_{\rm kin}$ is constant, the region would evolve down the diagram with time, as $E_{\rm W}$ grows linearly while $E_{\rm kin}$ does not.
The tracks in Figure~\ref{fig:energy-age} assume that $E_{\rm kin}$ remains constant after the burst.
In reality, a burst may not cause the coupling to change so abruptly, and instead change over a few $10^5$ years if limited by the sound speed in the photoionized gas.
Other expansion drivers may soften this decline, though a burst would also reduce the coupling between gas motion and thermal pressure from photoionized \hii.

The shell's total kinetic energy will decline after the burst during the momentum-conservation phase as more gas is swept up into the shell, which we do not model in the evolutionary tracks in Figure~\ref{fig:energy-age}.
The shell kinetic energy in this Figure only accounts for neutral gas mass, so any kinetic energy carried by gas which becomes photoionized on the interior of the shell is not accounted for and will also represent ``removal'' of $E_{\rm kin}$.

In summary, the evolutionary trend of bubbles on the diagram in Figure~\ref{fig:energy-age} is to the right while the bubble is intact and then to the bottom right after the bubble bursts.
Other avenues through which kinetic energy is added to the shell, such as radiation pressure, will push the region upwards on the diagram, and kinetic energy losses push the region downwards.
An earlier burst has a similar effect to an inefficient coupling phase (in which the pre-burst horizontal evolution occurs at $E_{\rm kin}/E_{\rm W} < 1$) in regions can end up lower on the diagram at earlier times.

We include in Figure~\ref{fig:energy-age} other Galactic \hii\ regions using shell masses, velocities, and ages from the literature (M42: \citealt{Pabst2020A&A...639A...2P}; RCW~36: \citealt{Bonne2022ApJ...935..171B_RCW36}; RCW~49: \citealt{Tiwari2021ApJ...914..117T}; RCW~120: \citealt{Luisi2021SciA....7.9511L}).
Kinematic shell ages are used where possible, and stellar cluster ages used otherwise.
The kinematic shell ages tend to be $\gtrsim$$\expo{5}$~yr younger than the stellar ages, which suggests that shells ``break out'' of the dense gas and expand on a multi-parsec scale after a few hundred thousand years.
In the cases of M16 and RCW~49, the asymmetric expansion or clear burst signatures make kinematic ages difficult to determine.
In M16, the anisotropic expansion renders the spherical-expansion analytical models used for N19 no more accurate than a simple constant-velocity assumption of $r / v = t = 20~{\rm pc} / 10~\kms \approx 2~{\rm Myr}$ which is consistent with the age of the cluster.
In RCW~49, \citet{Tiwari2021ApJ...914..117T} estimate a kinematic age of 0.5~Myr and conclude that it is too incongruous with the $\sim$2~Myr stellar age, so expansion must have accelerated more recently, perhaps due to increased mechanical feedback from evolved stars.
Uncertainty on the M16 shell kinetic energy is driven primarily by the uncertain shell mass, upon which we have placed the upper limit of $\expo{4}~M_\odot$, and also by uncertainty of order $\sim$2~\kms\ on the expansion velocity.
We use a 50\% uncertainty on the mass and a 1~\kms\ uncertainty on the shell expansion velocity of N19.
The uncertainty on the kinematic age of the N19 shell is dominated by the assumption of initial density $n_0$.

\subsubsection{Comparison to Observations of the Eagle Nebula}
NGC~6611 has swept up a thin, extended shell of some $10^4~M_\odot$ and imparted up to 1\% of its available wind energy to this neutral shell and up to 10\% to the plasma which fills the cavity inside the shell.
A considerable $\sim 10^5~M_\odot$ \citep{Zhan2016RAA....16...56Z} of molecular gas and a few $10^4~M_\odot$ of PDR gas remains within a few parsecs of the cluster, while the shell has expanded in some directions out to nearly 20~pc away and broken open towards others.
Wind energy is largely channeled away from the nearby reservoir of dense gas (see Table~\ref{tab:energetics}).

Shell expansion of 10~\kms\ at an age of 2~Myr is too fast for a thermally-driven Spitzer expansion according to Figure~\ref{fig:spitzer-exp} \citep{Spitzer1978ppim.book.....S}, and there is plenty of wind energy available from the cluster, so the greater M16 shell must be wind-driven.
Its location on the diagram in Figure~\ref{fig:energy-age} places it in the company of RCW~49, a region powered by another high-mass cluster with even more wind and radiative energy available, and RCW~36, a bipolar \hii\ region.
RCW~49 and RCW~36 have both broken open \citep{Tiwari2021ApJ...914..117T, Bonne2022ApJ...935..171B_RCW36}, as we suspect M16 has.

W584, the O9 V star powering the N19 bubble, has swept up a shell of $\sim 650~M_\odot$ of PDR gas and $\sim 3700~M_\odot$ of molecular gas.
Its winds have carried a modest $\sim 10^{47}$~erg over the last 0.5~Myr.
The PDR shell, expanding at 4~\kms, contains a very similar $\sim 10^{47}$~erg in kinetic energy.

We do not detect a shell expansion signature towards N19 in the molecular gas tracers, and the $3700~M_\odot$ molecular shell we see in projection must not cover the entire bubble.
The shell may expand faster toward the observer, unburdened by dense gas in this direction, than it does on the plane of the sky where we see plenty of dense gas.
The expansion velocity of the dense molecular gas is therefore unknown, but we can logically bound it below $\lesssim 4$~\kms\ as discussed in Section~\ref{sec:energ-n19}.
On the low end, it adds little kinetic energy to the total budget and the available wind energy is sufficient to drive shell expansion.
On the high end, kinetic energy is $\sim 6\times$ the available wind energy.
The molecular shell must be constraining the expansion velocity of N19 in the plane of the sky.
We place N19 in Figure~\ref{fig:spitzer-exp} under the assumption that the dense gas has hardly accelerated and kinetic energy is only imparted on the observed expanding PDR shell.

The analysis in Section~\ref{sec:paper2-discussion-expansion} of N19's observed radius and expansion velocity using the thermal pressure-driven and wind-driven expansion analytical calculations \citep{Spitzer1978ppim.book.....S, Weaver1977ApJ...218..377W} indicate that expansion is consistent with thermally-driven expansion due to pressure from the photoionized gas for an initial cloud density $\sim$300--400~\cc\ and an age of $\sim$0.5~Myr.
However, the mechanical energy input from W584 is similar to the kinetic energy in the shell and the thermal energy in the photoionized gas, so it is likely that winds contribute to the expansion.
Winds on their own are insufficient to accelerate the shell to the observed 4~\kms\ for a reasonable (not too low) initial cloud density.
Theoretical work by \citet{Raga2012RMxAA..48..199R_winds} and \citet{Ngoumou2015ApJ...798...32N} on the combined effects of ionization and stellar winds on \hii\ region expansion indicates that winds can increase the radius and expansion velocity of the bubble, though ionization tends to dominate the overall bubble structure.
N19's expansion must be driven by a combination of both thermal and wind energies.

\subsubsection{Comparison to Observations of Other \hii\ Regions}
In Figure~\ref{fig:energy-age}, the six regions are grouped into two distinct groups of 3.
One group, including N19, is typically younger and has a ratio $E_{\rm kin}/E_{\rm W}$ close to unity while the other group, which includes M16, is typically older and has a ratio $E_{\rm kin}/E_{\rm W} \sim 0.01$.
There are a handful of characteristics which might create these groups.

First, this may be an evolutionary track.
The diagonal overlays show idealized evolutionary tracks through the diagram assuming that the bubble around the star or cluster bursts and mechanical energy input decouples from the kinetic energy of the swept-up shell.
It is possible that the younger group represents intact or very recently burst regions while the older group represents evolved burst regions.

Second, RCW~120, M42, and N19 are all driven by a small number of O stars.
N19 is driven by a single O9 V, M42 is driven by $\theta^1$ Ori C which is an O7 V along with a couple early B-type stars \citep{Pabst2020A&A...639A...2P}, and RCW~120 is driven by a single O8 V \citep{Luisi2021SciA....7.9511L}.
In contrast, M16 and RCW~49 are both driven by massive clusters of multiple O-type stars; the earliest member of NGC~6611 is an O3, while RCW~49 hosts $>$30 O stars and a WR binary of two 80~$M_\odot$ members \citep{Tiwari2021ApJ...914..117T, Zeidler2015AJ....150...78Z}.
RCW~36 is driven by an O9.5 V and an O9 V \citep{Bonne2022ApJ...935..171B_RCW36}, which would associate it more with the former group.

Third, the environment in which each region formed could affect the coupling.
We show in this work that M16's cavity is shaped by the filament in which NGC~6611 was born and that N19 formed in a nearby cloud externally illuminated by NGC~6611.
M42 is a blister \hii\ region expanding out of the surface of the OMC1 core \citep{Pabst2020A&A...639A...2P}.
RCW~36 formed in a sheet and broke out of either side of the sheet almost simultaneously \citep{Bonne2022ApJ...935..171B_RCW36}.
RCW~120 moves through a cloud at 4~\kms\ and creates a bow shock and cometary \hii\ region \citep{Luisi2021SciA....7.9511L}.

Given the small number of regions, it is premature to conclude what primarily governs the energetic coupling efficiency between stellar feedback and the kinetic energy of the shell.
From this short discussion, we suggest that all three of these factors (age, central stars, and gas environment) may be important.
More studies are needed to fill in this diagram and build statistics.

\subsection{Effectiveness of Feedback on Clearing Away Gas}
Feedback from the NGC~6611 cluster was effective in clearing a cavity within its natal molecular cloud.
However, this cavity was significantly constrained by the dense filament embedded in the molecular cloud.
Though ionizing radiation reaches outside of the cluster and we find evidence that the photoionized \hii\ gas and collisionally-ionized wind-shocked plasma have escaped the region, there is plenty of dense molecular gas within a few parsecs of the cluster which has survived the intense feedback.

We consider the large-scale feedback of NGC~6611.
The $\sim$2~Myr old cluster has vented $10^4$~K and $10^6$~K gas into the wider environment.
Ionizing and FUV radiation freely escapes tens of parsecs away in multiple directions.
The feedback is directionally constrained by surrounding dense gas.
We note the slight upturn to $+b$ of the infrared lobes, which might arise from a Galactic plane density gradient acting as a second-order effect to the local molecular cloud density structure.
The vented gas might preferentially escape upwards; sensitive, degree-scale observations of diffuse X-ray emission around M16 may be able to determine whether such a gradient influences plasma distribution.

NGC~6611, after 2~Myr, has not evaporated or dispersed all the molecular gas that surrounds it.
Rather, we find a number of dense ridges which may harbor future generations of stars.
\citet{Karim2023AJ....166..240K} estimated that the Pillars of Creation would survive for another $\sim$1~Myr and may fragment and collapse to form more stars due to weakening magnetic support.
We extend this conclusion to the other dense clouds surrounding M16, such as the Spire and the Bright Northern Ridge.
There are also sections of unilluminated filament to $\pm b$, traced by 500~\micron\ and 870~\micron\ but not 70~\micron\ or 160~\micron\ in Figure~\ref{fig:m16_finder}.
N19 appears to be a younger cavity, so it may have formed when NGC~6611 was already a Myr old.
At least two active star forming regions (SFO~30/two IRAS sources in Bright Northern Ridge, and MYSO/IRAS source in Northern Cloud; Section~\ref{sec:results}) are observed towards M16, and one is embedded within dense, hot molecular gas and PDR rings inside the Bright Northern Ridge, between M16 and N19.
It is possible that compression from either/both M16 and N19 created a dense, massive clump from which this YSO formed.
We cannot determine if star formation was triggered by feedback from NGC~6611, but we conclude that NGC~6611 left plenty of dense gas intact for future generations of stars.

In numerical simulations by \citet{Gritschneder2010ApJ...723..971G, Walch2012MNRAS.427..625W, Dale2012MNRAS.424..377D, Dale2014MNRAS.442..694D_inhomogeneous}, inhomogeneous initial density structure is magnified by ionizing feedback and can mitigate the disruption of dense gas reservoirs which may become sites of continuing star formation, since ionizing photons and photoionized gas can leak out along low column density lines of sight.
Inhomogeneous initial structure comprises both turbulent or fractal structure \citep{Gritschneder2010ApJ...723..971G, Walch2012MNRAS.427..625W, Dale2014MNRAS.442..694D_inhomogeneous} as well as organized density structure such as accretion flows \citep{Dale2012MNRAS.424..377D, Dale2014ASSP...36..195D_accretionflowpillar} or filaments \citep{Fukuda2000ApJ...533..911F, Zamora-Aviles2019MNRAS.487.2200Z, Whitworth2021MNRAS.504.3156W, Watkins2019A&A...628A..21W}.

M16 lies above the Galactic plane and does not appear to have broken open towards the plane.
Feedback from M16, such as ionizing radiation, is probably not very impactful to star formation within the plane.
We predict that the inevitable sequence of supernovae will similarly expand cylindrically outwards and be constrained along the filament (bipolar HII region studies show this for sheets).
The cavity has already been evacuated, but the explosions may still impact the nearby clouds such as the one which obscures the optical $+l$ half of M16.

N19 is driven by a single O9 V star, so its capacity for large-scale feedback is limited compared to the nearby NGC~6611 powering M16.
N19's relative youth compared to M16 means that it may represent the future of either of the YSOs observed towards M16.

Protostellar outflow activity represents a crucial but less explored mechanism contributing to the complexity of stellar feedback and cloud evolution within M16.
Our \cii\ and CO observations illustrate clear interactions between stellar winds and the surrounding molecular gas, shaping distinct cavity structures.
However, protostellar jets and outflows originating from embedded YSOs such as IRAS 18152--1346 likely contribute significantly to the local kinematic structure and turbulence observed in M16.
Similar effects have been studied in the Orion Nebula by \citet{2022a_AA...663A.117K}, who demonstrated that active protostellar outflows can carve noticeable cavities into their parental clouds, while fossil outflows continue to influence local dynamics even after active accretion has ceased \citep{2022b_AA...660A.109K}.
Given the asymmetric and wide CO line profiles associated with IRAS 18152--1346 \citep{Indebetouw2007ApJ...666..321I, Xu2019A&A...627A..27X}, protostellar activity in M16 likely enhances local turbulence, modifies the ambient gas density structure, and complicates velocity patterns.
Future observations employing targeted tracers such as SiO, HCO$^{+}$, and higher-J CO transitions would be essential for quantifying these protostellar contributions and further clarifying their role within the broader framework of stellar feedback in the region.

\section{Conclusion}
We determine, by analyzing velocity-resolved \cii\ and CO line observations and wide-field continuum images, the geometry and physical conditions in the Eagle Nebula and the response of the gas to energetic feedback from the massive NGC~6611 cluster.
The dense gas structure of the region is dominated by a $\sim 10^5~M_\odot$ filament near $\vlsr \sim 25~\kms$.
From this filament the $\sim$10$^4~M_\odot$ cluster was born and displaced no more than $\sim$10$^4~M_\odot$ of gas in the form of a wind-driven shell.
The shell is highly elliptical and extends away from the filament's major axis, indicating that the filament significantly constrained the evolution of the cavity.
The shell's projected size ($\sim$20~pc radius) is consistent with its $\sim$10~\kms\ expansion velocity as traced by foreground and background shell fragments detected in \cii.
Unobstructed optical emission indicates that the shell is very thin or broken open towards the observer.

A smaller cloud hangs in front of NGC~6611 and the filament.
The smaller Northern Cloud hangs in front of NGC~6611 and hosts an independent \hii\ region cavity, N19, powered by the O9 V star W584.
The cavity is surrounded by a bright PDR shell and, displaced radially outwards, a molecular gas shell.
\cii\ observations toward N19 contain the signature of an expanding foreground PDR shell.
The Northern Cloud is also illuminated from below and behind by NGC~6611.
The few $\sim$\expo{5}~yr old N19 cavity is dynamically younger than M16 by at least one million years and is likely driven by a combination of mechanical wind energy and thermal pressure from photoionized gas.

Significant gas mass has been displaced by stellar feedback within the Eagle Nebula, but its massive filamentary skeleton holds firm against the erosion caused by multiple generations of star formation.
It shows, at multiple size scales from its collection of 0.1~pc scale pillars to the 10~pc scale filament, that the pre-existing density structure of the interstellar medium has significant influence on how energy is put back into the gas and how long dense gas remains in the presence of massive stars.
A new generation of young stellar objects has been observed in the region, some of them with the capability to drive more energetic feedback, and plenty of dense gas remains to form future generations should the conditions be right.

The growing body of observational \hii\ region literature forms a context for further analysis.
Two groups emerge in diagnostic figures: large, broken-open bubbles driven by winds from massive clusters, like M16; and small, intact bubbles driven by either winds or thermal pressure from gas ionized by one or a few late-O stars, like N19.
Continued effort to build rich, velocity-resolved far-infrared and sub-mm datasets around a diverse set of Galactic star-forming regions will clarify these relationships and their underlying mechanics.

\begin{acknowledgements}
We thank the anonymous referee for providing constructive comments that improved the quality of this article.

This work is based on observations made with the NASA/DLR Stratospheric Observatory for Infrared Astronomy (SOFIA).
SOFIA is jointly operated by the Universities Space Research Association, Inc. (USRA), under NASA contract NAS2-97001, and the Deutsches SOFIA Institut (DSI) under DLR contract 50 OK 0901 to the University of Stuttgart.
Financial support for the SOFIA Legacy Program, FEEDBACK, at the University of Maryland was provided by NASA through award SOF070077 issued by USRA.
The FEEDBACK project is supported by the BMWI via DLR, project Nos. 50 OR 2217 (FEEDBACK-plus).

This work was supported by the CRC1601 (SFB 1601 sub-project B2) funded by the DFG (German Research Foundation) -- 500700252.

This publication makes use of data from FUGIN, FOREST Unbiased Galactic plane Imaging survey with the Nobeyama 45-m telescope, a legacy project in the Nobeyama 45-m radio telescope.
The FUGIN data were retrieved from the JVO portal (\url{http://jvo.nao.ac.jp/portal/}) operated by ADC/NAOJ.

Based on data acquired with the Atacama Pathfinder Experiment (APEX).
APEX is a collaboration between the Max-Planck-Institut für Radioastronomie, the European Southern Observatory (ESO), and the Onsala Space Observatory.
Based on data obtained from the ESO Science Archive Facility with DOI(s): \doi{10.18727/archive/20}.
The ATLASGAL project is a collaboration between the Max-Planck-Gesellschaft, the European Southern Observatory and the Universidad de Chile.
It includes projects E-181.C-0885, E-078.F-9040(A), M-079.C-9501(A), M-081.C-9501(A) plus Chilean data.

The Digitized Sky Surveys were produced at the Space Telescope Science Institute under U.S. Government grant NAG W-2166. The images of these surveys are based on photographic data obtained using the Oschin Schmidt Telescope on Palomar Mountain and the UK Schmidt Telescope. The plates were processed into the present compressed digital form with the permission of these institutions.

The UK Schmidt Telescope was operated by the Royal Observatory Edinburgh, with funding from the UK Science and Engineering Research Council (later the UK Particle Physics and Astronomy Research Council), until 1988 June, and thereafter by the Anglo-Australian Observatory. The blue plates of the southern Sky Atlas and its Equatorial Extension (together known as the SERC-J), as well as the Equatorial Red (ER), and the Second Epoch [red] Survey (SES) were all taken with the UK Schmidt.

This research made use of Regions, an Astropy package for region handling \citep{Bradley2022zndo...7259631B}.
This research has made use of the VizieR catalogue access tool, CDS, Strasbourg, France (DOI: \doi{10.26093/cds/vizier}).
The original description of the VizieR service was published by \cite{2000A&AS..143...23O}.
\end{acknowledgements}

\facilities{SOFIA(upGREAT), APEX(LAsMA, LABOCA), Herschel(PACS, SPIRE), Spitzer(IRAC, MIPS), WISE, Chandra(ACIS), Purple Mountain Observatory, Nobeyama Radio Observatory, VLA}

\software{SPEX \citep{kaastra_2024_12771915},
    Radex \citep{vanderTak2007A&A...468..627V},
    astropy \citep{astropy2013, astropy2018, astropy2022},
    Spectral Cube (\url{https://doi.org/10.5281/zenodo.3558614}),
    regions (\url{https://doi.org/10.5281/zenodo.7259631}),
    pvextractor (\url{https://github.com/radio-astro-tools/pvextractor}),
    numpy \citep{numpy_harris2020array},
    scipy \citep{2020SciPy-NMeth},
    pandas \citep{pandas_McKinney_2010, pandas_McKinney_2011},
    matplotlib \citep{matplotlib_Hunter:2007}
    }

\appendix
\section{CO analysis with Radex} \label{sec:appendix-radex}
A grid of observable CO line parameters is generated by varying column density \nht\ and \htwo\ density $n$ along the two axes with $T_{\rm K} = 30$~K fixed based on derived dust temperatures and CO line intensities and full-width half-maximum line width $\Delta v = 2~\kms$ fixed based on observed line widths.
Values of \nht\ and $n$ are spaced logarithmically from ${\rm log}_{10} (\nht / \cmsq) = 20\Endash24$ and ${\rm log}_{10} (n / \cc) = 2\Endash6$ in steps of 0.05 dex; values outside these ranges are unrealistic in this context at the resolution of our observations.

\begin{figure*}
    \centering
    \includegraphics[width=\textwidth]{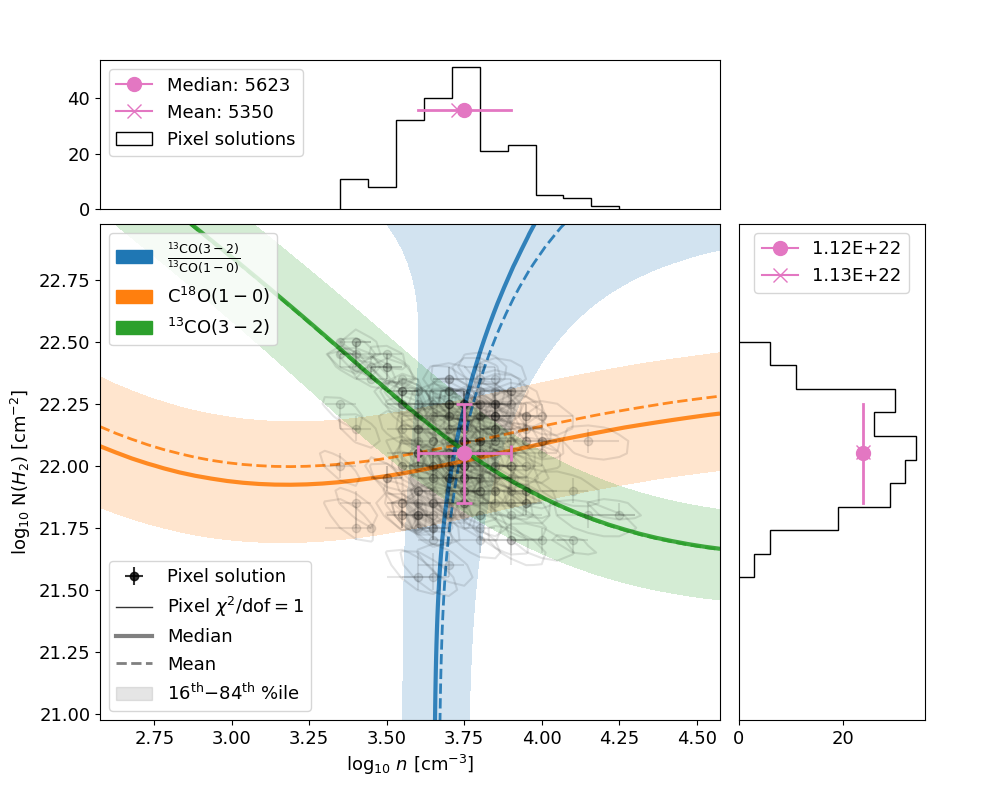}
    \caption{The CO line model grid solutions for N19. Grey points in the central panels mark individual pixel solutions and their $\chi^2 = 1$ contours, which we use as error ellipses. Each pixel's error bars are calculated from the error ellipse. The curves show the median (solid) and mean (dashed) line measurements. The shaded regions around the curves show the 16$^{\rm th}$ to 84$^{\rm th}$ percentile ranges for the measurements. Histograms along each axis show the pixel solution distributions for each parameter. Their medians (circle) and means (X) are marked in pink along with the 16$^{\rm th}$ to 84$^{\rm th}$ percentile range, which is used as uncertainty. The median solutions and uncertainties are overlaid onto the central grid in pink.}
    \label{fig:radex-n19}
\end{figure*}

\begin{figure*}
    \centering
    \includegraphics[width=\textwidth]{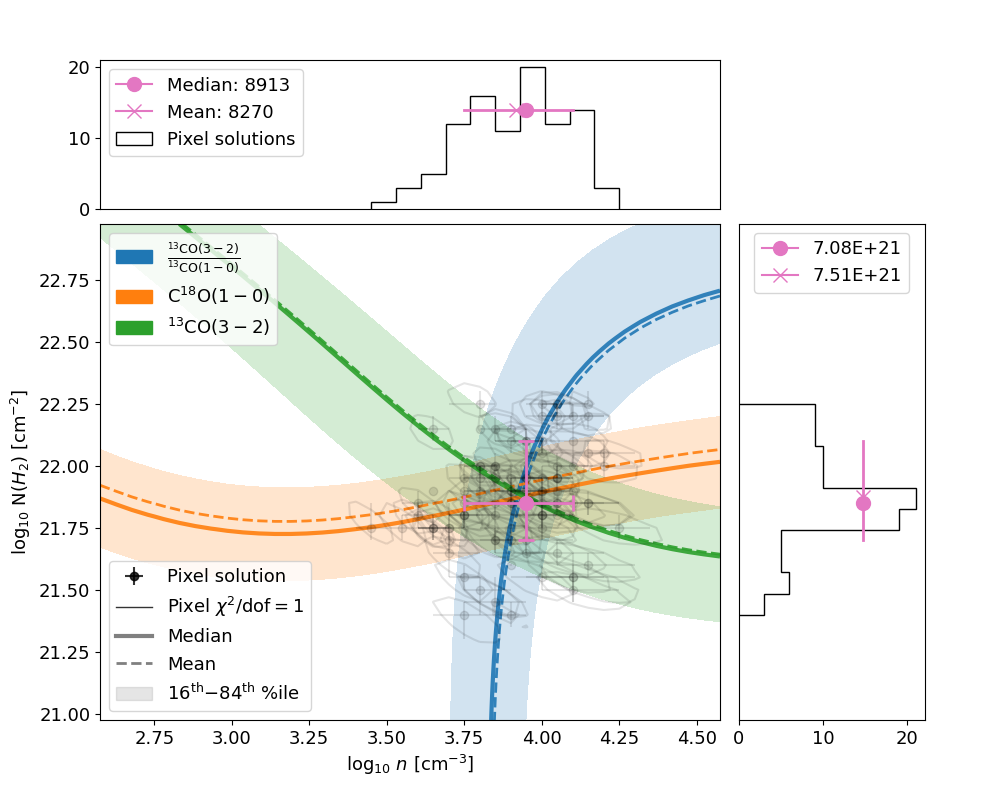}
    \caption{Same as Figure~\ref{fig:radex-n19} for the Bright Northern Ridge.}
    \label{fig:radex-bnr}
\end{figure*}

The peak intensity of an observed CO transition appears as a contour over the grid, and a collection of observations of different CO transitions is visualized as an overlay plot (see the colored lines in Figures~\ref{fig:radex-n19} and \ref{fig:radex-bnr}).
Line ratios are valuable as they are less sensitive to the absolute brightness and filling factor of a source, so we use the ratio \thcott/\thco, which is sensitive to density where both transitions are not optically thick.

\begin{figure*}
    \centering
    \includegraphics[width=0.75\textwidth]{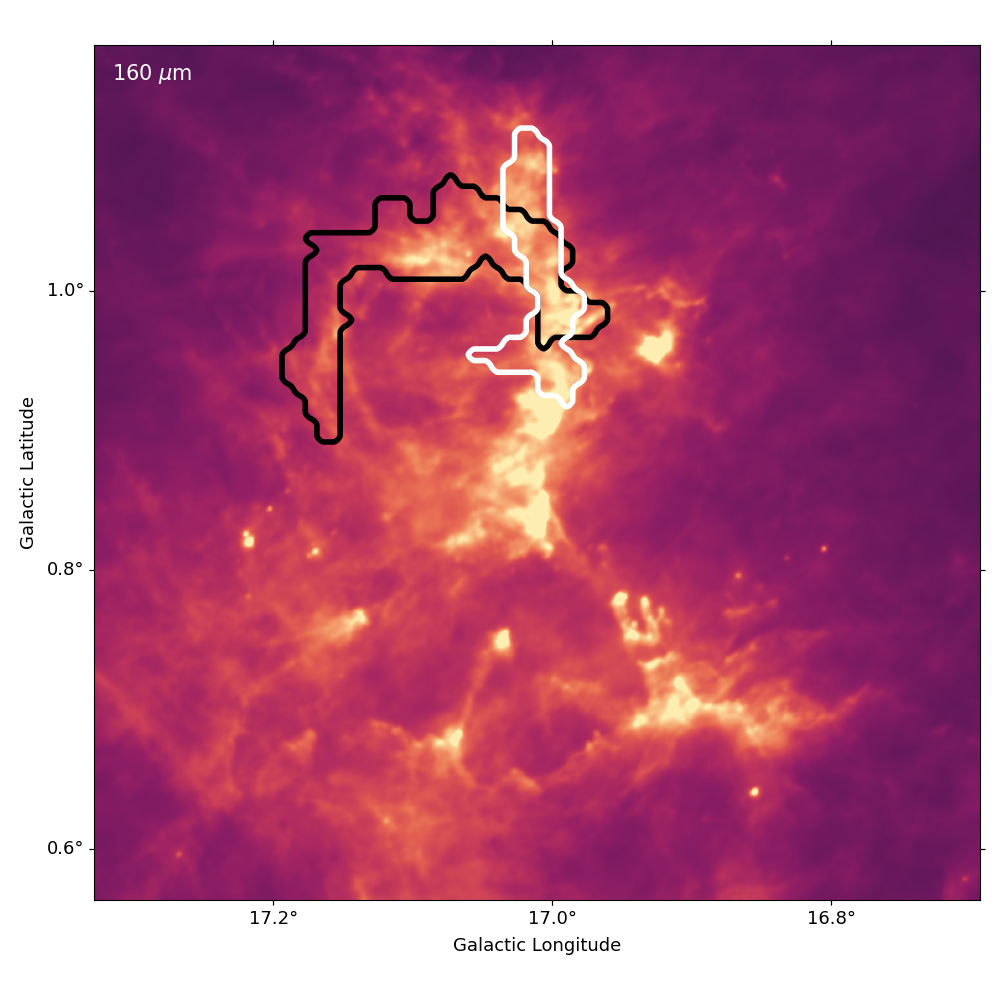}
    \caption{The N19 (black) and Bright Northern Ridge (white) masks contoured over the 160~\micron\ image.}
    \label{fig:radex-mask}
\end{figure*}

The \thcott\ observations are convolved to the 55\arcsec\ PMO beam and regridded to the 30\arcsec\ pixel grid to match the CO (\jmton{1}{0}) observations.
We create masks for N19 and the Bright Northern Ridge, the two regions for which we apply this method, using integrated \twcott\ intensities in the velocity intervals $\vlsr = 10\Endash21~\kms$ for N19 and 23--27~\kms\ for the Bright Northern Ridge and show the masked regions over the 160~\micron\ image in Figure~\ref{fig:radex-mask}.
We use peak line intensities in these intervals as the measurements and take the ratio described above.
We use the RMS noise for each line as statistical uncertainty and adopt a 10\% systematic uncertainty for each measurement to account for differences in the absolute calibrations.

We calculate the $\chi^2$ at each model gridpoint, representing an (\nht, $n$) pair, using all three measurements and their uncertainties towards each pixel (one line of sight) under the masks.
The model whose \nht\ and $n$ gridpoint holds the minimum $\chi^2$ value is considered the solution for that pixel.
We use as an error ellipse the $\chi^2 = 1$ contour and plot the results for all pixels in Figures~\ref{fig:radex-n19} and \ref{fig:radex-bnr} as black points.
We are fitting 2 parameters to 3 measurements, so we have one degree of freedom and the reduced $\chi^2$ is the same as $\chi^2$.

To obtain typical \nht\ and $n$ values for N19 and the Bright Northern Ridge, we take the median of the pixel solutions.
We use the 16$^{\rm th}$ and 84$^{\rm th}$ percentile values as the lower and upper error bounds, respectively.
This is visualized with the histograms along each axis in Figures~\ref{fig:radex-n19} and \ref{fig:radex-bnr} and their medians and error bounds marked in pink.
The solution for each region is overlaid in pink on the cluster of point solutions in the central panel.
The solutions and their uncertainties are listed in Table~\ref{tab:cocoldens}.

\section{NGC~6611 stars} \label{sec:appendix-stars}
\cite{Hillenbrand1993AJ....106.1906H} and \cite{Stoop2023AA...670A.108S} both present catalogs of O and B stars in NGC~6611.
The former is a well-established catalog which has been used in studies such as \citet{Karim2023AJ....166..240K} and the latter is a compilation of types from more recent literature \citep{Hillenbrand1993AJ....106.1906H, Evans2005A&A...437..467E, Wolff2007AJ....133.1092W, 2008A&A...489..459M, Sana2009MNRAS.400.1479S}.
We take stars from Table 3A by \cite{Hillenbrand1993AJ....106.1906H} and Table C1 by \cite{Stoop2023AA...670A.108S}.

The most massive cluster member is listed by \citet{Hillenbrand1993AJ....106.1906H} as O5 V((f*)) (No. 205) and by \citet{Stoop2023AA...670A.108S} as a binary O3.5 V((f)) + O7.5 V (No. 142); both of these refer to the same cluster member.
This and 2--3 other early O stars near the cluster core dominate the FUV radiation and stellar wind production of the entire cluster.
Notably, the more recent catalog includes a handful of stars re-identified as binaries.

We filter and analyze each catalog independently to understand the effect on the feedback capacity estimates of catalog choice.
Using the \texttt{scoby} software, we associate each star with a stellar model from the PoWR grid \citep{powr_Sander2015A&A...577A..13S, powr_Hainich2019A&A...621A..85H} so that it has an associated FUV luminosity $L_{\rm FUV}$, which is tied to its ability to illuminate PDRs, and an H-ionizing photon emission rate $Q_0$.
Both quantities are integrated from theoretical spectra.
Feedback capacities from multiple-star systems are summed over the individual stars.
We select systems with $log_{10}(L_{\rm FUV} / L_{\odot}) > 4.49$, which is equivalent to stars O9 V and brighter, and which are within a projected distance of 2.5~pc (5\arcmin) of the cluster center $(\alpha, \delta) = (274\fdg67, -13\fdg78)$ (J2000) determined by \citet{Stoop2023AA...670A.108S}.
Selected stars are listed in Table~\ref{tab:stars}.

\begin{deluxetable*}{cc ll rr c}
    % Hillenbrand_t3A_idx Stoop_tC1_idx
    \label{tab:stars}
    \tablecaption{NGC~6611 early-type cluster members.}
    \tablewidth{0pt}
    \tablehead{
        \colhead{RA} & \colhead{DE} & \colhead{H} & \colhead{S} & \colhead{H} & \colhead{S} & \colhead{Within} \\
        \colhead{(J2000)} & \colhead{(J2000)} & \colhead{Type} & \colhead{Type} & \colhead{Index} & \colhead{Index} & \colhead{Filter?}
    }
    \decimalcolnumbers
    \startdata
         \\
        M16 \\ \hline
        274.6290 & -13.7190 & O8.5 V & O8.5 V  & 161 & 15 & H, S \\
        274.6343 & -13.8134 & O8.5 V & O9 V & 166 & 10 & H, S \\
        274.6364 & -13.7533 & O5.5 V((f)) & O4 V((f)) + O7.5 V & 175 & 8 & H, S \\
        274.6502 & -13.7935 & O7 V((f)) & O6.5 V((f)) + B0-1 V & 197 & 1 & H, S \\
        274.6518 & -13.8007 & O5 V((f*)) & O3.5 V((f)) + O7.5 V & 205 & 142 & H, S \\
        274.6541 & -13.7980 & B1 III & B1 V & 210 & 17 & H \\
        274.6562 & -13.7276 & O7 III((f)) & O7 V((f)) & 222 & 35 & H, S \\
        274.6671 & -13.7552 & O7 II(f) & O7 II(f) & 246 & 13 & H, S \\
        274.6910 & -13.7753 & B0 V & O7 V + B0.5 V + B0.5 V & 314 & 6 & S \\
        274.7341 & -13.8086 & O8.5 V & O7 V + O8 V & 401 & 2 & H, S \\ \hline \\
        N19 \\ \hline
        274.5985 & -13.6078 & O9 V & O9 V & 584 & 22 & \\
    \enddata
    \tablecomments{Positions and types of stars/systems selected from the \cite{Hillenbrand1993AJ....106.1906H} (H) and \cite{Stoop2023AA...670A.108S} (S) catalogs which are within 5\arcmin\ (2.5~pc) of the cluster center and are above the FUV luminosity threshold according to their type in either catalog. Coordinates are ICRS at epoch J2000. Type and index columns are marked with ``H'' and ``S'' to indicate which catalog they reference. Indices are the ``ID'' value in Table~3A from \cite{Hillenbrand1993AJ....106.1906H} and row numbers in Table~C1 from \cite{Stoop2023AA...670A.108S}. The last column states whether each member fulfills the filter criteria in each catalog: the letter ``H'' indicates that the system is above the FUV luminosity threshold according to the type in the \cite{Hillenbrand1993AJ....106.1906H} catalog, and the letter ``S'' indicates the same for the \cite{Stoop2023AA...670A.108S} catalog type. All systems appear in both catalogs, but the type variation causes 1 system from each catalog to drop below the FUV luminosity threshold. The last row, separated with horizontal lines, lists the information for W584, the star powering N19. All of these stars, including W584, are considered NGC~6611 members.}
\end{deluxetable*}

\section{M16 X-Ray Plasma} \label{sec:appendix-xray}
\begin{figure*}
    \centering
    \includegraphics[width=\textwidth]{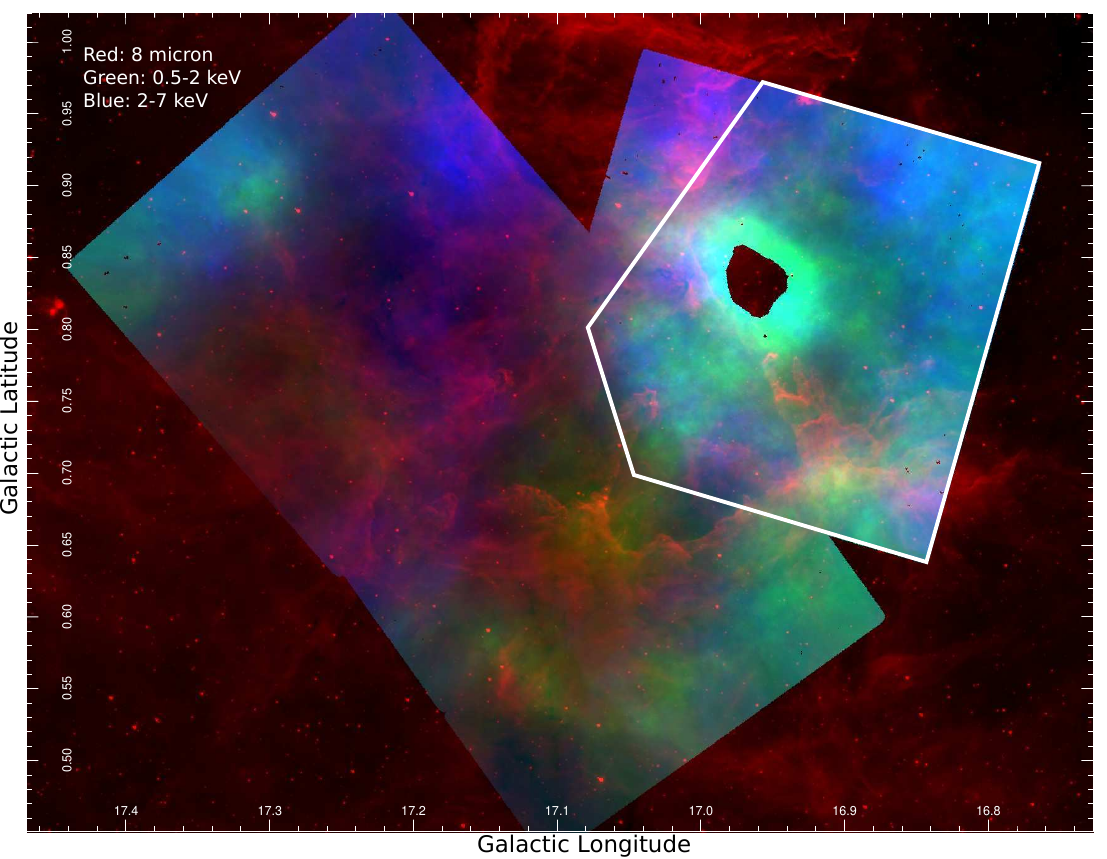}
    \caption{Color composite showing the diffuse emission extracted from Chandra ACIS X-ray observations over the 8~\micron\ image. 8~\micron\ is shown in red, the 0.5--2~keV band in green, and the 2--7~keV band in blue. The extraction region for the analyzed spectrum is outlined in white. The harder band is less susceptible to extinction. The brightest X-ray emission surrounds NGC~6611, whose immediate area is masked out since it is dominated by point sources. The X-ray-dark region to the left of NGC~6611 is spatially correlated with Northern Cloud CO emission. The X-ray emission at the top-left lies towards the inside of the eastern lobe and emission to the right of the Pillars lies inside the western lobe.}
    \label{fig:xray}
\end{figure*}

M16 was observed with Chandra ACIS between 0.5--7~keV \citep{Linsky2007ApJ...654..347L, Guarcello2010A&A...521A..61G}, and the diffuse emission spectrum separated from the point source emission and extracted by \cite{Townsley2014ApJS..213....1T}.
We analyze the spectrum extracted from the region outlined in white in Figure~\ref{fig:xray} using the SPEX package \citep{Kaastra1996uxsa.conf..411K_spex, kaastra_2024_12771915}, applying optimal binning \citep{Kaastra2016A&A...587A.151K_binning} and optimizing the model fit using C-statistics \citep{Kaastra2017A&A...605A..51K_cstat}.
The best-fit model requires one ``hard'' ($T \sim \expo{7}$--\expo{8}~K) and two ``soft'' ($T \sim \expo{6}$--\expo{7}~K) collisionally-ionized components and a foreground absorption of $\ntot \sim \nexpo{1.1}{22}~\cmsq \sim 5~A_V$, which is comparable to the optical extinction towards the cluster stars \citep{Stoop2023AA...670A.108S}.
The softest component, $T = \nexpo{1.7}{6}$~K, is associated with a volume emission measure ${\rm EM} = \nexpo{1.4}{58}~\cc$ over the 246 square arcminute (63~pc$^2$) extraction area (Figure~\ref{fig:xray}).
Following \citet{Tiwari2021ApJ...914..117T}, we assume the emitting plasma fills a sphere whose projected (circular) area is equal to the extraction area.
Electron density, assuming $n_e = n_{\rm H}$, relates to emission measure and emitting volume as ${\rm EM} = n_e n_{\rm H} V = n_e^2 V$, and we derive $n_e = 1.1~\cc$ which is similar to the values found for RCW~49 by \citet{Tiwari2021ApJ...914..117T} and RCW~36 by \citet{Bonne2022ApJ...935..171B_RCW36}.
We estimate the thermal pressure in the plasma to be $\pressure{therm} = \nexpo{1.9}{6}~\Kcc$, and, using the same volume used to derive electron density, estimate the thermal energy contained in the extraction region to be $E_{\rm therm} = \nexpo{3}{48}$~erg.
Figure~\ref{fig:xray} shows X-ray emission towards the interiors of the eastern and western lobe cavities at the edges of the ACIS fields, so we expect that plasma fills these lobes, though there would be a negative gradient in density away from the cluster.
We use the ellipsoidal cavity dimensions (two equal 20~pc semimajor, one 6~pc semiminor) to estimate the volume in the M16 cavity and place, using the thermal plasma pressure measured towards the center, an upper limit $E_{\rm therm,\,tot} < \nexpo{8}{49}$~erg on the thermal energy in plasma inside the entire cavity.
This is $\sim$1/10$^{\rm th}$ the total available mechanical wind energy from NGC~6611, though it may approach the total wind energy if winds are clumpy.

\bibliography{main}{}
\bibliographystyle{aasjournalv7}

\end{document}